\documentclass[12pt,notitlepage,a4paper]{article}
\usepackage{color,graphicx}
\usepackage{amssymb}
\usepackage{delarray,amsmath,bbm}
\usepackage[latin1]{inputenc}
\usepackage[american]{babel}
\usepackage{cite}

\def\ie{{\emph{i.e.}}}

\newcommand{\rc}{\nonumber\\}
\newcommand{\beq}{\begin{equation}}
\newcommand{\eeq}{\end{equation}}
\newcommand{\bear}{\begin{eqnarray}}
\newcommand{\eear}{\end{eqnarray}}
\numberwithin{equation}{section}

\newfont{\namefont}{cmr10}
\newfont{\addfont}{cmti7 scaled 1440}
\newfont{\boldmathfont}{cmbx10}
\newfont{\headfontb}{cmbx10 scaled 1728}
\renewcommand{\theequation}{{\rm\thesection.\arabic{equation}}}

\begin{document}
\baselineskip=15.5pt
\pagestyle{plain}
\setcounter{page}{1}

\begin{center}
\vspace{0.1in}

\renewcommand{\thefootnote}{\fnsymbol{footnote}}

\begin{center}
\Large \bf Unquenched massive flavors and flows\\ in Chern-Simons matter theories 
\end{center}
\vskip 0.1truein
\begin{center}
\bf{Yago Bea,${}^1$\footnote{yago.bea@fpaxp1.es} 
Eduardo Conde,${}^2$\footnote{econdepe@ulb.ac.be} 
Niko Jokela,${}^1$\footnote{niko.jokela@usc.es} 
and Alfonso V. Ramallo${}^1$\footnote{alfonso@fpaxp1.usc.es}} \\
\end{center}
\vspace{0.5mm}

\begin{center}\it{
${}^1$Departamento de  F\'\i sica de Part\'\i  culas \\
Universidade de Santiago de Compostela \\
and \\
Instituto Galego de F\'\i sica de Altas Enerx\'\i as (IGFAE)\\
E-15782 Santiago de Compostela, Spain}
\end{center}

\begin{center}\it{
${}^2$Physique Th\'eorique et Math\'ematique and International Solvay Institutes\\
Universit\'e Libre de Bruxelles \\ Campus Plaine - CP 231, B-1050 Bruxelles,
Belgium
}
\end{center}

\setcounter{footnote}{0}
\renewcommand{\thefootnote}{\arabic{footnote}}

\vspace{0.4in}

\begin{abstract}
\noindent 
We construct a holographic dual to the three-dimensional ABJM Chern-Simons matter theory with unquenched massive flavors. The flavor degrees of freedom are introduced by means of D6-branes extended along the gauge theory directions and delocalized in the internal space.  To find the solution we have to solve the supergravity equations of motion with the source terms introduced by the D6-branes.  The background we get is a running solution representing the renormalization group flow between two fixed points, at the IR and the UV, in both of which the geometry is of the form $AdS_4\times {\cal M}_6$, where ${\cal M}_6$ is a six-dimensional compact manifold. Along the flow, ${\cal N}=1$ supersymmetry is preserved and the flavor group is Abelian. The flow is generated by changing the quark mass $m_q$. When $m_q\to \infty$ we recover the original unflavored ABJM solution, while for $m_q\to 0$ our solution becomes asymptotically equivalent to the one found recently for massless smeared flavors.  We study the effects of the dynamical quarks as their mass is varied on different observables, such as the holographic entanglement entropy, the quark-antiquark potential, the two-point functions of high dimension bulk operators, and the mass spectrum of mesons.

\end{abstract}

\smallskip
\end{center}

\newpage

\tableofcontents

\section{Introduction}
Three-dimensional  Chern-Simons matter theories have been studied extensively in the last few years due to their rich mathematical structure and their connection with different systems of condensed matter physics. In particular, the Aharony-Bergman-Jafferis-Maldacena (ABJM) theory \cite{Aharony:2008ug} has provided a highly non-trivial example of the AdS/CFT correspondence \cite{jm,Aharony:1999ti}. The ABJM theory is an ${\cal N}=6$ supersymmetric $U(N)\times U(N)$  gauge theory with Chern-Simons levels $k$ and $-k$, coupled to matter fields which transform in the bifundamental representations $(N,\bar N)$ and $(\bar N, N)$ of the gauge group. The ABJM construction was based on the analysis of \cite{BL, Gustavsson:2007vu}, in which the supersymmetric Chern-Simons theories were proposed as the low energy theories of multiple M2-branes. When $N$ and $k$ are large the ABJM theory admits a gravity dual in type IIA supergravity in ten dimensions. The corresponding background is a geometry of the form $AdS_4\times {\mathbb C}{\mathbb P}^3$ with fluxes (see refs. \cite{Klebanov:2009sg,Klose:2010ki,Marino:2011nm, Bagger:2012jb} for reviews of different aspects of the ABJM theory).  

One of the possible generalizations of the ABJM theory is the addition of flavor fields transforming in the fundamental representations $(N,1)$ and $(1,N)$ of the gauge group. In the supergravity description these flavors can be added by considering D6-branes extended along the $AdS_4$ directions and wrapping a three-dimensional submanifold of ${\mathbb C}{\mathbb P}^3$.  
By imposing the preservation of ${\cal N}=3$ supersymmetry one finds that the D6-brane must wrap a ${\mathbb R}{\mathbb P}^3$ submanifold of the internal space \cite{Hohenegger:2009as,Gaiotto:2009tk}. When the number of flavors is small one can treat the D6-branes as probes, which is equivalent to the quenched approximation on the field theory side. This is the approach followed in refs.  \cite{Hikida:2009tp,Jensen:2010vx,Ammon:2009wc,Zafrir:2012yg} (see also \cite{Jokela:2012dw}). 

In order to go beyond the quenched approximation, one must be able to solve the supergravity equations of motion  including the backreaction induced by  the source terms generated by  the flavor branes. The sources modify the Bianchi identities satisfied by the forms and the Einstein equations satisfied by the metric.  These equations with sources are, in general, very difficult to solve, since they contain Dirac $\delta$-functions whose support is the worldvolume of the branes. In order to bypass this difficulty we will follow here the approach proposed in \cite{Bigazzi:2005md} in the context of non-critical holography, which consists of considering a continuous distribution of flavor branes. When the branes are smeared in this way there are no $\delta$-function sources  in the equations of motion and  they become more tractable. Substituting a discrete set of branes by a continuous distribution of them is only accurate if the number $N_f$ of flavors is very large. Therefore, this approach is valid in the so-called Veneziano limit \cite{Veneziano:1976wm}, in which both $N$ and $N_f$ are large and their ratio $N_f/N$ is fixed. The smearing procedure was successfully  applied to obtain supergravity solutions that include flavor backreaction in several systems \cite{CNP, conifold,D3-D7} (see \cite{Nunez:2010sf} for a detailed review and further references). 

A holographic dual to ABJM with unquenched {\it massless} flavors in the Veneziano limit was found in \cite{Conde:2011sw}. In this setup the flavor branes  fill the $AdS_4$ and are  smeared in the internal  ${\mathbb C}{\mathbb P}^3$  space in such a way that ${\cal N}=1$ supersymmetry is preserved. 
Notice, that since the flavor branes are not coincident, the flavor symmetry is $U(1)^{N_f}$ rather than $U(N_f)$. A remarkable feature  of the solution found in \cite{Conde:2011sw} is its simplicity and the fact that the ten-dimensional geometry is of the form $AdS_4\times {\cal M}_6$, where ${\cal M}_6$ is a compact six-dimensional manifold whose metric is a squashed version of the unflavored Fubini-Study metric of ${\mathbb C}{\mathbb P}^3$. The radii and squashing factors of this metric  depend non-linearly on the flavor deformation parameter 
${N_f\over N}\,\lambda$, where $\lambda=N/k$ is the 't Hooft coupling of the theory. Moreover, the dilaton is also constant and, since the metric contains an $AdS_4$ factor, the background is the gravity dual of a three-dimensional conformal field theory with flavor. Actually, it was checked in perturbation theory  in \cite{Bianchi} that the ABJM theory has conformal fixed points even after the addition of flavor. This solution captures rather well many of the effects due to loops of the fundamentals in  several observables \cite{Conde:2011sw}. Its generalization at non-zero temperature in \cite{Jokela:2012dw}
leads to  thermodynamics which pass several non-trivial tests required to a flavored black hole.

Contrary to other backgrounds with unquenched flavors, the supergravity solutions dual to ABJM with smeared sources are free of pathologies, both at the IR and the UV. This fact offers us a unique opportunity to study different flavor effects holographically in a well-controlled setup. In this paper, we will study such  effects when {\it massive} flavors are considered. The addition of massive flavors breaks conformal invariance explicitly and, therefore, the corresponding dual geometry should not contain an Anti-de Sitter factor anymore.  Actually, for massive flavors the quark mass is an additional parameter at our disposal which we can vary and see what is the effect on the geometry and observables. Indeed, let $m_q$ denote the quark mass.  In the IR limit in which $m_q$ is very large we expect the quarks to be integrated out and their effects to disappear from the different observable quantities. Thus, in the IR limit we expect to find a geometry which reduces to the unflavored ABJM background. On the contrary, when $m_q\to 0$,  we are in the UV regime and we should recover the deformed Anti-de Sitter background of 
\cite{Conde:2011sw}. The important point to stress here is that the quark mass triggers a non-trivial renormalization group flow between two fixed points and that we can vary $m_q$ to enhance or suppress  the effects due to the loops of the fundamentals.

To find the supergravity solutions along the flow, we will adopt an ansatz with brane sources in which the metric and forms are squashed as in \cite{Conde:2011sw}. By imposing the preservation of ${\cal N}=1$ supersymmetry, the different functions of the ansatz must satisfy a system of first-order BPS equations, which reduce to a single second-order master equation. The full background can be reconstructed from the solution to the master equation. 

The flavor branes corresponding to massive flavors do not extend over the full range of the holographic coordinate. Indeed, their tip should lie at a finite  distance (related to the quark mass)  from the IR end of the geometry.  Moreover, in  the asymptotic UV region, the geometry we are looking for should reduce to the one in \cite{Conde:2011sw}, since the quarks should be effectively massless in that region. Therefore, we have to solve the BPS equations without sources at the IR and match this solution with another one in which the D6-brane charge is non-vanishing and such that it reduces to the massless flavored solution of \cite{Conde:2011sw} in the deep UV. Amazingly, we have been able to find an analytic solution in the region without sources which contains a free parameter which can be tuned in such a way that the background reduces to the massless flavored geometry in the asymptotic UV. This semi-analytic solution interpolates between two different  conformal $AdS$ geometries and contains the quark mass and the number of flavors as control parameters.

With the supergravity dual at our disposal, we can study the holographic flow for different observables. The general picture we get from this analysis is the following. Let $l$ be a length scale characterizing the observable. Then, the relevant parameter to explore the flow is the dimensionless quantity $m_q\,l$.  When $m_q\,l$ is very large (small) the observable is dominated by the IR unflavored (UV massless flavored) conformal geometry, whereas for intermediate values of $m_q\,l$ we move away from the fixed points. We will put a special emphasis on the study of the holographic entanglement entropy, following the prescription of \cite{Ryu}. In particular, we study the refined entanglement  entropy for a disk proposed in \cite{Liu:2012eea}, which can be used as a central function for the F-theorem \cite{Ftheorem}.  We check the monotonicity of the refined entropy along the flow (see \cite{Casini:2012ei} for a general proof of this monotonic character in three-dimensional theories).  Other observables we analyze are the Wilson loop and quark-antiquark potential, the two-point functions of high-dimension bulk operators, and the mass spectrum of quark-antiquark bound states. 

The rest of this paper is divided into two parts. The first part starts in Section \ref{review-ABJM} with a brief review of the ABJM solution. In Section \ref{squashed_solutions} we introduce the squashed ansatz, write the master equation for the BPS geometries with sources, and classify its solutions according to their UV behavior. In Section \ref{unflavored_running} we write the analytic solution of the unflavored system that was mentioned above while, in Section \ref{interpolating} we construct solutions which interpolate between an unflavored IR region and a UV domain with D6-brane sources. The backgrounds corresponding to ABJM flavors with a given mass are studied in Section \ref{massive_flavor}. 

In the second part of the paper we study the different observables. In Section \ref{Holographic_entanglement_entropy} we analyze the holographic entanglement entropy for a disk. Section \ref{Wilson} is devoted to the calculation of the quark-antiquark potential from the Wilson loop. In Section \ref{Two-point_section} we study the two-point functions of bulk operators with high mass, while the meson spectrum is obtained in Section \ref{mesons}. Section \ref{conclu} contains a summary of our results and some conclusions. The paper is completed with several appendices with detailed calculations and extensions of the results of the main text.

\section{Review of the ABJM solution}
\label{review-ABJM}

The ten-dimensional metric of the ABJM solution in string frame is given by:
\beq
ds^2\,=\,L_{ABJM}^2\,ds^2_{AdS_4}\,+\,4\,L_{ABJM}^2\,ds^2_{{\mathbb C}{\mathbb P}^3}\,\,,
\label{ABJM-metric}
\eeq
where $ds^2_{AdS_4}$ and $ds^2_{{\mathbb C}{\mathbb P}^3}$ are respectively
 the  $AdS_4$  and ${\mathbb C}{\mathbb P}^3$ metrics. The former,  in Poincar\'e coordinates,  is given by:
\beq
ds^2_{AdS_4}\,=\,r^2\,d x_{1,2}^2\,+\,{dr^2\over r^2}\,\,,
\label{AdS4metric}
\eeq
where $d x_{1,2}^2$ is the Minkowski metric in 2+1 dimensions.  In (\ref{ABJM-metric})  $L_{ABJM}$ is the radius of the $AdS_4$ part of the metric and is given, in string units, by:
\beq
L_{ABJM}^4\,=\,2\pi^2\,{N\over k}\,\,,
\label{ABJM-AdSradius}
\eeq
where $N$ and $k$  are two integers which correspond, in the gauge theory dual, to the rank of the gauge groups and the Chern-Simons level, respectively. The ABJM background contains a constant dilaton, which can be written in terms of $N$ and $k$ as:
\beq
e^{\phi_{ABJM}}\,=\,{2L_{ABJM}\over k}\,\,=\,\,2\sqrt{\pi}\,\Big(\,{2N\over k^5}\,\Big)^{{1\over 4}}\,\,.
\label{ABJMdilaton}
\eeq
Apart from the metric and the dilaton written above, the ABJM solution of type IIA supergravity contains a RR two-form $F_2$ and a RR four-form  $F_4$, whose expressions can be written as:
\beq
F_2\,=\,2k\,J\,\,,\qquad\qquad
F_4\,=\,{3\over 2}\,k\,L_{ABJM}^2\,\Omega_{AdS_4}\,=\,{3\pi\over \sqrt{2}}\,\,
\sqrt{kN}\,\Omega_{AdS_4}\,\,,
\label{F2-F4-ABJM}
\eeq
where $J$ is the K\"ahler form of ${\mathbb C}{\mathbb P}^3$ and $\Omega_{AdS_4}$ is the volume element of the $AdS_4$ metric (\ref{AdS4metric}).  It follows from (\ref{F2-F4-ABJM}) that $F_2$ and $F_4$ are closed forms (\ie, $dF_2=dF_4=0$).

The metric of the ${\mathbb C}{\mathbb P}^3$ manifold in ({\ref{ABJM-metric}) is the canonical Fubini-Study metric.  Following the approach of \cite{Conde:2011sw}, we will regard  ${\mathbb C}{\mathbb P}^3$ as an ${\mathbb S}^2$-bundle over ${\mathbb S}^4$, where the fibration is constructed by using the self-dual $SU(2)$ instanton on the four-sphere. This representation of  
${\mathbb C}{\mathbb P}^3$ is the one obtained when it is constructed as the twistor space of the four-sphere. As in \cite{Conde:2011sw}}, this ${\mathbb S}^4$-${\mathbb S}^2$ representation will allow us to deform the ABJM background by squashing appropriately the metric and forms, while keeping some amount of supersymmetry. More explicitly, we will write $ds^2_{{\mathbb C}{\mathbb P}^3}$  as:
\beq
ds^2_{{\mathbb C}{\mathbb P}^3}\,=\,{1\over 4}\,\,\Big[\,
ds^2_{{\mathbb S}^4}\,+\,\big(d x^i\,+\, \epsilon^{ijk}\,A^j\,x^k\,\big)^2\,\Big]\,\,,
\label{CP3=S4-S2}
\eeq
where $ds^2_{{\mathbb S}^4}$ is the standard metric for the unit four-sphere, $x^i$ ($i=1,2,3$) are Cartesian coordinates that parameterize  the unit two-sphere ($\sum_i (x^i)^2\,=\,1$) and $A^i$ are the components of the non-Abelian one-form connection corresponding to the $SU(2)$ instanton. 
Let us now introduce  a specific system of  coordinates  to represent the metric (\ref{CP3=S4-S2}) and the two-form $F_2$. First of all,  let $\omega^i$ ($i=1,2,3$) be a set of  $SU(2)$ left-invariant one-forms satisfying $d\omega^i={1\over2}\,\epsilon_{ijk}\,\omega^j\wedge\omega^k$. Together with a new coordinate $\xi$, the $\omega^i$'s can be used to parameterize the metric of  the four-sphere ${\mathbb S}^4$ as:
\beq
ds^2_{{\mathbb S}^4}\,=\,
{4\over(1+\xi^2)^2}
\left[d\xi^2+{\xi^2\over4}\left((\omega^1)^2+(\omega^2)^2+(\omega^3)^2
\right)\right]\,\,,
\label{S4metric}
\eeq
where $0\le \xi<+\infty$ is a non-compact coordinate. The $SU(2)$ instanton one-forms $A^i$ can be written in these coordinates as:
\beq
A^{i}\,=\,-{\xi^2\over 1+\xi^2}\,\,\omega^i\,\,. 
\label{A-instanton}
\eeq
Let us next parameterize the $x^i$ coordinates of the unit ${\mathbb S}^2$ by two angles $\theta$ and $\varphi$ ($0\le\theta<\pi$, $0\le\varphi<2\pi$),
\beq
x^1\,=\,\sin\theta\,\cos\varphi\,\,,\qquad\qquad
x^2\,=\,\sin\theta\,\sin\varphi\,\,,\qquad\qquad
x^3\,=\,\cos\theta\,\,.
\label{cartesian_S2}
\eeq
Then, it is straightforward to demonstrate that the ${\mathbb S}^2$ part of the Fubini-Study metric can be written as:
\beq
\big(d x^i\,+\, \epsilon^{ijk}\,A^j\,x^k\,\big)^2\,=\,(E^1)^2\,+\,(E^2)^2\,\,,
\eeq
where  $E^1$ and $E^2$ are the following one-forms:
\bear
&&E^1=d\theta+{\xi^2\over1+\xi^2}\left(\sin\varphi\,\omega^1-\cos\varphi\,\omega^2\right)\,,
\rc\rc
&&E^2=\sin\theta\left(d\varphi-{\xi^2\over1+\xi^2}\,\omega^3\right)+{\xi^2\over1+\xi^2}\,
\cos\theta\left(\cos\varphi\,\omega^1+\sin\varphi\,\omega^2\right)\,.
\label{Es}
\eear
Therefore,  the  $ {\mathbb C}{\mathbb P}^3$ metric can be written in terms of the one-forms defined above as:
\beq
ds^2_{{\mathbb C}\mathbb{P}^3}\,=\,{1\over 4}\,\Big[\,ds^2_{{\mathbb S}^4}\,+\,
(E^1)^2\,+\,(E^2)^2\,\Big]\,\,.
\label{CP3-metric}
\eeq
We will now write the expression of $F_2$ in such a way that the ${\mathbb S}^4$-${\mathbb S}^2$ split structure is manifest. Accordingly, we define three new one-forms  $S^i$ $(i=1,2,3)$ as:
\bear
&&
S^1=\sin\varphi\,\omega^1-\cos\varphi\,\omega^2\,,\rc\rc
&&
S^2=\sin\theta\,\omega^3-\cos\theta\left(\cos\varphi\,\omega^1+
\sin\varphi\,\omega^2\right)\,,\rc\rc
&&
S^3=-\cos\theta\,\omega^3-\sin\theta\left(\cos\varphi\,\omega^1+
\sin\varphi\,\omega^2\right)\,.
\label{rotomega}
\eear 
Notice that the $S^i$ are just the  $\omega^i$ rotated by the  two angles $\theta$ and $\varphi$.
In terms of the forms defined in (\ref{rotomega})
 the line element  of the four-sphere is obtained by substituting $\omega^i\to S^i$ in (\ref{S4metric}). Let us next define the one-forms ${\cal S}^{\xi}$   and ${\cal S}^{i}$ as:
\beq
{\cal S}^{\xi}\,=\,{2\over 1+\xi^2}\,d\xi\,\,,\qquad\qquad
{\cal S}^{i}\,=\,{\xi\over 1+\xi^2}\,S^i \,\,,\qquad(i=1,2,3)\,\,,
\label{calS}
\eeq
in terms of which the metric of the four-sphere is  
$ds^2_{{\mathbb S}^4}=({\cal S}^{\xi})^2+\sum_i({\cal S}^{i})^2$.  Moreover, the RR two-form $F_2$  in (\ref{F2-F4-ABJM}) can be written in terms of the one-forms defined in
(\ref{Es}) and (\ref{calS}) as:
\beq
F_2\,=\,{k\over 2}\,\Big(\,E^1\wedge E^2\,-\,\big(
{\cal S}^{\xi}\wedge {\cal S}^{3}\,+\,{\cal S}^1\wedge {\cal S}^{2}\big)\,\Big)\,\,.
\label{F2-ansatz}
\eeq
The solution of type IIA supergravity reviewed above is a good gravity dual of the $U(N)_k\times U(N)_{-k}$ ABJM field theory when the $AdS$ radius $L_{ABJM}$  is large in string units and when the string coupling constant $e^{\phi}$ is small. From (\ref{ABJM-AdSradius}) and (\ref{ABJMdilaton}) it is straightforward to prove that these conditions are satisfied if $k$ and $N$ are in the range  $N^{{1\over 5}}\ll k\ll N$.

\section{Squashed solutions}
\label{squashed_solutions}

Let us consider the deformations of the ABJM background which preserve the ${\mathbb S}^4$-${\mathbb S}^2$ splitting. These deformed backgrounds will solve the equations of motion of type IIA supergravity (with sources) and will preserve at least two supercharges. We will argue below that some of these backgrounds are dual to Chern-Simons matter theories with fundamental massive flavors.

The general ansatz for the ten-dimensional metric of our solutions in string frame  takes the form:
\begin{equation}
d s^2_{10}=h^{-1/2}d x^2_{1,2}+h^{1/2}\left[d r^2+e^{2f}d s^2_{\mathbb{S}^4}+e^{2g}\left(\left(E^1\right)^2+\left(E^2\right)^2\right) \right]\,,
\label{metric_ansatz}
\end{equation}
where the warp factor $h$ and the functions $f$ and  $g$ depend on the holographic coordinate $r$. Notice that $f$ and $g$ determine the sizes of the ${\mathbb S}^4$ and ${\mathbb S}^2$ within the internal manifold.  Actually, their difference $f-g$ determines the squashing of the 
${\mathbb C}{\mathbb P}^3$  and will play an important role in characterizing our solutions. We will measure this squashing by means of the function $q$, defined as:
\beq
q\,\equiv\,e^{2f-2g}\,\,.
\eeq
Clearly, the ABJM solution has $q=1$. A departure from this value would signal a non-trivial deformation of the metric. Similarly, the RR two- and four-forms will be given by:
\begin{align}
F_4&=K\,d^3x\wedge d r\,,\\
F_2&=\frac{k}{2}\left(E^1\wedge E^2-\eta\left({\cal S}^{\xi}\wedge{\cal S}^3+{\cal S}^1\wedge{\cal S}^2\right)\right)\,,
\label{eqn:F2}
\end{align}
where $k$ is a constant and  $K=K(r)$, $\eta=\eta(r)$ are new functions. The background  is also endowed with  a dilaton $\phi=\phi(r)$. As compared with the ABJM value (\ref{F2-ansatz}), the expression of $F_2$ in our ansatz contains the function $\eta(r)$ which generically introduces an asymmetry between the ${\mathbb S}^4$ and ${\mathbb S}^2$  terms.  Moreover, when $\eta\not=1$ the two-form $F_2$ is no longer closed and the corresponding Bianchi indentity is violated. Indeed, one can check that:
\bear
&&dF_2\,=\,-{k\over 2}\,\,(\eta-1)\,\,
\Big[\,
E^1\wedge ({\cal S}^{\xi}\wedge {\cal S}^{2}\,-\,{\cal S}^1\wedge {\cal S}^{3}\big)\,+\,
E^2\wedge ({\cal S}^{\xi}\wedge {\cal S}^{1}\,+\,{\cal S}^2\wedge {\cal S}^{3}\big)\,
\Big]\,\,-\rc\rc
&&\qquad\qquad\qquad\qquad
-{k\over 2}\,\,\eta'\,\,dr\wedge 
\Big({\cal S}^{\xi}\wedge {\cal S}^{3}\,+\,{\cal S}^1\wedge {\cal S}^{2}\Big)\,\,.
\label{massive-Omega}
\eear
The violation of the Bianchi identity of $F_2$ means that we have D6-brane sources in our model. 
Indeed, since $F_2=* \,F_8$, if  $dF_2\not= 0$ then  the Maxwell equation of $F_8$ contains a source term, which is due to the presence of D6-branes since the latter are electrically charged with respect to $F_8$. The charge distribution of the D6-brane sources is determined by the function $\eta$, which we  will call the profile function.

The function $K$ of the RR four-form can be related to the other functions of the ansatz by using 
its equation of motion $d*F_4=0$ and  the flux quantization condition for the integral of $*F_4$ over the internal manifold. The result is \cite{Conde:2011sw}:
\beq
K\,=\,3\pi^2\,N\,h^{-2}\,e^{-4f-2g}\,\,,
 \label{K-N}
 \eeq
where the integer $N$ is identified with the ranks of the gauge groups in the gauge theory dual (\ie, with the number of colors).

It is convenient  to introduce a new radial variable $x$, related to $r$ through the differential equation:
\beq
x\,{dr\over dx}\,=\,e^{g}\,\,.
\label{r-x-diff-eq}
\eeq
From now on, all functions of the holographic variable  are considered as functions of $x$, unless otherwise specified. 
The ten-dimensional metric in this new variable takes the form:
\beq
ds^2_{10}\,=\,h^{-{1\over 2}}\,dx^2_{1,2}\,+\,h^{{1\over 2}}\,
\Big[\,e^{2g}\,\,{dx^2\over x^2}\,+\,e^{2f}\,ds_{{\mathbb S}^4}^2\,+\,
e^{2g}\,\Big(\,\big(E^1\big)^2\,+\,\big(E^2\big)^2\Big)\,\Big]\,\,.
\eeq

It was shown in \cite{Conde:2011sw} that the background given by the ansatz written above preserves ${\cal N}=1$ supersymmetry in three dimensions if the functions satisfy a system of first-order differential equations.  It turns out that this BPS system can be reduced to a unique second-order differential equation for a particular combination of the functions of the ansatz. The details of this reduction are given in Appendix \ref{BPS}. Here we will just present the final result of this analysis. First of all, let us define the function $W(x)$ as:
\beq
W(x)\,\equiv\,{4\over k}\,h^{{1\over 4}}\,e^{2f-g-\phi}\,x\,\,.
\label{W_definition}
\eeq
Then, the BPS system can be reduced to the following second-order non-linear differential equation for $W(x)$:
\beq
W''\,+\,4\eta'\,+\,(W'+4\eta)\,\Bigg[{W'+10\eta\over 3W}\,-\,
{W'+4\eta+6\over x (W'+4\eta)}\Bigg]\,=\,0\,\,.
\label{master_eq_W}
\eeq
We will refer to (\ref{master_eq_W}) as the master equation and to $W(x)$ as the master function. Interestingly, the BPS equations do not constrain the profile function $\eta$. Therefore,  we can choose $\eta(x)$ (which will fix the type of supersymmetric sources of our system) and afterwards we can solve (\ref{master_eq_W}) for $W(x)$. Given $\eta(x)$ and $W(x)$ one can obtain the other functions that appear in the metric. Indeed, as proved in Appendix \ref{BPS}, $g(x)$ and $f(x)$ are given by:
\bear
&&e^{g(x)}\,=\,{x\over W^{{1\over 3}}}\,\exp\Big[{2\over 3}\int^x\,{\eta(\xi)d\xi\over W(\xi)}\Big]
\,\,,\rc\rc
&&e^{f(x)}\,=\,\sqrt{{3x\over W'+4\eta}}\,\,W^{{1\over 6}}\,
\exp\Big[{2\over 3}\int^x\,{\eta(\xi)d\xi\over W(\xi)}\Big]\,\,,
\label{g-f-W}
\eear
while the warp factor $h$ can be written as:
\beq
h(x)\,=\,4\pi^2\,{N\over k}\,e^{-g}\,
(W'+4\eta)\Bigg[\,\int_{x}^{\infty}\,
{\xi\,e^{-3 g(\xi)}\over W(\xi) ^2}
\,d\xi\,+\,\beta\,\Bigg]\,\,
\,\,,
\label{warp_factor_x}
\eeq
where $\beta$ is a constant that determines the behavior of $h$ as $x\to\infty$ ($\beta=0$ if we impose that  $h\to 0$ as $x\to\infty$). Finally,  the dilaton is given by:
\beq
e^{\phi(x)}\,=\,{12\over k}\,
{x \,h^{{1\over 4}}\over W^{{1\over 3}} (W'+4\eta)}\,\exp\Big[{2\over 3}\int^x\,{\eta(\xi)d\xi\over W(\xi)}\Big]\,\,.
\label{dilaton_x}
\eeq
From the expression of $f$ and $g$ in (\ref{g-f-W})
it follows   that the squashing function $q$ can be written in terms of the master function $W$ and its derivative as:
\beq
q\,=\,{3 W\over x(W'+4\eta)}\,\,.
\label{q_W_Wprime}
\eeq

\subsection{Classification of solutions}
Let us study the behavior of the solutions of the master equation in the UV region $x\to\infty$. This analysis will allow us to have a classification of the different solutions. We will assume that the profile function $\eta(x)$ reaches a constant value as $x\to\infty$, and we will denote:
\beq
\lim_{x\to\infty} \eta(x)\,=\,\eta_0\,\,.
\eeq
Let us restrict ourselves to the case in which $\eta_0\not=0$. We will  assume that $W(x)$ behaves for large $x$ as:
\beq
W(x)\,\approx A_0\,x^{\alpha}\,\,,
\qquad\qquad x\to\infty\,\,,
\label{Asym_W}
\eeq
where $A_0$ and $\alpha$ are constants. 
It is easy to check that this type of behavior is consistent only when  the exponent $\alpha\ge 1$ or, in other words, when $W(x)$ grows at least as a linear function of $x$ when  $x\to\infty$. 

We will also characterize the different solutions by the asymptotic value of the squashing function $q$, which determines the deformation of the internal manifold in the UV. Let us denote
\beq
q_0\,=\,\lim_{x\to\infty} q(x)\,\,. 
\eeq
It follows from (\ref{q_W_Wprime}) that the asymptotic value of the squashing function and that of the profile function are closely related.  Actually, this relation depends on whether the exponent $\alpha$ in (\ref{Asym_W}) is strictly greater or equal to one. Indeed, plugging  the asymptotic behavior (\ref{Asym_W}) in (\ref{q_W_Wprime}) one immediately proves that:
\beq
q_0\,=\,
\begin{cases}
{3\over \alpha}\,\,,\qquad\qquad
{\rm for\,\,} \alpha>1 \,\,,\cr\cr
{3A_0\over A_0+4\eta_0}\,\,,\qquad
{\rm for\,\,}\alpha=1\,\,.
\end{cases}
\label{q_0_alpha}
\eeq
This result indicates that we have to study separately the cases $\alpha>1$ and $\alpha=1$. As we show in the next two subsections these two different asymptotics correspond to two qualitatively different types of solutions.

\subsubsection{The asymptotic  $G_2$ cone}
\label{G2-cone}

Let us assume that the master function behaves as in (\ref{Asym_W}) for some $\alpha>1$.  By plugging this asymptotic form in the master equation (\ref{master_eq_W}) and keeping the leading terms  as $x\to\infty$, one readily verifies that the coefficient $A_0$ is  not constrained and that the exponent $\alpha$ takes the value:
\beq
\alpha\,=\,{3\over 2}\,\,.
\eeq
Therefore, it follows from (\ref{q_0_alpha}) that the asymptotic squashing is:
\beq
q_0=2\,\,.
\eeq
Let us evaluate the asymptotic form of all the functions of the metric. From  (\ref{g-f-W}), we get, at leading order:
\beq
e^{g}\approx C\,\sqrt{x}\,\,,
\eeq
where $C$ is a constant of integration. Moreover, since $q_0=2$, the asymptotic value of the function $f$ is:
\beq
e^{f}\approx \sqrt{2}\, C\,\sqrt{x}\,\,.
\eeq
Let us now evaluate the warp factor $h$ from  (\ref{warp_factor_x}). Clearly, we have to compute the integral:
\beq
\int_{x}^{\infty}\,
{\xi\,e^{-3 g(\xi)}\over W(\xi)^2}
\,d\xi\sim x^{-{5\over 2}}\,\,,
\label{asymp_integral_h}
\eeq
which vanishes when $x\to\infty$. Therefore, 
by choosing  the constant $\beta$ in (\ref{warp_factor_x}) to be non-vanishing we can neglect the integral (\ref{asymp_integral_h}) and, since $e^{g}\,W'\to{\rm constant}$, then the warp factor $h$ becomes also a constant when $x\to\infty$. To 
clarify the nature of the asymptotic metric, let us change variables, from $x$ to a new radial variable $\rho$, defined as $\rho=2C\sqrt{x}$. Then, after some constant rescalings of the coordinates the metric becomes:
\beq
ds^2_{10}\approx dx^2_{1,2}\,+\,ds^2_7\,\,,
\eeq
where $ds^2_7$ is:
\beq
ds^2_7= d\rho^2\,+\,{\rho^2\over 4}\,
\Big[2 ds^2_{{\mathbb S}^4}\,+\,
\left(E^1\right)^2+\left(E^2\right)^2\Big]\,\,.
\label{G2_cone_metric}
\eeq
The metric (\ref{G2_cone_metric}) is a Ricci flat cone with $G_2$ holonomy, whose principal orbits at fixed $\rho$ are ${\mathbb C}{\mathbb P}^3$ manifolds with a squashed Einstein metric. In the asymptotic region of large $\rho$ the line element (\ref{G2_cone_metric})  coincides with the metric of the resolved  Ricci flat cone found in \cite{G2cone}, which was constructed from the bundle of  self-dual two-forms over ${\mathbb S}^4$ and is topologically ${\mathbb S}^4\times {\mathbb R}^3$ (see \cite{Atiyah:2001qf} for applications of this manifold to the study of the dynamics of M-theory).

\subsubsection{The asymptotic  $AdS$ metric}
\label{AdS_UV}

Let us now explore the second possibility for the exponent $\alpha$ in (\ref{Asym_W}), namely $\alpha=1$. In this case the coefficient $A_0$ cannot be arbitrary. Indeed, by analyzing the master equation as $x\to\infty$ we find that $A_0$ and $\eta_0$ must be related as:
\beq
A_0^2\,+\,(9-\eta_0)\,A_0\,-\,20\,\eta_0^2\,=\,0\,\,.
\label{A_0_eq}
\eeq
On the other hand, $A_0$ should be related to the asymptotic squashing $q_0$ as in (\ref{q_0_alpha}), which we now write as:
\beq
A_0\,=\,{4q_0\over 3-q_0}\,\,\eta_0\,\,.
\label{A0_q0}
\eeq
By plugging (\ref{A0_q0}) into (\ref{A_0_eq}) we arrive at the following quadratic relation between $q_0$ and $\eta_0$:
\beq
q_0^2\,-3(1+\eta_0)\,q_0\,+\,5\eta_0\,=\,0\,\,.
\label{q0_eta0}
\eeq
Using this equation we can re-express $A_0$ as:
\beq
A_0\,=\,{q_0(\eta_0+q_0)\over 2-q_0}\,\,.
\eeq
Moreover, we can solve (\ref{q0_eta0}) for $q_0$ and obtain the following  two possible  asymptotic squashings in terms of $\eta_0$:
\beq
q_0^{\pm}\,=\,{1\over 2}\,\Big[\,3+3\eta_0\,\mp\,\sqrt{9\eta_0^2-2\eta_0+9}\,\Big]\,\,.
\label{q_0-pm}
\eeq
Thus, there are two possible branches in this case, corresponding to the two signs in (\ref{q_0-pm}). In this paper we will only consider the $q_0^{+}$ case, since this is the one which has the same asymptotics as the ABJM solution  when there are no D6-brane sources. Indeed, (\ref{q_0-pm}) gives $q_0^{+}=1$ when $\eta_0=1$, which means that the internal manifold in the deep UV is just the un-squashed ${\mathbb C}{\mathbb P}^3$ (when $\eta_0=1$ there are no D6-brane sources in the UV, see (\ref{massive-Omega})).

Let us now study in detail the  asymptotic metric in the UV  corresponding to the $x\to\infty$ squashing  $q_0^{+}$ (which from now on we simply denote as $q_0$). By substituting  $\eta\to \eta_0$ and $W\to A_0\, x$ in (\ref{g-f-W}) and performing the integral, we get:
\beq
e^{g(x)}\,\approx C\,x^{{2\over 3}\,\big(1+{\eta_0\over A_0}\big)}\,\,,
\eeq
where $C$ is a constant of integration. Using (\ref{A0_q0}) this expression can be rewritten as:
\beq
e^{g(x)}\,\approx C\,x^{{1\over b}}\,\,,
\eeq
where  $b$ is given by:
\beq
b\,=\,{2\,q_0\over q_0+1}\,\,.
\label{b_new}
\eeq
The remaining functions of the metric can be found in a similar way. We get for $f$ and $h$ the following asymptotic expressions:
\beq
 e^{f(x)}\,\approx\,C\,\sqrt{q_0}\,x^{{1\over b}}\,\,,
 \qquad\qquad
 h(x)\,\approx\,4\pi^2\,{N\over k}\,{2-b\over 
 C^4\,A_0}\, \,{1\over x^{{4\over b}}}\,\,.
 \eeq
Let us write the above expressions in terms of the original $r$ variable, which can be related to $x$ by integrating the equation:
\beq
x\,{dr\over dx}\,=\,e^{g}\,\approx\,C\,x^{{1\over b}}\,\,.
\eeq
For large $x$ we get:
\beq
r\approx b\,C\,x^{{1\over b}}\,\,,
\eeq
and the functions $g$, $f$, and $h$ can be written  in terms of $r$ as:
\beq
e^{g}\approx {r\over b}\,\,,\qquad\qquad
e^{f}\approx {\sqrt{q_0}\over b}\,r\,\,,\qquad\qquad
h\approx {L_0^4\over r^4}\,\,,
\eeq
where $L_0$ is given by:
\beq
L_0^4\,=\,4\pi^2\,{N\over k}\,{(2-b)\,b^4\over A_0}\,\,.
\eeq
In terms of the asymptotic values $\eta_0$ and $q_0$,  $L_0$ can be written as:
\beq
L_0^4\,=\,128\pi^2\,{N\over k}\,
{(2-q_0)\,q_0^3\over (\eta_0+q_0)\,(q_0+1)^5}\,\,.
\label{L0_explicit}
\eeq
Using these results we find that the asymptotic metric takes the form:
\beq
ds^2\,\approx\,L_0^2\,\,ds^2_{AdS_4}\,+\,
{L_0^2\over b^2}\,\Big[\,q_0\,ds^2_{{\mathbb S}^4}\,+\,
(E^1)^2\,+\,(E^2)^2\,\Big]\,\,,
\label{metric-AdS-asymp}
\eeq
where we have rescaled the Minkowski  coordinates as $x^{\mu}\to L_0^2\,x^{\mu}$. The metric (\ref{metric-AdS-asymp}) corresponds to the product of $AdS_4$  space with radius $L_0$ and a squashed ${\mathbb C}{\mathbb P}^3$. The parameter $b$ will play an important role in the following. Its interpretation is rather clear from  (\ref{metric-AdS-asymp}): it represents the relative squashing of the 
${\mathbb C}{\mathbb P}^3$ part of the asymptotic metric with respect to the
$AdS_4$  part.

It is now straightforward to show that in the UV 
the dilaton reaches a constant value $\phi_0$, related to $q_0$ and $\eta_0$ as:
\beq
e^{\phi_0}\,\approx\,4\sqrt{2\pi}\,\,
\Bigg[\,
{(2-q_0)^{5}\over q_0\,(q_0+1)\,(\eta_0+q_0)^5}
\Bigg]^{{1\over 4}}\,\,
\Bigg({2N\over k^5}\Bigg)^{{1\over 4}}\,\,,
\label{dilaton-AdS_asymp}
\eeq
while the RR four-form approaches the value:
\beq
F_4\,\approx\,12\sqrt{2}\,\pi\,
\Bigg[\,
{q_0^5\,(\eta_0+q_0)\over 
(2-q_0)\,(q_0+1)^7}\Bigg]^{{1\over 2}}
\,\,
\sqrt{k\,N}\,\,\,\Omega_{AdS_4}\,\,,
\label{F4-AdS_asymp}
\eeq
where  $\Omega_{AdS_4}$ is the volume element of $AdS_4$. 

Interestingly, when the profile function $\eta$ is constant and equal to $\eta_0$, the metric, dilaton, and forms written above solve the BPS equations not only in the UV, but also in the full domain of the holographic coordinate. Equivalently, $W=A_0\,x$ is an exact solution to the master equation (\ref{master_eq_W}) if $\eta$ is constant and equal to $\eta_0$ and $A_0$ is given by (\ref{A0_q0}). Actually, when $\eta_0=1$ one can check that $q_0=b=1$ and the asymptotic background becomes the ABJM solution ($W=2x$ for this case). Moreover, when $\eta=\eta_0>1$ the background corresponds\footnote{
Notice that the  expression for $b$  written in (\ref{b_new}) is equivalent to the one obtained in \cite{Conde:2011sw}, namely:
\beq
b\,=\,{q_0(\eta_0+q_0)\over 2(q_0+\eta q_0-\eta_0)}\,\,.
\label{b-squashing}
\nonumber
\eeq
In order to check this equivalence it is convenient to use the following relation between $q_0$ and $\eta_0$:
$
q_0+\eta_0 q_0-\eta_0\,=\,(q_0+1)\,(\eta_0+q_0)/4\,\,.
$\
}  to the one found in \cite{Conde:2011sw} for the ABJM model  with unquenched massless flavors, if one identifies $\eta_0$ with $1+{3N_f\over 4k}$, where $N_f$ is the number of flavors.

The main objective of this paper is the construction of solutions which interpolate between the $\eta=1$ ABJM background in the IR and the $AdS_4$ asymptotics with $\eta_0>1$ in the UV. Equivalently, we are looking for backgrounds such that the squashing function $q(x)$ varies from the value $q=1$ when $x\to 0$ to $q=q_0>1$ for $x\to\infty$. These backgrounds naturally correspond to gravity duals of Chern-Simons matter models with massive unquenched flavors. Indeed, in such models, when the energy scale is well below the quark mass the fundamentals are effectively integrated out and one should recover the unflavored ABJM model. On the contrary, if the energy scale is large enough the quarks can be taken to be massless and the corresponding gravity dual should match the one found in \cite{Conde:2011sw}. In  the next section we present a one-parameter family of analytic unflavored solutions which coincide with the ABJM background in the deep IR and that have a squashing function $q$ which grows as we move towards the UV. In Sections \ref{interpolating} and \ref{massive_flavor} we show that these running solutions can be used to construct the gravity duals to massive flavor that we are looking for.

\section{The unflavored system}
\label{unflavored_running}
In this section we will consider the particular case in which the profile is $\eta=1$. In this case $dF_2=0$ and there are no flavor sources. It turns out that one can find a particular analytic solution of the BPS system written in Appendix \ref{BPS}. This solution was found in \cite{Conde:2011sw} in a power series expansion around the IR. Amazingly, this series can be summed exactly and a closed analytic form can be written for all functions. Let us first write them in the coordinate $r$. The functions $f$ and $g$ are given by:
\beq
e^{f}\,=\,r\,\sqrt{{1+c\,r\over 1+ 2c\,r}}\,\,,
\qquad\qquad
e^{g}\,=\,r\,{1+c\,r\over 1+ 2c\,r}\,\,,
\label{running_unflavor_in_r}
\eeq
where $c$ is a constant. For $c=0$ this solution is ABJM without flavor (\ie, $AdS_4\times {\mathbb C} {\mathbb P}^3$ with fluxes), while for $c\not=0$ it is a running background which reduces to ABJM in the IR, $r\to0$. The squashing function $q$ can be immediately obtained from  (\ref{running_unflavor_in_r}):
\beq
q\,=\,{1+2c\,r\over 1+ c\,r}\,\,.
\eeq
For $c\not=0$ the squashing function $q$ interpolates between the ABJM value $q=1$ in the IR and the UV value:
\beq 
q_0\,=\,2\,\,.
\eeq
The warp factor for this solution is:
\bear
&&h(r)\,=\,{2\pi^2\,N\over k}\,\,
{(1+ 2c\,r)^2\over r^4\,(1+c\,r)^2}\,\Bigg[1+2c\,r\Big(3cr(1+2c\,r)\,-\,1\Big)\,+\,\rc\rc
&&\qquad\qquad\qquad\qquad
+\,12 c^3\,(1+ c\,r)\,r^3\,\Big(
\log\Big[{cr\over 1+cr}\Big]\,+\,\alpha\Big)\,\Bigg]\,\,,
\eear
where $\alpha$ is a constant which has to be fixed by adjusting the behavior of the metric in the UV. Finally,  the dilaton can be related to the warp factor as:
\beq
e^{\phi}\,=\,{2\over k}\,{1+c\,r\over (1+ 2c\,r)^2}\,r\,h^{{1\over 4}}\,
\,.
\eeq

Let us now re-express this running analytic solution in terms of the variable $x$, related to $r$ by (\ref{r-x-diff-eq}), which in the present case  becomes:
\beq
{1\over x}\,{dx\over dr}\,=\,{1+2c\,r\over (1+ c\,r)r}\,\,.
\eeq
This equation can be easily integrated:
\beq
\gamma\,x\,=\,c\,r(1+c\,r)\,\,,
\label{x-r}
\eeq
where $\gamma$ is a constant of integration which parameterizes the freedom from passing to  the $x$ variable. By solving (\ref{x-r}) for $r$ we get:
\beq
r\,=\,{1\over  2c}\,\Big[\sqrt{1+4\,\gamma\, x}-1\,\Big]\,\,.
\eeq
It is straightforward to write the functions $f$ and $g$ in terms of $x$:
\bear
&&e^{f}\,=\, {\gamma\over c}\,x\,\Big[{2\over \sqrt{1+4\,\gamma\, x}\,
(\sqrt{1+4\,\gamma\, x}+1)}\Big]^{{1\over 2}}\,\,,\rc\rc
&&e^{g}\,=\,  {\gamma\over c}\,{x\over \sqrt{1+4\,\gamma\, x}}\,\,,
\label{functions_running_x}
\eear
while the squashing function is:
\beq
q\,=\,2\,{\sqrt{1+4\,\gamma\, x}\over 1+\sqrt{1+4\,\gamma\, x}}\,\,.
\eeq

The warp factor $h$  in terms of the $x$ variable is:
\bear
&&h\,=\,{8\pi^2 N c^4\over k}\,\Big(1+{1\over 4\gamma \,x}\Big)\,
\Bigg[\Big({1\over 2}+ 6\gamma\,x\,+\,{1+
(1-6\gamma\,x)\sqrt{1+4\,\gamma\, x}\over 4\gamma\,x}\Big)
{\sqrt{1+4\,\gamma\, x}+1\over \gamma^2\,x^2}\,+\,\rc\rc
&&\qquad\qquad\qquad\qquad
+24\,\log\Big[{\sqrt{4\,\gamma\, x}\over \sqrt{1+4\,\gamma\, x}+1}\Big]\,+\,\alpha
\Bigg]\,\,.
\label{h_running-x}
\eear
By choosing appropriately the constant $\alpha$ in (\ref{h_running-x}) this running solution behaves as the $G_2$-cone in the UV region $x\to\infty$. 
The dilaton  as a function of $x$ is:
\beq
e^{\phi}\,=\,{2\over k}\,{\gamma\over c}\,{x\over 1+4\,\gamma\, x}\,h^{{1\over 4}}\,\,.
\eeq

Working in the variable $x$, it is very interesting to find the function $W(x)$. For the solution described above, $W$ can be found  by plugging the different  functions   in the definition (\ref{W_definition}). We find:
\beq
W(x)\,=\,{4\,(1+4\,\gamma\, x)\,x\over 1\,+\,\sqrt{1+4\,\gamma\, x}}\,\,.
\label{W_running_unflavored}
\eeq
One can readily check that the function written in (\ref{W_running_unflavored}) solves the master equation (\ref{master_eq_W}) for $\eta=1$.  For large $x$, the function $W(x)$ behaves as:
\beq
W\sim 8\sqrt{\gamma}\,\,x^{{3\over 2}}\,\,,
\eeq
which corresponds to an exponent $\alpha=3/2$ in (\ref{Asym_W}). This is consistent with the asymptotic value $q_0=2$ of the squashing found above. 

Let us finally point out that we have checked explicitly that the geometry discussed in this section is free of curvature singularities. 

\section{Interpolating solutions}
\label{interpolating}

Let us now construct solutions to the BPS equations which interpolate between an IR region in which there are no D6-brane sources (\ie, with $\eta=1$) and a UV region in which $\eta>1$ and, therefore, the Bianchi identity of $F_2$ is violated.  In the $r$ variable the profile $\eta(r)$ will be such that $\eta(r)=1$ for $r\le r_q$, while $\eta(r)>1$ for $r>r_q$. In the region $r\le r_q$ our interpolating solutions will reduce to  the unflavored running solution of Section \ref{unflavored_running} for some value of the constant $c$. In order to match this solution with the one in the region $r>r_q$ it is convenient to work in the $x$ coordinate (\ref{x-r}). The point $r=r_q$ will correspond to some $x=x_q$. Notice, however, that we have some freedom  in performing the $r\to x$ change of variables. This freedom is parameterized by the constant $\gamma$ in (\ref{x-r}). We will fix this freedom by requiring that $x_q=1$, \ie, that the transition between the unflavored and flavored region takes place at the point $x=1$. Then,  (\ref{x-r}) immediately implies that $\gamma$ is given in terms of $c$ and $r_q$:
\beq
\gamma\,=\,c\,r_q\,(1+c\,r_q)\,\,.
\label{gamma_rq}
\eeq
We will use (\ref{gamma_rq}) to eliminate the constant $c$ in favor of $r_q$ and $\gamma$. Actually, if we define $\hat\gamma$ as:
\beq
\hat\gamma\,\equiv\,\sqrt{1+4\gamma}\,\,,
\label{hat_gamma}
\eeq
then $c$ is given by
\beq
c\,=\,{\hat\gamma-1\over 2\,r_q}\,\,.
\label{c_gamma}
\eeq
In this running solution the squashing factor $q$ is equal to one in the deep IR at 
$x=0$. When $x>0$ the function $q(x)$ grows monotonically until it reaches a certain value $\hat q$ at $x=1$, which is related to the parameter $\hat\gamma$ as:
\beq
\hat \gamma\,=\,{\hat q\over 2-\hat q}\,\,.
\eeq
In the region $x\ge 1$ we have to solve the master equation (\ref{master_eq_W}) with $\eta(x)>1$ and initial conditions given by the values of $W$ and $W'$ attained by the unflavored running solution at $x=1$. These values depend on the parameter $\hat\gamma$. They can be straightforwardly found by taking $x=1$ in the function (\ref{W_running_unflavored}) and in its derivative. We find:
\beq
W(x=1)\,=\,{4\,\hat\gamma^2\over 1+\hat \gamma}\,\,,
\qquad\qquad
W'(x=1)\,=\,6\,\hat\gamma-4\,\,.
\label{initial_W_Wprime}
\eeq
Let us now write the different functions of these interpolating solutions in the two regions 
$x\le 1$ and $x\ge 1$.  For $x\le 1$ we have to rewrite (\ref{functions_running_x}) after eliminating the constant $c$ by using (\ref{c_gamma}) (which implies that $\gamma/c=(\hat\gamma+1)\,r_q/2$). For the functions $f$, $g$, and the dilaton $\phi$ we get:
\bear
&&e^{f}\,=\,r_q \,{\hat \gamma+1\over \sqrt{2}}\,
 {x\over\Big[ \sqrt{1+4\,\gamma\, x}\,
(\sqrt{1+4\,\gamma\, x}+1)\Big]^{{1\over 2}}}
\,\,,\rc\rc
&&e^{g}\,=\,  r_q\,{\hat \gamma+1\over 2}\,{x\over \sqrt{1+4\,\gamma\, x}}\,\,,
\qquad\qquad\qquad\qquad (x\le 1)\,\,,\rc\rc
&&e^{\phi}\,=\, r_q\,{\hat \gamma+1\over k}\,{x\over 1+4\,\gamma\, x}\,
h^{{1\over 4}}\,\,,
\label{background_functions_xle1}
\eear
where $h$ is the function written in (\ref{h_running-x}) for $c=(\hat\gamma-1)/(2\,r_q)$. 
By using the general equations of Section \ref{squashed_solutions}, the solution  for $x\ge 1$ can be written in terms of $W(x)$, which can be obtained by numerical integration of the master equation with initial conditions (\ref{initial_W_Wprime}). 
This defines a solution in the full range of $x$ for every $\gamma$ and $r_q$. 
Notice that (\ref{g-f-W}), (\ref{warp_factor_x}), and (\ref{dilaton_x}) contain arbitrary multiplicative constants, which we will fix by imposing continuity of $f$, $g$,  and $h$ at $x=1$. We get for $f$, $g$ and $\phi$:
\bear
&&e^{f}\,=\,r_q\,\Big[{(\hat \gamma+1)^2\over 2\,\hat \gamma}\Big]^{{1\over 3}}\,\,
\sqrt{{3x\over W'+4\eta}}\,\,W^{{1\over 6}}\,
\exp\Big[{2\over 3}\int^x_1\,{\eta(\xi)d\xi\over W(\xi)}\Big]
\,\,,\rc\rc
&&e^{g}\,=\, r_q\,\Big[{(\hat \gamma+1)^2\over 2\,\hat \gamma}\Big]^{{1\over 3}}\,\,
{x\over W^{{1\over 3}}}\,\exp\Big[{2\over 3}\int^x_1\,{\eta(\xi)d\xi\over W(\xi)}\Big]\,\,,
\qquad\qquad\qquad\qquad (x\ge 1)\,\,,\rc\rc
&&e^{\phi}\,=\, 
r_q\,\Big[{(\hat \gamma+1)^2\over 2\,\hat \gamma}\Big]^{{1\over 3}}\,\,{12\over k}\,
{x\,h^{{1\over 4}}\over W^{{1\over 3}} (W'+4\eta)}\,\exp\Big[{2\over 3}\int^x_1\,{\eta(\xi)d\xi\over W(\xi)}\Big]\,
\,\,.
\label{background_functions_xge1}
\eear
The warp factor $h(x)$ for $x\ge 1$ is given by (\ref{warp_factor_x}), where the integration constant $\beta$  is related to the constant $\alpha$ of (\ref{h_running-x}) by the following matching condition at $x=1$:
\beq
\lim_{x\to 1^-}\,h(x)\,=\,\lim_{x\to 1^+}\,h(x)\,\,.
\label{matching_h}
\eeq
For a given profile function $\eta(x)$, the solution described above depends on the parameter $\gamma$, which determines $W(x)$ for $x\le 1$ through (\ref{W_running_unflavored}) and sets the initial  conditions (\ref{initial_W_Wprime}) needed to integrate the master equation in the 
$x\ge 1$ region. The solution $W(x)$ obtained numerically in this way grows generically as $x^{3/2}$ for large $x$ which, according to our analysis in Section \ref{G2-cone}, gives rise to the geometry of the $G_2$-cone in the UV. We are, actually, interested in obtaining solutions with the $AdS$ asymptotics discussed in Section \ref{AdS_UV}, for a set of profiles that correspond to flavor D6-branes with a non-zero quark mass. In order to get these geometries we have to fine-tune the parameter $\gamma$ to some precise value which depends on the number of flavors. This analysis is presented in the next section.

\section{Massive flavor}
\label{massive_flavor}

We now apply the formalism developed so  far to find supergravity backgrounds representing massive flavors in ABJM. These solutions will depend on a deformation parameter $\hat \epsilon$, related to the total number of flavors $N_f$ and the Chern-Simons level $k$ as:
\beq
\hat\epsilon\,\equiv\,{3N_f\over 4k}\,\,,
\eeq
where the factor $3/4$ is introduced for convenience and $N_f/k$ is just ${N_f\over N}\,\lambda$ with $\lambda=N/k$ being the 't Hooft coupling. The profile $\eta$,  which corresponds to a set of smeared flavor D6-branes ending at $r=r_q$,  has been found in \cite{Conde:2011sw}. The main technique employed in \cite{Conde:2011sw} was the comparison between the smeared brane action for the distribution of flavor branes and the action for a fiducial embedding in a background of the type studied here.  This fiducial embedding was determined by using kappa symmetry. With our present conventions,\footnote{ The variable $\rho$ used in Section 8 of 
\cite{Conde:2011sw} is related to $x$ by $x=e^{\rho}$.}  assuming as above that $r=r_q$ corresponds to $x=x_q=1$, the function $\eta(x)$ is given by:
\beq
\eta(x)\,=\,1\,+\,\hat \epsilon\,\Big(1-{1\over x^2}\Big)\,\Theta(x-1)\,\,,
\label{eta-flavor}
\eeq
where $\Theta(x)$ is the Heaviside step function. 
It follows from (\ref{eta-flavor}) that the asymptotic value $\eta_0$ of the profile is:
\beq
\eta_0\,=\,1+\hat \epsilon\,\,.
\label{eta_0-epsilon}
\eeq
We want to find interpolating solutions of the type studied in Section \ref{interpolating} which have the AdS asymptotic behavior in the UV corresponding to the value of $\eta_0$ written in (\ref{eta_0-epsilon}). These solutions have an asymptotic squashing (corresponding to $q_0^{+}$ in (\ref{q_0-pm}))  given in terms of $\hat\epsilon$ as:
\beq
q_0\,=\,3\,+\,{3\over 2}\hat \epsilon\,-\,2\,
\sqrt{1\,+\,\hat\epsilon\,+\,{9\over 16}\,\hat\epsilon^{\,2}}\,\,.
\label{q0-epsilon}
\eeq
It follows from (\ref{q0-epsilon}) that the asymptotic squashing $q_0$ grows with the deformation parameter $\hat\epsilon$. Indeed, $q_0=1$ for $\hat\epsilon=0$, whereas for $\hat\epsilon\to\infty$ the squashing reaches its maximum  value: $q_0\to 5/3$. By using the relation between $b$ and $q_0$  (eq. (\ref{b_new})) we also conclude that $b\to 5/4$  when $\hat\epsilon\to\infty$. 
\begin{figure}[ht]
\center
\includegraphics[width=0.7\textwidth]{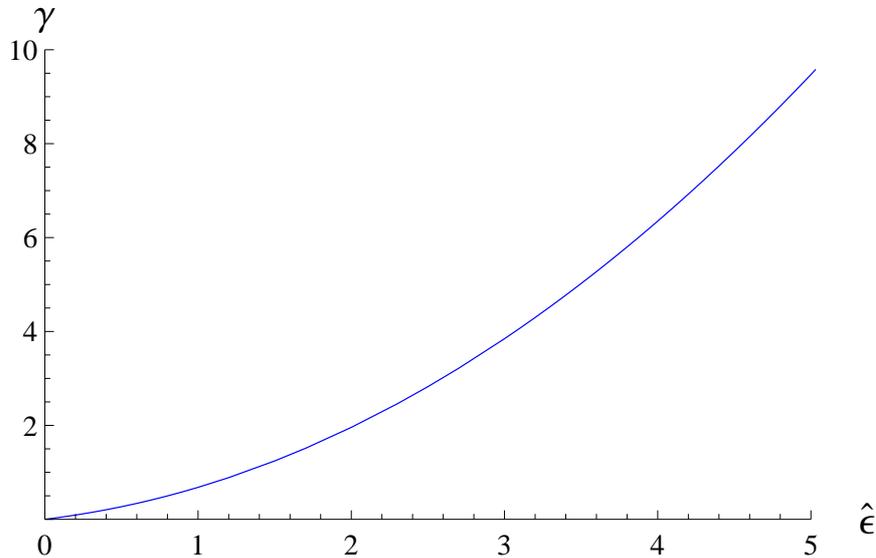}
\caption{Representation of $\gamma(\hat\epsilon)$.} 
\label{fig:gamma}
\end{figure}

\begin{figure}[ht]
\begin{center}
\includegraphics[width=0.44\textwidth]{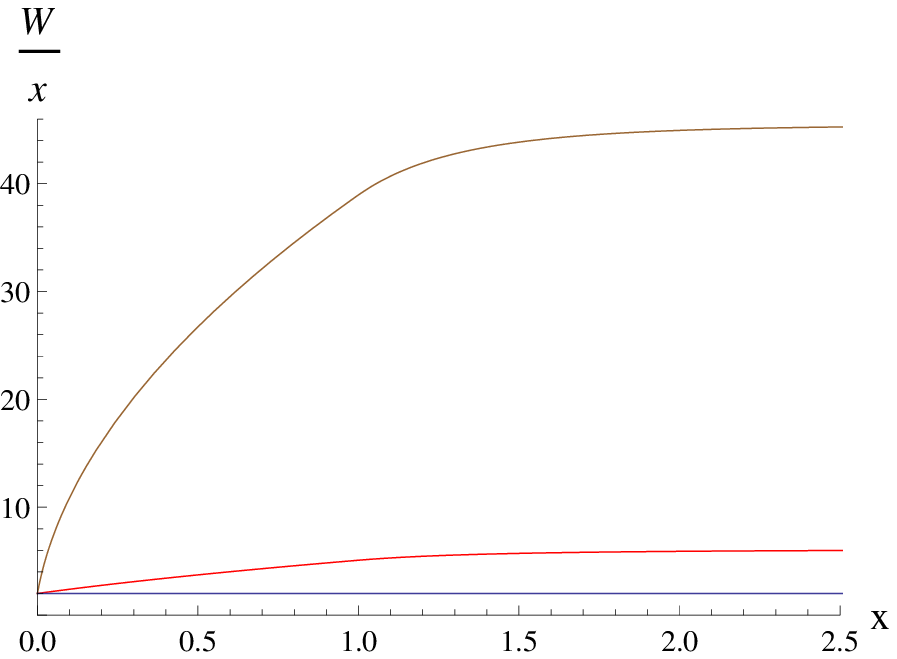}
\qquad\qquad
\includegraphics[width=0.44\textwidth]{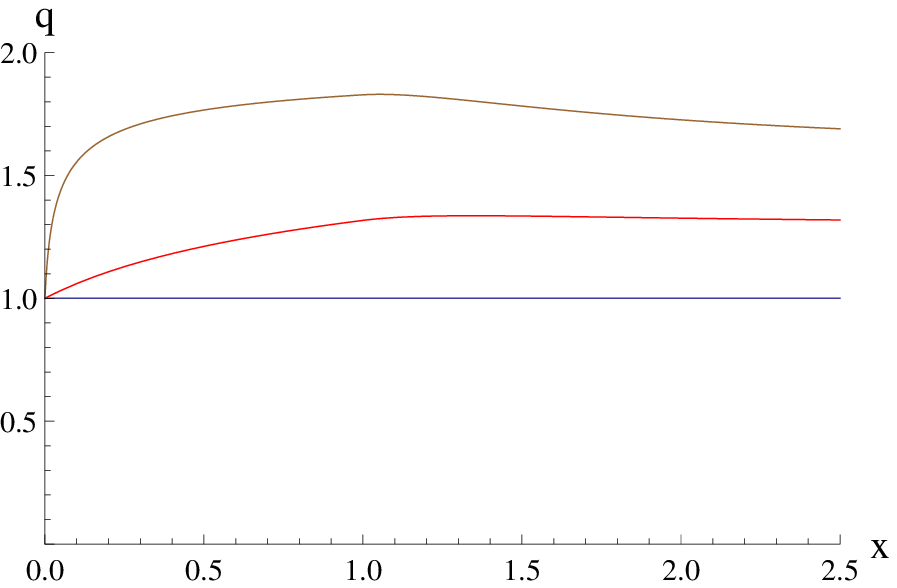}
\end{center}
\caption{Representation of $W(x)/x$ and $q(x)$ for different numbers of  flavors ($\hat\epsilon=0$ (bottom, blue), $\hat\epsilon=1$ (middle, red), and $\hat\epsilon=9$ (top, brown)). The plots on the left show that  $W(x)/x$ becomes constant as we approach the IR and UV conformal points at $x=0$ and $x=\infty$, respectively. Similarly, the squashing function $q(x)$ interpolates between $q=1$ at $x=0$ and $q=q_0$ at large $x$.} 
\label{fig:W_and_q}
\end{figure}

To find the solution for $x\ge 1$ we have to solve numerically the BPS system in this region.  The most efficient way to proceed is by looking at the master equation  for $W(x)$ with the initial conditions (\ref{initial_W_Wprime}). For a generic value of $\gamma$ the numerical solution either gives  rise to negative values of $W(x)$ (which is unphysical  for $k>0$, see  the definition (\ref{W_definition})) or behaves in the UV as $W(x)\sim x^{{3\over 2}}$, which corresponds to the $G_2$-cone asymptotics with $q_0=2$ discussed in Section \ref{G2-cone}. Only when $\gamma$ is fine-tuned to some particular value (which depends on $\hat \epsilon$) we get in the UV that $W(x)\sim x$  and that $q_0$ is given by (\ref{q0-epsilon}).  To determine this critical value of $\gamma$ we have to perform a numerical shooting for every value of $\hat\epsilon$. In what follows we understand that $\gamma=\gamma(\hat\epsilon)$ is the function of the deformation parameter which results of this shooting. The  function $\gamma(\hat\epsilon)$ is plotted in Fig. \ref{fig:gamma}, where we notice that $\gamma(\hat\epsilon=0)=0$ and, therefore, we recover the unflavored ABJM background when the deformation parameter vanishes. In the opposite limit $\hat\epsilon\to\infty$, the function  $\gamma(\hat\epsilon)$ grows as 
$\hat\epsilon^2$. Actually,  $\gamma(\hat\epsilon)$ can be accurately represented by a function of the type:
\beq
 \gamma(\hat\epsilon)\,=\,\gamma_1\,\hat\epsilon\,+\,\gamma_2\,\hat\epsilon^{\,2}\,\,,
 \eeq
with $\gamma_1\,=\,0.351$ and $\gamma_2\,=\,0.309$. In Fig. \ref{fig:W_and_q} we plot  the function $W(x)$ and  the squashing function $q(x)$ for some selected values of $\hat\epsilon$.

From the function $W(x)$ we can obtain $f$ and $g$ by performing the integrals in (\ref{background_functions_xge1}). The whole metric is determined if $h$ is known. We will compute $h$ from (\ref{warp_factor_x}) with $\beta=0$, which corresponds to requiring that $h\to 0$ in the UV. In the $x\le 1$ region the warp factor $h(x)$ is given by (\ref{h_running-x}), with the constant $\alpha$ determined by the matching condition (\ref{matching_h}).  The limit on the left-hand side of (\ref{matching_h}) can be determined  explicitly from (\ref{h_running-x}):
\bear
\lim_{x\to 1^-}h(x)=
{\pi^2\over 8 \,r_q^4}\,{N\over k}\,
{\big(\hat\gamma-1\big)^4\,(1+4\gamma)\over \gamma}
\Bigg[\Big({1\over 2}+6\gamma+{1+(1-6\gamma)\hat\gamma\over 4\gamma}\,\Big)
{\hat\gamma+1\over \gamma^2}+24\log {\sqrt{4\gamma}\over\hat\gamma+1}+
\alpha\,\Bigg],\rc
\eear
whereas $h(x\to 1^+)$ is given by:
\beq
\lim_{x\to 1^+}\,h(x)\,=\,
{48\,\pi^2\over r_q}\,{N\over k}\,
{\hat\gamma^2\over \hat\gamma+1}\,\,
\int_{1}^{\infty}\,{\xi\,e^{-3g(\xi)}\over  \big[ W(\xi)\big]^2}\,d\xi\,\,.
\eeq

Notice that $r_q$ (the value of the  $r$ coordinate at  the tip of the flavor branes) appears as a free parameter in eqs. (\ref{background_functions_xle1}) and (\ref{background_functions_xge1}). Actually, $r_q$ can be easily related to the mass $m_q$ of the quarks which deform the geometry. Indeed, by computing the Nambu-Goto action of a fundamental string stretching along the holographic direction between $r=0$ and $r=r_q$ at fixed value of all the other spacelike coordinates in the geometry (\ref{metric_ansatz}), we get that $m_q$ and $r_q$ are linearly related as:
\beq
m_q\,=\,{r_q\over 2\pi\alpha'}\,\,,
\label{rq_mq}
\eeq
where $\alpha'$ is the Regge slope (which we will take to be equal to one in most of our equations). Equivalently, we can relate $m_q$ to the constant $c$ appearing in the solution in the $x\le 1$ region:
\beq
m_q\,=\,{\sqrt{1+4\gamma(\hat\epsilon)}-1\over 4\pi}\,\,{1\over c}\,\,,
\label{mq-gamma-c}
\eeq
where $\gamma(\hat\epsilon)$ is the function obtained by the shooting and only depends on the deformation parameter $\hat\epsilon$.  

We have computed the curvature invariants for the flavored metric and we have checked that the geometry is regular both in the IR ($x\to 0$) and UV ($x\to\infty$). However, the curvature has a finite discontinuity at $x=1$, as can be directly concluded by inspecting Einstein's equations (see Appendix \ref{BPS}). This ``threshold" singularity occurs at the point where the sources are added and could be avoided by smoothing the introduction of brane sources with an additional smearing (see the last article in \cite{CNP} for a similar analysis in other background). 

\subsection{UV asymptotics}
\label{UV_mass_corrections}

The full background in the $x\ge 1$ region must be found by numerical integration and shooting, as described above. However, in the UV region $x\to\infty$ one can solve the master equation (\ref{master_eq_W}) in power series for large $x$. Indeed, one can find 
a solution where $W(x)$ is represented as:
\beq
W(x)\,=\,x\,\,\sum_{i=0}^{\infty}\,{A_{2i}\over x^{2i}}\,\,,
\label{W_asymp_UV}
\eeq
where the coefficients $A_{2i}$ can be obtained recursively. The coefficient $A_0$ of the leading term was written in (\ref{A0_q0}). The next two coefficients are:
\beq
A_2\,=\,-{40\,\eta_0\,-\,11\,A_0\over 9+13\,\eta_0\,-\,2\,A_0}\,\hat\epsilon
\,\,,\qquad\qquad
A_4\,=\,{5A_2^2\,+\,20\,\hat\epsilon^2\,+\,25\,\hat\epsilon\,A_2\over 
9(1+3\eta_0-2A_0)}\,\,.
\eeq
Notice that a linear behavior of $W(x)$ with $x$ corresponds to a conformal $AdS$ background,  whereas the deviations from conformality are encoded in the non-linear corrections. 

From the result written above one can immediately  obtain the asymptotic behavior of the squashing function for large $x$. Indeed, let us use in the expression  of $q$ in terms of $W$ and $W'$  (eq. (\ref{q_W_Wprime})) the following large $x$ expansion: 
\beq
x\,(\,W'+4\eta\,)\,=\,(A_0+4\eta_0)\,x\,-\,{A_2+4(\eta_0-1)\over x}\,+\,\cdots\,\,.
\label{Wprime-eta-expansion}
\eeq
We get:
\beq
q(x)\,=\,q_0\,+\,{q_2\over x^2}\,+\,\cdots\,\,,
\eeq
where $q_0$ is the asymptotic value of the squashing (see (\ref{q0-epsilon}))  and $q_2$ is given by:
\beq
q_2\,=\,{2b\over 3(2-b)^2}\,\,\Big[\,(3-2b)\,{\eta_0-1\over \eta_0}\,+\,(3-b)\,{A_2\over A_0}\,\Big]\,\,,
\eeq
where $b$ is related to $q_0$ and $\hat\epsilon$ by  (\ref{b_new}) and (\ref{q0-epsilon}). Similarly, we can find analytically  the first corrections to the UV conformal behavior. The details of these calculations  are given in Appendix \ref{UV_asymptotics}.  In this section we just present the final results.  First of all,  let us define the constant $\kappa$ (depending on the deformation $\hat\epsilon$) as:
\beq
\kappa\,\equiv\,b\,
\Big[{(\hat \gamma+1)^2\over 2\,\hat \gamma\,A_0}\Big]^{{1\over 3}}\,\,
\exp\Bigg[{2\over 3}\int^{\infty}_1\,\Big[{\eta(\xi)\over W(\xi)}\,-\,
{\eta_0\over A_0\,\xi}\Big]
d\xi\,\Bigg]\,\,.
\label{kappa_def}
\eeq
Then the functions $g$ and $f$ can be expanded for large $x$ as:
\beq
e^{g}\,=\, {\kappa\,r_q\over b}\,\,x^{{1\over b}}\,\,
\Big[\,1\,+\,{g_2\over x^2}\,+\,\cdots\Big]\,\,,
\qquad\qquad
e^{f}\,=\, \sqrt{q_0}\,\,{\kappa\,r_q\over b}\,\,x^{{1\over b}}\,\,
\Big[\,1\,+\,{f_2\over x^2}\,+\,\cdots\Big]\,\,,
\label{g_f_UVexpansions_x}
\eeq
where the coefficients  $g_2$  and $f_2$ are:
\bear
&&g_2\,=\,
{3-2b \over 6b}\,{\eta_0-1\over \eta_0}\,+\,{3-4b\over 6b}\,{A_2\over A_0}
\,\,,\rc\rc
&&f_2\,=\,{1\over 3}\,\Big(\,{3\over 2b}\,+\,{1\over 2-b}\,-1\,\Big)\,{A_2\over A_0}\,+\,
{(2+b)(3-2b)\over 6b(2-b)}{\eta_0-1\over \eta_0}\,\,.
\label{g2_f2}
\eear
Moreover, the  UV expansion of the warp factor $h$ and the dilaton is:
\beq
h(x)\,=\,{L_0^4\over \kappa^4\,r_q^4}\,
x^{-{4\over b}}
\,\Big[1\,+\,{h_2\over x^2}\,+\,\cdots\Big]\,\,,
\qquad\qquad
e^{\phi}\,=\, e^{\phi_0}\,\Big(1\,+\,{\phi_2\over x^2}\,+\,\cdots\Big)\,\,,
\label{UVexpansion_h_phi}
\eeq
with the coefficients $h_2$ and $\phi_2$ given by:
\bear
&&h_2\,=\,{3-2b\over 2b}\,
\Big(\,1+{8\over 3(b-2)}\,-\,{3\over 3+2b}\,\Big)\,{\eta_0-1\over \eta_0}\,+\,
\Big(\,1\,+\,{2\over 3(b-2)}\,-\,{2\over b}\,+\,{3\over 3+2b}\,\Big)\,{A_2\over A_0}
\,\,,\rc\rc
&&\phi_2\,=\,{3-2b\over 8b}\,\,\Big(\,1+{4b\over 2-b}-{3\over 3+2b}\,\Big)\,
{\eta_0-1\over \eta_0}-{3\over 4}\,\Big(\,1-{2\over 3(2-b)}-{1\over 3+2b}\,\Big)\,
{A_2\over A_0}\,\,.
\label{h_2_phi_2}
\eear
It is also interesting to write the previous expansions in terms of the $r$ variable. Again, the details are worked out in Appendix \ref{UV_asymptotics} and the final result is:
\bear
&&e^{g(r)}= {r\over b}\,\big[1\,+\,\tilde g_2\,\Big({ r_q\over r}\Big)^{2b}\,+\,\cdots
\big]\,\,,
\qquad\qquad\,\,\,\,
e^{f(r)}= {\sqrt{q_0}\,\,r\over b}\,\big[1\,+\,\tilde f_2\,\Big({ r_q\over r}\Big)^{2b}\,+\,\cdots
\big]\,\,,\cr\cr
&&h(r)= \Big[{L_0\over r}\Big]^{4}\,\,
\Big[\,1\,+\,\tilde h_2\,\Big({r_q\over r}\Big)^{2b}\,+\,\cdots\Big]\,\,,
\qquad\,\,
e^{\phi}\,=\, e^{\phi_0}\,\Big(1\,+\,\tilde\phi_2\,\Big({ r_q\over r}\Big)^{2b}
\,+\,\cdots\Big)\,\,,\qquad\qquad
\label{All_UV_expansions_r}
\eear
where the coefficients $\tilde g_2$, $\tilde f_2$,  $\tilde h_2$,  and $\tilde \phi_2$ are related to the ones in (\ref{g2_f2}) and (\ref{h_2_phi_2}) as:
\bear
&& \tilde g_2\,=\,{2b\over 2b-1}\,\kappa^{2b}\,g_2\,\,,
\qquad\qquad\qquad\qquad
 \tilde f_2\,=\,  \kappa^{2b}\,\Big(\,f_2\,+\,{g_2\over 2b-1}\,\Big)\,\,,\rc\rc
&& \tilde h_2\,=\,  \kappa^{2b}\,\Big(\,h_2\,-\,{4g_2\over 2b-1}\,\Big)\,\,,
\qquad\qquad\,\,\,\,\,
\tilde \phi_2\,=\,\kappa^{2b}\,\phi_2\,\,.
\label{tilde_UV_coeff}
\eear
\begin{figure}[ht]
\center
\includegraphics[width=0.7\textwidth]{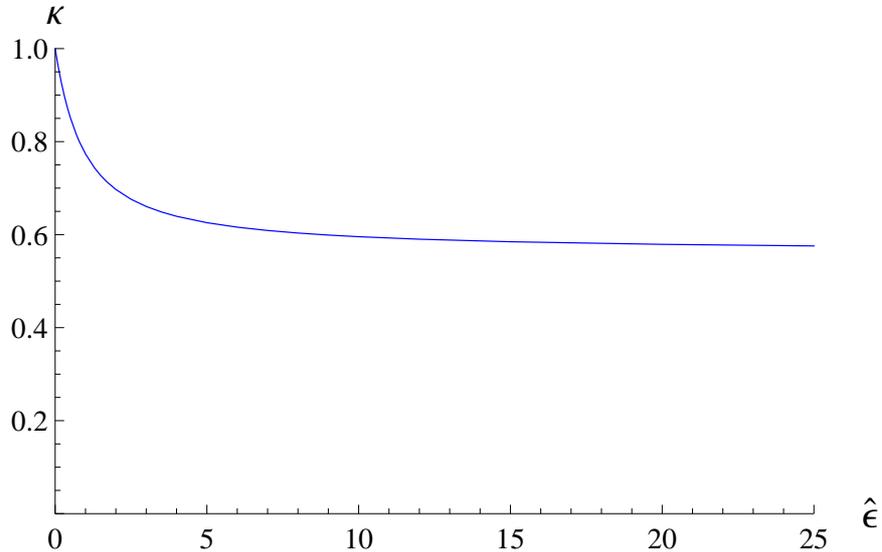}
\caption{Plot of  $\kappa$ as a function of the deformation parameter $\hat\epsilon$. The $\kappa$ asymptotes to some positive constant as $\hat\epsilon \to \infty$.} 
\label{fig: kappa}
\end{figure}
Recalling (see (\ref{rq_mq})) that $r_q=2\pi \,m_q$, it is clear from (\ref{All_UV_expansions_r}) that the deviation from conformality is controlled  by the quark mass and that the parameter $b$ determines the power of the first mass corrections. In our holographic context this is quite natural if one takes into account that $b$ determines the dimension $\Delta$ of the quark-antiquark bilinear operator in the theory with dynamical quarks ($\Delta=3-b$, see \cite{Conde:2011sw,Jokela:2012dw} and below).  The coefficients of these mass corrections depend on the constants $g_2$, $f_2$, $h_2$, and $\phi_2$ (whose analytic expressions we know from eqs. (\ref{g2_f2}) and (\ref{h_2_phi_2})), as well as on the constant $\kappa$, which must be determined numerically.   $\kappa$ as a function of the deformation parameter is plotted in Fig. \ref{fig: kappa}.  From this plot we notice that $\kappa(\hat\epsilon)$ interpolates continuously between $\kappa=1$ for $\hat\epsilon=0$ and some positive  constant value at large $\hat\epsilon$. 

\section{Holographic entanglement entropy}
\label{Holographic_entanglement_entropy}

In a quantum theory the entanglement entropy $S_{A}$ between a spatial region $A$ and its complement is defined as the entropy seen by an observer in $A$ which has no access to the degrees of freedom living in the complement of $A$. It can be computed as the von Neumann entropy for the reduced density matrix obtained by taking the trace over the degrees of freedom of the complement of $A$. For quantum field theories admitting  a gravity dual, Ryu and Takayanagi proposed in \cite{Ryu} a simple prescription to compute $S_A$ from the corresponding supergravity background. The holographic entanglement entropy between $A$ and its complement in the proposal of \cite{Ryu}  is obtained by finding the eight-dimensional spatial surface $\Sigma$ whose boundary coincides with the boundary of $A$  and is such that it minimizes the functional:
\beq
S_{A}\,=\,{1\over 4\,G_{10}}\,\int_{\Sigma}\,
d^8\xi\,e^{-2\phi}\,\sqrt{\det g_8}\,\,,
\eeq
where  the $\xi$'s are a system of eight coordinates of $\Sigma$,  $G_{10}$ is the ten-dimensional Newton constant ($G_{10}=8\pi^6$ in  our units) and $g_8$ is the induced metric on $\Sigma$ in the string frame. The functional $S_{A}$ evaluated on the minimal surface $\Sigma$ is precisely the entanglement entropy between the region $A$ and its complement. 

\begin{figure}[ht]
\center
\includegraphics[width=0.5\textwidth]{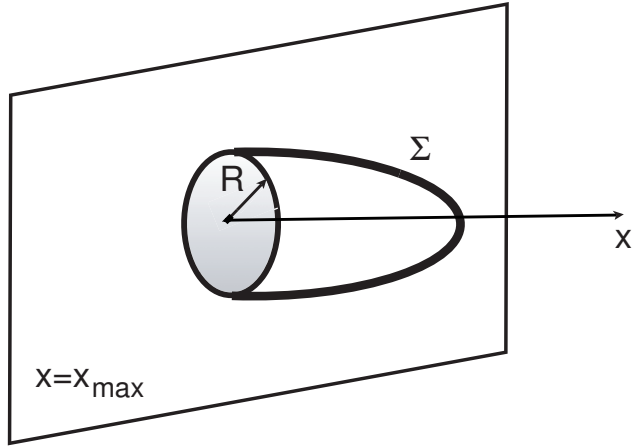}
\caption{The surface  $\Sigma$ ends on the disk of radius $R$ at the boundary $x=x_{max}\to \infty$.} 
\label{Sigma_disk}
\end{figure}

In our case $A$ is a region of the $(x^1,x^2)$-plane. In this section we will study in detail the case in which the region $A$  is a disk with radius $R$ as depicted in Fig. \ref{Sigma_disk} (see Appendix \ref{entanglement_appendix} for the analysis of the entanglement entropy of a strip in the $(x^1,x^2)$-plane).  In order to deal with the disk case it is convenient to choose a system of polar coordinates for the plane:
\beq
(dx^1)^2\,+\,(dx^2)^2\,=\,d\rho^2\,+\,\rho^2\,d\Omega_1^2\,\,.
\eeq
We will describe the eight-dimensional fixed time surface $\Sigma$ by a function $\rho=\rho(x)$ with $\rho$ being the radial coordinate of the boundary plane and $x$ the holographic coordinate of the bulk. The eight-dimensional induced metric is:
\beq
ds^2_{8}\,=\,h^{-{1\over 2}}\,\big[\,\rho\,'^{\,2}+G(x)\,\big]\, dx^2\,+\,
h^{-{1\over 2}}\,\rho^2\,d\Omega_1^2\,
+h^{{1\over 2}}\,\Big[\,
e^{2f}\,ds_{{\mathbb S}^4}^2+
e^{2g}\,\Big(\,\big(E^1\big)^2\,+\,\big(E^2\big)^2\Big)\,\Big]\,\,,
\eeq
where $\rho\,'$ denotes the derivative with respect to  the holographic coordinate $x$  and   the function $G(x)$ is defined as:
\beq
G(x)\equiv {e^{2g}\,h\over x^2}\,\,.
\label{G_def}
\eeq
Let us next define a new function $H(x)$ as:
\beq
H(x)\,\equiv\,h^2\,e^{-4\phi}\,e^{8f+4g}\,\,.
\label{H_def}
\eeq
Then, the entanglement entropy as a function of $R$ is given by:
\beq
S(R)\,=\,{2\pi\,V_6\over 4 G_{10}}\,\int_{x_*}^{\infty} dx
\,\rho\,\sqrt{H(x)}\,\sqrt{\rho\,'^{\,2}+G(x)}\,\,,
\label{entropy_total_disk}
\eeq
where   $V_6=32\,\pi^3/3$ is the volume of the internal manifold
and $x_*$ is the $x$ coordinate of the turning point of $\Sigma$. 
The Euler-Lagrange equation of motion derived from the entropy  functional
 (\ref{entropy_total_disk})  is:
\beq
{d\over dx}\,\Bigg[\,
\sqrt{H(x)}\,{\rho\,\rho'\over 
\sqrt{\rho\,'^{\,2}+G(x)}}\,\Bigg]-\,\sqrt{H(x)}\,\,\sqrt{\rho\,'^{\,2}+G(x)}\,=\,0\,\,.
\label{Euler_Lagrange_disk}
\eeq
Notice that  the integrand in (\ref{entropy_total_disk}) depends on the independent variable $x$ and we therefore  cannot find a first-integral for the second-order  differential equation (\ref{Euler_Lagrange_disk}). Thus, we have to deal directly with 
(\ref{Euler_Lagrange_disk}), which must be solved with the following boundary conditions at the tip of $\Sigma$:
\beq
\rho(x=x_*)\,=\,0\,\,,
\qquad\qquad
\rho'(x=x_*)\,=\,+\infty\,\,.
\label{bc_x_tip}
\eeq
Notice also that the radius $R$ of the disk  at the boundary is just the UV limit of $\rho$:
\beq
\rho(x\to\infty)=R\,\,.
\eeq
The integral (\ref{entropy_total_disk}) for $S(R)$ diverges due to the 
contribution of the  UV region of  large $x$ . In order to characterize this divergence and to extract the finite part, let us study the behavior of the integrand in  (\ref{entropy_total_disk}) as $x\to\infty$. From the definitions of the functions $H(x)$ and $G(x)$  and the UV behavior written in  (\ref{g_f_UVexpansions_x}) and (\ref{UVexpansion_h_phi}), it follows that $H$ and $G$  display a power-like behavior as  $x\to\infty$, 
\beq
H(x)\approx H_{\infty}\,\,x^{{4\over b}}\,\,,
\qquad\qquad
G(x)\approx G_{\infty}\,\,x^{-2-{2\over b}}\,\,,
\qquad\qquad
(x\to\infty)\,\,,
\label{H_G_UV}
\eeq
where  the coefficients  $H_{\infty}$ and $G_{\infty}$ are
\beq
H_{\infty}\,=\,{L_0^8\,\kappa^4\,r_q^4\,q_0^4\,e^{-4\phi_0}\over b^{12}}\,\,,
\qquad\qquad
G_{\infty}\,=\,{L_0^4\over b^2\,r_q^2\,\kappa^2}\,\,.
\label{HG_infinity}
\eeq
 By taking the $x\to\infty$ values of $\rho$ and $\rho'$ ($\rho=R$ and  $\rho'=0$) inside the integral in (\ref{entropy_total_disk}), as well as the asymptotic form of $H(x)$ and $G(x)$ (eq. (\ref{H_G_UV})), we get:
\beq
S_{div}(R)\,=\,{\pi\,V_6\over 2 G_{10}}\,\,
R\,\sqrt{H_{\infty}\,G_{\infty}}\,
\int^{x_{max}}\,x^{{1\over b}-1}\,dx\,\,,
\eeq
where $x_{max}$ is the maximum value of the holographic coordinate $x$ (which acts as a UV regulator).  After performing the integral, we obtain:
\beq
S_{div}(R)\,=\,{\pi\,V_6\over 2 G_{10}}\,\,
R\,b\,\sqrt{H_{\infty}\,G_{\infty}}\,\,x_{max}^{{1\over b}}\,\,.
\label{S_div_first}
\eeq
Let us rewrite (\ref{S_div_first}) in terms of physically relevant quantities. First of all, we notice that:
\beq
b\,\sqrt{H_{\infty}\,G_{\infty}}\,=\,
{L_0^6\,q_0^2\,e^{-2\phi_0}\over b^6}\,\kappa\, r_q\,=\,
{3\pi^2\over 2}
{F_{UV} ({\mathbb S}^3)\over L_0^2}\,\kappa\, r_q\,\,,
\label{bHG_F}
\eeq
where $F_{UV} ({\mathbb S}^3)$ is the free energy\footnote{When the field theory is formulated on a three-sphere, its free energy is defined as:
\beq
F({\mathbb S}^3)\,=\,-\log |\,Z_{{\mathbb S}^3}\,|\,\,,
\eeq
where $Z_{{\mathbb S}^3}$ is the Euclidean path integral. For a CFT whose gravity dual is of the form $AdS_4\times {\cal M}_6$, the holographic calculation of $F({\mathbb S}^3)$  gives \cite{Emparan:1999pm}:
\beq
F({\mathbb S}^3)\,=\,{\pi L^2\over 2 G_N}\,\,,
\eeq
where $L$ is the $AdS_4$ radius and $G_N$ is the effective four-dimensional Newton's constant. } of the massless flavored theory on the three-sphere:
\beq
F_{UV} ({\mathbb S}^3)\,=\,{2\pi\over 3}\,{N^2\over \sqrt{2\lambda}}\,\,
\xi\big({N_f\over k}\big)\,\,,
\label{F_UV}
\eeq
where the function $\xi\Big({N_f\over k}\Big)$ is given by:
\beq
\xi\Big({N_f\over k}\Big)\equiv{1\over 16}\,\,
{q_0^{{5\over 2}}\,(\eta_0+q_0)^4\over (2-q_0)^{{1\over 2}}\,\,(q_0+\eta_0 q_0\,-\,\eta_0)^{{7\over 2}}}\,\,.
\label{xi-smeared}
\eeq
In (\ref{xi-smeared})  $\eta_0=1+\hat \epsilon$ and $q_0$ is written  in (\ref{q0-epsilon}) in  terms of the deformation parameter.  For the unflavored ABJM theory the free energy is given by (\ref{F_UV}) with $\xi=1$. This formula displays the famous $N^{{3\over 2}}$ behavior. The function 
$\xi(N_f/k)$ encodes the corrections to this behavior due to the smeared massless flavors. It was first computed in \cite{Conde:2011sw}, where it was shown that it is remarkably close to  the value found in \cite{Gaiotto:2009tk} for localized embeddings. 
The function $\xi$ is a monotonic function of the deformation parameter which grows as 
$\sqrt{\hat \epsilon}$ for large values of $\hat\epsilon$.

Using (\ref{bHG_F}) and the fact that, in the deep UV region of large $x$, $r_{max}=\kappa\, r_q\,x_{max}^{{1\over b}}$ (see (\ref{r-x_relation})), we can rewrite $S_{div}(R)$ as:
\beq
S_{div}(R)\,=\,{F_{UV} ({\mathbb S}^3)\over L_0^2}\,r_{max}\,R\,\,.
\label{Sdiv_disk}
\eeq
We notice in (\ref{Sdiv_disk}) that $S_{div}(R)$ diverges linearly with $r_{max}$. The coefficient of this divergent term is linear in the disk radius $R$ and in $F_{UV} ({\mathbb S}^3)$. The latter is a measure of the effective number of degrees of freedom of the flavored theory in the high-energy UV limit in which the flavors can be considered to be massless. The appearance of $F_{UV} ({\mathbb S}^3)$ in (\ref{Sdiv_disk}) is thus quite natural.

The separation between the divergent and finite parts of $S(R)$ has ambiguities. In order to solve these ambiguities, Liu and Mezei proposed in \cite{Liu:2012eea} to consider the function ${\cal F}(R)$, defined as:
\beq
{\cal F}(R)\,\equiv\,R\,{\partial S\over \partial R}\,-\,S\,\,.
\label{calF_definition}
\eeq
It was argued in \cite{Liu:2012eea} that ${\cal F}(R)$ is finite and a monotonic function of $R$ which provides a measure of the number of degrees of freedom of a system at a scale $R$. 

In a 3d CFT the entanglement entropy for a disk of radius $R$ has the form:
\beq
S_{CFT}(R)\,=\,\alpha \,R\,-\,\beta\,\,,
\label{CFT-entropy}
\eeq
where $\alpha$ is a UV divergent  non-universal part and $\beta$ is finite and independent of $R$.  It was shown in \cite{Casini:2011kv} that the finite part $\beta$ is equal to the free energy of the theory on ${\mathbb S}^3$. Notice that ${\cal F}=\beta$ when $S$ is of the form (\ref{CFT-entropy}). Therefore, for a conformal fixed point the function ${\cal F}$ is constant and equal to the free energy on the three-sphere of the corresponding CFT.  In the next subsection we will obtain the UV and IR values of  ${\cal F}$ and we will show that  they coincide  with the free energies on ${\mathbb S}^3$ of the massless flavored theory and of the unflavored ABJM model, respectively. 

It is interesting to point out that the entanglement entropy of a disk in a (2+1)-dimensional system at large $R$ can also be written in the form (\ref{CFT-entropy}), if one neglects terms  which vanish in the $R\to\infty$ limit.  In a gapped system, the $R$-independent part $\beta$ of the right-hand side of (\ref{CFT-entropy}) is the so-called topological entanglement entropy \cite{Kitaev:2005dm,Levin:2006zz} and serves to characterize topologically ordered many-body  states which contain non-local entanglement due to non-local correlations (examples of such states are the Laughlin states of the fractional quantum Hall effect or the ${\mathbb Z}_2$ fractionalized states). The  topological entanglement entropy $\beta$ is related to the so-called total quantum dimension 
${\cal D}$  of the system as  $\beta=\log {\cal D}$. In general ${\cal D}>1$ (or $\beta>0$) signals a topological order (for example ${\cal D}=\sqrt{q}$ for the quantum Hall system with filling fraction $\nu=1/q$, with $q$ an odd integer). 
\begin{figure}[ht]
\center
\includegraphics[width=0.7\textwidth]{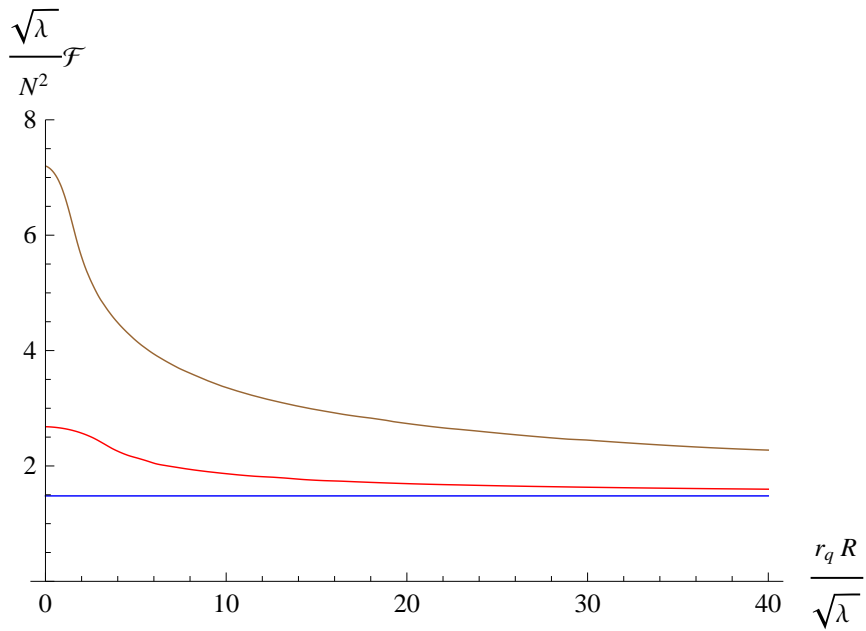}
\caption{Plot of the running free energy ${\cal F}$ as a function of $r_q\,R$ for $\hat\epsilon=0$ (bottom, blue), $1$ (middle, red), and $9$ (top, brown). } 
\label{curlyF}
\end{figure}

For our system, we can obtain the embedding function $\rho(x)$ by numerical integration of the differential equation (\ref{Euler_Lagrange_disk}) and then we can get the functions $S(R)$ and ${\cal F}(R)$ by using (\ref{entropy_total_disk}) and  the definition of ${\cal F}(R)$ in (\ref{calF_definition}). The results for the latter are plotted as a function of 
$R\,r_q\propto R\, m_q$ in Fig. \ref{curlyF}. We notice that  ${\cal F}(R)$ is a monotonically decreasing function that interpolates smoothly between  the two limiting values at $R\,m_q=0$ and $R\,m_q\to\infty$. The UV limit of ${\cal F}$ at small $R\,m_q$ equals the free energy (\ref{F_UV}) of the massless flavored theory, whereas for large $R\,m_q$ the function ${\cal F}$ approaches the free energy of the unflavored ABJM model (\ie,  the value in (\ref{F_UV}) with $\xi=1$). This behavior is in agreement with the general expectation in \cite{Liu:2012eea,Casini:2012ei}  and corresponds to a smooth decoupling of the massive flavors as their mass $m_q$ is increased continuously.  

We will study the UV and IR  limits of ${\cal F}$ analytically in the next two subsections.  Some details of these calculations are deferred to Appendix \ref{entanglement_appendix}, where we also study the entanglement entropy for the strip geometry.

\subsection{UV limit}
\label{UV_limit_disk_entanglement}
In order to study the UV limit of the disk entanglement entropy, 
let us write the Euler-Lagrange equation (\ref{Euler_Lagrange_disk}) when $H(x)$ and $G(x)$ are given by their asymptotic values (\ref{H_G_UV}):
\beq
{d\over dx}\,\Bigg[\,
x^{{2\over b}}\,{\rho\,\rho'\over 
\sqrt{\rho\,'^{\,2}+G_{\infty}\,
x^{-2-{2\over b}}}}\,\Bigg]-\,x^{{2\over b}}\,\,\sqrt{\rho\,'^{\,2}+G_{\infty}\,
x^{-2-{2\over b}}}\,=\,0\,\,.
\label{profile_eq_UV}
\eeq
This equation can be solved exactly by the function:
\beq
\rho_{UV}(x)\,=\,\sqrt{R^2\,-\,b^2\,G_{\infty}\,x^{-{2\over b}}}\,\,,
\label{rho_exact}
\eeq
which clearly satisfies the initial conditions (\ref{bc_x_tip}), with the following value of  the turning point coordinate $x_*$:
\beq
x_{*}^{2/ b}\,=\,{b^2\,G_{\infty}\over R^2}\,\,.
\label{xstar}
\eeq
Since $G_{\infty}\propto r_q^{-2}$, it follows from (\ref{xstar}) that $x_{*}^{2/ b}\propto (r_q\,R)^{-2}$. Therefore, the turning point coordinate $x_*$ is large if $r_q$ or $R$ are small. In this case it would be justified to use the asymptotic UV values of the functions $G$ and $H$, since the minimal surface $\Sigma$ lies entirely in the large $x$ region.  Notice also that (\ref{rho_exact}) can be written  in terms of $x_*$ as:
\beq
\rho_{UV}\,=\,R\,\sqrt{1-\Big({x_*\over x}\Big)^{2/b}}\,\,.
\eeq
In order to calculate the entropy in this UV limit it is very useful to use the following relation satisfied by the function $\rho_{UV}(x)$ written in (\ref{rho_exact}):
\beq
\rho_{UV}\,\sqrt{\big(\rho\,'_{UV}\big)^{2}+G_{\infty}\,
x^{-2-{2\over b}}}\,=\,R\,\sqrt{G_{\infty}}\,\, x^{-1-{1\over b}}\,\,.
\label{rhoUV_relation}
\eeq
Making use of (\ref{rhoUV_relation})  in (\ref{entropy_total_disk}), we find the following expression for the entanglement entropy:
\beq
S_{UV}(R)\,=\,{\pi\,V_6\over 2 G_{10}}\,\,
R\,\sqrt{H_{\infty}\,G_{\infty}}\,
\int^{x_{max}}_{x_*}\,x^{{1\over b}-1}\,dx\,\,.
\eeq
The divergent part of this integral is due to its upper limit and is just given by (\ref{Sdiv_disk}). The finite part of $S$ is:
\beq
S_{finite,UV}\,=\,-{\pi\,V_6\over 2 G_{10}}\,\,b\,
\sqrt{H_{\infty}\,G_{\infty}}\,\,R\,x_*^{1\over b}\,=\,-{\pi\,V_6\over 2 G_{10}}\,\,
b^2\,G_{\infty}\,\sqrt{H_{\infty}}\,\,,
\label{Sfinite_UV_disk}
\eeq
where, in the second step, we used (\ref{xstar}) to eliminate $x_*$. Notice that  the right-hand side of (\ref{Sfinite_UV_disk}) is independent  of the disk radius $R$. Moreover, by using (\ref{HG_infinity})  and (\ref{bHG_F}) we find that:
\beq
S_{finite,UV}\,=\,-F_{UV} ({\mathbb S}^3)\,\,.
\eeq
Therefore, in this UV limit, the dependence on $R$ of the entanglement entropy takes the form (\ref{CFT-entropy}), where $\beta$ is just the free energy of the massless flavored theory on the three-sphere. It follows trivially from this form of $S_{UV}$ and  the definition (\ref{calF_definition}) that ${\cal F}_{UV}=\beta$ and therefore:
\beq
{\cal F}_{UV}\,\equiv\,
\lim_{r_qR\to 0}\,{\cal F}(R)\,=\,
F_{UV} ({\mathbb S}^3)\,\,.
\label{calF_UV}
\eeq
It is also possible to compute analytically the first correction to  (\ref{calF_UV}) for small values of  $r_q\,R$. The details of this calculation are given in Appendix 
\ref{entanglement_appendix}. Here we will just present the final result, which can be written as:
\beq
{\cal F}(R)\,=\,F_{UV} ({\mathbb S}^3)\,+\,c_{UV}\,(r_q\,R)^{2(3-\Delta_{UV})}
\,+\,\cdots
\,\,,
\label{calF_nearUV}
\eeq
where $c_{UV}$ is a constant coefficient depending on the deformation parameter (see eq. (\ref{c_UV})) and $\Delta_{UV}\,=\,3-b$ is the dimension of the quark-antiquark bilinear operator in the UV flavored theory (this dimension was found  in Section 7.3 of \cite{Conde:2011sw} from the analysis of the fluctuations of the flavor branes, see also \cite{Jokela:2012dw}). It is interesting to point out that (\ref{calF_nearUV}) is the  behavior expected \cite{Liu:2012eea} for a flow caused by a source deformation with a relevant operator of dimension $\Delta_{UV}$. Moreover, one can verify that $c_{UV}$ is negative for all values of the deformation parameter $\hat \epsilon$, which confirms that the UV fixed point is a local maximum of  ${\cal F}$.

\subsection{IR limit}
\label{IR_limit_disk_entanglement}

\begin{figure}[ht]
\center
\includegraphics[width=0.75\textwidth]{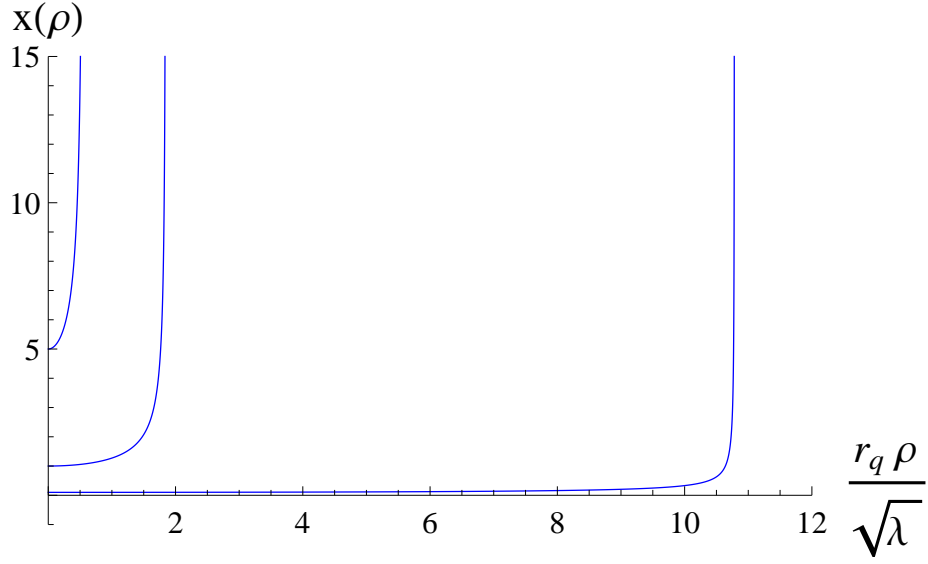}
\caption{Representation of the  embedding function $\rho(x)$  for different values of the turning point $x_*=0.1, 1, 5$ and $\hat\epsilon = 9$.} 
\label{profiles}
\end{figure}

Let us now analyze the IR limit of the entanglement entropy $S(R)$ and of the function ${\cal F}(R)$. This limit occurs when the 8d surface $\Sigma$ penetrates deeply into the geometry and, therefore, when the coordinate $x_*$ of the
turning point is small ($x_*\ll 1$). This happens when either the disk radius $R$  or the quark mass $m_q=r_q/2\pi$ are large.  Looking at the embedding function  $\rho(x)$ obtained by numerical integration of (\ref{Euler_Lagrange_disk}) one notices that, when $x_*$ is small, the function $\rho(x)$ is approximately constant and equal to its asymptotic value $\rho=R$ in the region $x\ge 1$ (see Fig. \ref{profiles}). Therefore, the dependence of $\rho$ on the holographic coordinate $x$ is determined by the integral of (\ref{Euler_Lagrange_disk})  in the region 
$x\le 1$, where the background is given by the unflavored running solution. Actually, when $x_*$ is small it is a good approximation to consider (\ref{Euler_Lagrange_disk}) for the unflavored background, \ie, when the constants $c,\gamma\to 0$, with $\gamma/c$ fixed and given by $\gamma/c=r_q$. In this limit the different functions of the background are:
\beq
e^{f}\approx e^{g}\approx r_q\,x\,\,,
\qquad\qquad
h\approx {1\over r_q^4}\,\,{2\pi^2\,N\over k}\,{1\over x^4}\,\,.
\label{functions_deep_IR}
\eeq
It follows that $G(x)$ and $H(x)$, as defined in (\ref{G_def}) and (\ref{H_def}), are then given by:
\beq
G_{IR}(x)\,=\,{1\over r_q^2}\,{2\pi^2\,N\over k}\,{1\over x^4}\,=\,G_{0}\,x^{-4}\,\,,
\qquad\qquad
H_{IR}(x)\,=\,r_q^4\,{4\pi^4\,N^2\over k^2}\,e^{-4\phi_{IR}}\,x^4\,=\,H_{0}\,x^{4}\,\,,
\label{G_H_IR}
\eeq
where, in the last step, we have defined the constants $G_0$ and $H_0$, and 
$\phi_{IR}$ is the constant dilaton corresponding to the unflavored background,
\beq
e^{\phi_{IR}}\,=\,2\sqrt{\pi}\,\Big({2N\over k^5}\Big)^{{1\over 4}}\,\,.
\eeq
Let us now elaborate  on the expression (\ref{entropy_total_disk}) for the entanglement entropy. We split the integration interval of the variable $x$ as 
$[x_*,\infty]=[x_*,1]\cup [1,\infty]$ and take  into account that one can put $\rho=R={\rm constant}$ in the region $x\ge 1$. We get:
\beq
S_{IR}(R)\,=\,
{\pi\,V_6\over 2 G_{10}}\,\int_{x_*}^{1} dx
\,\rho\,\sqrt{H_{IR}(x)}\,\sqrt{\rho\,'^{\,2}+G_{IR}(x)}\,+\,
{\pi\,V_6\over 2 G_{10}}\,R\,\int_{1}^{x_{max}} \,\,
\sqrt{H(x)\,G(x)}
\,\,.
\label{entropy_disk_IR}
\eeq
The second term in (\ref{entropy_disk_IR}) is linear in $R$ and will not contribute to ${\cal F}(R)$. To evaluate the first integral in (\ref{entropy_disk_IR}) we must determine the embedding function $\rho(x)$ by integrating  (\ref{Euler_Lagrange_disk}) when $G(x)$ and $H(x)$ are given by their IR values (\ref{G_H_IR}). The resulting equation is just the same as (\ref{profile_eq_UV}) with $G_{\infty}$ and $H_{\infty}$  substituted by $G_0$ and $H_0$ and $b=1$. Then, the function $\rho(x)$  can be written as in (\ref{rho_exact}), 
\beq
\rho_{IR}(x)\,=\,\sqrt{\hat R^{\,2}\,-\,G_0\,x^{-2}}\,\,,
\label{rho_IR_sol}
\eeq
where $\hat R$ is a constant. By requiring that $\rho_{IR}(x=1)=R$, we get:
\beq
\hat R^{\,2}\,=\,R^2\,+\,G_0\,\,.
\eeq
It follows from (\ref{rho_IR_sol})  that the coordinate $x_*$ of the turning point is given by:
\beq
x_{*}^{2}\,=\,{G_0\over \hat R^2}\,=\,{G_0\over G_0+R^2}\,\,.
\label{x_star_IR}
\eeq
Notice that when $r_q\to\infty$ (and $G_0\to 0$) or $R$ is large one can neglect the $G_0$ in the denominator of (\ref{x_star_IR}) and then 
$x_{*}^{2}\,\approx\, G_0\,R^{-2}\propto (r_q\,R)^{-2}$, which is a small number. Moreover, by using the explicit form (\ref{rho_IR_sol}) of $\rho(x)$ in this IR region, we get:
\beq
\rho_{IR}\,\sqrt{(\rho'_{IR})^2+G_{IR}}\,=\,\hat R\,\sqrt{G_{0}}\,\, x^{-2}\,\,,
\eeq
and the first integral in (\ref{entropy_disk_IR}) can be explicitly evaluated:
\beq
\int_{x_*}^{1} dx
\,\rho_{IR}\,\sqrt{H_{IR}(x)}\,\sqrt{(\rho'_{IR})^2+G_{IR}(x)}\,=\,
\hat R\,\sqrt{G_0\,H_0}\,-\,G_0\,\sqrt{H_0}\,\,.
\label{S_IR_integral}
\eeq
Notice that, at leading order $\hat R\approx R$ and, thus, the first term in (\ref{S_IR_integral}) does not contribute to ${\cal F}(R)$. Then, the IR limit of ${\cal F}$ is determined by the second contribution in (\ref{S_IR_integral}) and given by:
\beq
{\cal F}_{IR}\,=\,{\pi\,V_6\over 2 G_{10}}\,\,G_0\,\sqrt{H_0}\,\,.
\eeq
Moreover, from the values of $G_0$ and $H_0$ written in (\ref{G_H_IR}) we get:
\beq
G_0\,\sqrt{H_0}\,=\,{\pi^3\over \sqrt{2}}\,N^{{3\over 2}}\,k^{{1\over 2}}\,=\,
{3\pi^2\over 2}\,F_{IR} ({\mathbb S}^3)\,\,,
\eeq
where $F_{IR} ({\mathbb S}^3)$ is the free energy on the three-sphere of the unflavored ABJM theory:
\beq
F_{IR} ({\mathbb S}^3)\,=\,{\pi\sqrt{2}\over 3}\,k^{{1\over 2}}\,N^{{3\over 2}}\,\,.
\eeq
It follows that the IR limit of the ${\cal F}$ function is:
\beq
{\cal F}_{IR}\,
\equiv\,
\lim_{r_qR\to \infty}\,{\cal F}(R)\,=\,
F_{IR} ({\mathbb S}^3)\,\,,
\label{calF_IR}
\eeq
as expected in the deep IR limit in which the flavors become infinitely massive and can therefore be integrated out. The corrections to the result (\ref{calF_IR}) near the IR fixed point could be obtained by applying the techniques recently introduced in \cite{Liu:2013una}. We will not attempt to perform this calculation here.

\section{Wilson loops and the quark-antiquark potential}
\label{Wilson}
In this section we evaluate the expectation values of  the Wilson loop and the corresponding quark-antiquark potential for our model. We will employ the standard holographic prescription of refs. \cite{Maldacena:1998im,Rey:1998ik}, in which one considers a fundamental string hanging from the UV boundary. Then, one computes the regularized Nambu-Goto action for this configuration, from which the $q\bar q$ potential energy can be extracted. In a theory with dynamical flavors this potential energy contains information about the screening of external charges by the virtual quarks popping out from the vacuum. In our case we expect  having a non-trivial flow connecting two conformal behaviors as we move from the UV regime of small  $q\bar q$ separation (in units of the quark mass $m_q$) to the IR regime of large $q\bar q$ distance. We will verify below that this expectation is indeed fulfilled by our model.

Let us denote by $(t,x^1,x^2)$ the Minkowski coordinates and consider a fundamental string for which we take $(t,x^1)$ as its worldvolume coordinates. If the embedding is characterized by a function $x=x(x^1)$, with $x$ being the holographic coordinate, the induced metric is:
\beq
ds_2^2\,=\,-h^{-{1\over 2}}\,dt^2\,+\,h^{{1\over 2}}\,\Big[\,
{e^{2g}\over x^2}\,x'^{\,2}\,+\,h^{-1}\,\Big]\,(dx^1)^2\,\,,
\eeq
where $x'$ denotes the derivative of $x$ with respect to $x^1$. The Nambu-Goto Lagrangian density takes the form:
\beq
{\cal L}_{NG}\,=\,{1\over 2\pi}\,\sqrt{-\det g_2}\,=\,
{1\over 2\pi}\,\sqrt{{e^{2g}\over x^2}\,x'^{\,2}\,+\,h^{-1}}\,\,.
\eeq
As ${\cal L}_{NG}$ does not depend on $x^1$, we have the following conservation law:
\beq
x'\,{\partial\,{\cal L}_{NG}\over \partial x'}\,-\,{\cal L}_{NG}\,=\,{\rm constant}\,\,.
\eeq
Therefore, if $x_*$ denotes the turning point of the string, we have the first integral of the equations of motion:
\beq
\sqrt{1+{e^{2g}\,h\over x^2}\,x'^{\,2}}\,=\,{\sqrt{h_*}\over \sqrt{h}}\,\,,
\label{first_integral}
\eeq
where $h_*\equiv h(x=x_*)$. Then $x'$ is given by:
\beq
x'\,=\,\pm {x \sqrt{h_*-h(x)}\over e^{g(x)}\,h(x)}\,\,,
\eeq
where the two signs correspond to the two branches of the hanging string. 
The $q\bar q$ separation $d$ in the $x^{1}$ direction is:
\beq
d\,=\,2\,\int_{x_*}^{\infty}\,
{e^{g(x)}\,h(x)\over x\,\sqrt{h_*-h(x)}}\,\,dx\,\,.
\label{d_x-integral}
\eeq
In order to compute the potential energy of the $q\bar q$ pair, let us evaluate the on-shell action. By using the first integral (\ref{first_integral}) it is straightforward to check that the on-shell value of the Nambu-Goto Lagrangian density  is:
\beq
{\cal L}_{NG} ({\rm on-shell})\,=\,{1\over 2\pi}\,{\sqrt{h_*}\over h}\,\,.
\eeq
Therefore, the on-shell action becomes:
\beq
S_{{\rm on-shell}}\,=\,{T\over \pi}\,\,\int_{x_*}^{\infty}\,
{e^{g(x)}\over x}\,\,{\sqrt{h_*}\over \sqrt{h_*-h(x)}}\,dx\,\,,
\label{unregularized_NGaction}
\eeq
where $T=\int dt$. The integral (\ref{unregularized_NGaction}) is divergent and must be regularized as in \cite{Maldacena:1998im,Rey:1998ik} by subtracting the action of two straight strings stretched  between the origin and the UV boundary, which corresponds to subtracting the (infinite) quark masses in the static limit. After applying this procedure we arrive at the following  expression for the regulated on-shell action:
\beq
S_{{\rm on-shell}}^{{\rm reg}}\,=\,S_{{\rm on-shell}}\,-\,{T\over \pi}
\int_{0}^{\infty}\,{e^{g(x)}\over x}\,\,dx\,\,,
\eeq
from which we get the $q\bar q$ potential energy:
\beq
E_{q\bar q}\,=\,{1\over \pi}\,\int_{x_*}^{\infty}\,\,
{e^{g(x)}\over x}\,\,\Big[\,{\sqrt{h_*}\over \sqrt{h_*-h(x)}}\,-\,1\Big]
dx\,-\,{r_*\over \pi}\,\,,
\label{E_x-integral}
\eeq
where $r_*$ is the $r$ coordinate of the turning point:
\beq
r_*\,=\,\int^{x_*}_{0}\,{e^{g(x)}\over x}\,dx\,\,.
\eeq

\begin{figure}[ht]
\center
\includegraphics[width=0.75\textwidth]{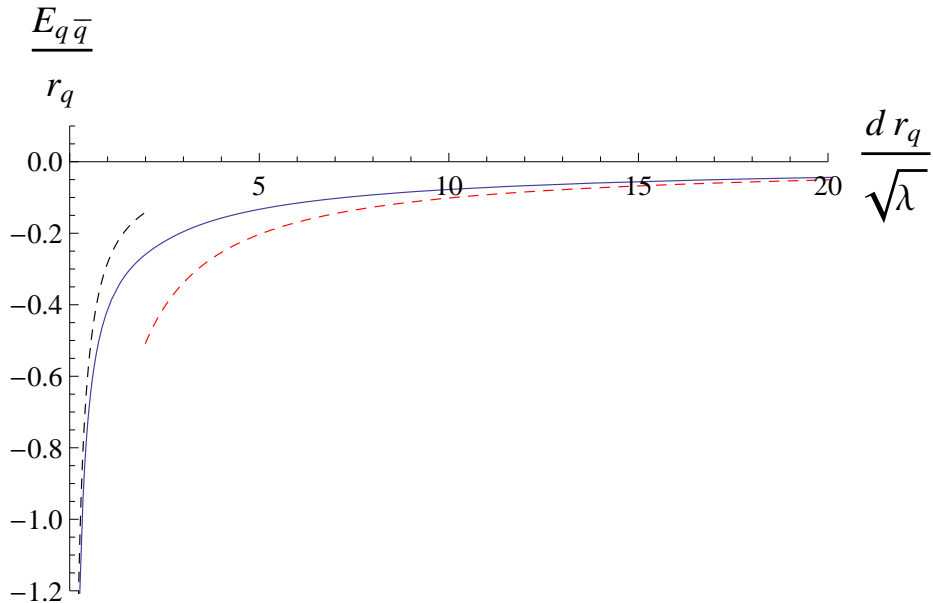}
\caption{Plot of the $q\bar q$ potential energy for $\hat\epsilon=9$. The numerical result is compared to the UV leading estimate (\ref{leading_UV_pot}) (black dashed curve  on the left) and to the leading IR potential (red dashed curve on the right). } 
\label{quark_potential}
\end{figure}

From (\ref{d_x-integral}) and (\ref{E_x-integral}) we have computed numerically the potential energy $E_{q\bar q}$ as a function of the $q\bar q$ distance $d$. The result of this numerical calculation is shown in Fig. \ref{quark_potential}.  As mentioned above, we expect to have a potential energy which interpolates between the  two conformal behaviors with $E_{q\bar q}\propto 1/d$ at the UV and IR. Actually, in the limiting cases in which  $r_q d$ is small or large the function $E_{q\bar q}(d)$ can be calculated analytically (see Appendix \ref{asymp_qq_potential}). In both cases $E_{q\bar q}\propto 1/d$, but with different coefficients. Indeed, in the UV limit $r_q\,d\to 0$, the $q\bar q$ potential can be approximated as:\footnote{Also, the first correction to the UV conformal behavior (\ref{leading_UV_pot})  is computed in Appendix \ref{asymp_qq_potential}.}
\beq
E_{q\bar q}^{UV}\approx-
{4\pi^3\,\sqrt{2\lambda}\over  \big[\Gamma\big({1\over 4}\big)\big]^4}\,\,\sigma\,\,
{1\over d}\,\,,
\qquad\qquad
( r_q\,d\to 0)\,\,,
\label{leading_UV_pot}
\eeq
where $\lambda=N/k$ is the 't Hooft coupling and $\sigma$ is the so-called screening function:
\beq
\sigma\,=\,\sqrt{
{2-q_0\over q_0(q_0+\eta_0 q_0-\eta_0)}}\,\,\,\,b^2\,=\,
{1\over 4}\,\,
{q_0^{{3\over 2}}\,\,(\eta_0+q_0)^2\,(2-q_0)^{{1\over 2}}
\over
(q_0+\eta_0 q_0-\eta_0)^{{5\over 2}}}\,\,.
\label{screening-sigma}
\eeq
Notice that $\sigma$ encodes all the dependence of the right-hand side of (\ref{leading_UV_pot}) on the number of flavors. Actually, the potential (\ref{leading_UV_pot}) is just the one corresponding to having massless flavors (which was first computed for this model in \cite{Conde:2011sw}), as expected in the high-energy UV regime in which all masses can be effectively neglected. The function $\sigma$ characterizes the corrections of the static $q\bar q$ potential due to the screening produced by the unquenched massless flavors ($\sigma\to 1$ for $N_f\to 0$, whereas $\sigma$ decreases as $\sigma\propto \sqrt{k/N_f}$ for $N_f$ large). In Fig.  \ref{quark_potential} we compare the leading UV result (\ref{leading_UV_pot}) with the numerical calculation in the small $r_q \,d$ region.

Similarly, one can compute analytically the $q\bar q$  potential in the region where $r_q\,d$ is large. At leading order the result is (see Appendix \ref{asymp_qq_potential}):
\beq
E_{q\bar q}^{IR}\approx-
{4\pi^3\,\sqrt{2\lambda}\over  \big[\Gamma\big({1\over 4}\big)\big]^4}\,\,
{1\over d}\,\,,
\qquad\qquad
( r_q\,d\to \infty)\,\,.
\label{leading_IR_pot}
\eeq
In Fig.  \ref{quark_potential} we compare the analytic expression (\ref{leading_IR_pot}) to the numerical result in the large distance region. 
Notice that the difference between (\ref{leading_UV_pot}) and (\ref{leading_IR_pot}) is that the screening function $\sigma$ is absent in (\ref{leading_IR_pot}). Therefore, in the deep IR the flavor effects on the $q\bar q$ potential disappear, which is consistent with the intuition that massive flavors are integrated out at low energies.  

\section{Two-point functions of high dimension operators}
\label{Two-point_section}

In this section we study the two-point functions of bulk operators with high dimension. The form of these correlators can be obtained semiclassically by analyzing the geodesics of massive particles in the dual geometry \cite{Balasubramanian:1999zv,Louko:2000tp,Kraus:2002iv},
\beq
\langle {\cal O}(x)\,{\cal O}(y)\rangle\,\sim \,e^{-m\,{\cal L}_r(x,y)}\,\,,
\label{OO-vev}
\eeq
where $m$ is the mass of the bulk field dual to ${\cal O}$. We are assuming that $m$ is large in order to apply a saddle point approximation in the calculation of the correlator. In (\ref{OO-vev}) ${\cal L}_r(x,y)$ is a regularized length along a spacetime geodesic connecting the boundary points $x$ and $y$. To find these geodesics, let us write the Einstein frame metric of our geometry as:
\beq
ds^2_{10}\,=\,e^{-{\phi\over 2}}\,h^{-{1\over 2}}\,dx^2_{1,2}\,+\,
e^{-{\phi\over 2}}\,h^{{1\over 2}}\,
\Big[\,e^{2g}\,\,{dx^2\over x^2}\,+\,e^{2f}\,ds_{{\mathbb S}^4}^2\,+\,
e^{2g}\,\Big(\,\big(E^1\big)^2\,+\,\big(E^2\big)^2\Big)\,\Big]\,\,.
\eeq
Then, the induced metric for a curve parametrized as $x=x(x^1)$ is:
\beq
ds^2_{1}\,=\,e^{-{\phi\over 2}}\,h^{-{1\over 2}}\,\,
\Big(1+G(x)\,x'{\,^2}\Big)\,(dx^1)^2\,\,,
\eeq
with $x'=dx/dx^1$ and  $G(x)$ is the function  defined in (\ref{G_def}). Therefore, 
the action of a particle of mass $m$ whose worldline is the curve  $x=x(x^1)$ is:
\beq
S\,=\,m\int ds_1\,=\,m\int e^{-{\phi\over 4}}\,h^{-{1\over 4}}\,
\sqrt{1+G(x)\,x'{\,^2}}\,dx^1\,\,.
\label{particle_action}
\eeq
The geodesics we are looking for are solutions of the Euler-Lagrange equation derived from the action (\ref{particle_action}). This equation has a  first integral which is given by:
\beq
e^{{\phi(x)\over 4}}\,\big[h(x)\big]^{{1\over 4}}\,
\sqrt{1+G(x)\,x'{\,^2}}\,=\,
e^{{\phi_*\over 4}}\,h_*^{{1\over 4}}\,\,,
\label{first_integral_particle}
\eeq
where $\phi_*\equiv\phi(x=x_*)$ and $h_*\equiv h(x=x_*)$, with $x_*$ being the $x$ coordinate of the turning point, \ie, the minimum value of $x$ along the geodesic. It follows from (\ref{first_integral_particle}) that:
\beq
x\,'\,=\,\pm {1\over \sqrt{ G(x)}}\,
\sqrt{e^{{1\over 2}\,(\phi_*-\phi(x))}\,\,\Big({h_*\over h(x)}\Big)^{1\over 2}\,-\,1}\,\,.
\eeq
The spatial separation $l$ of the two points in the correlator can be obtained by integrating $1/x'$. We get:
\beq
l\,=\,2\,\int_{x_*}^{\infty}\,dx
{\sqrt{G(x)}\over
\sqrt{e^{{1\over 2}\,(\phi_*-\phi(x))}\,\,\big({h_*\over h(x)}\big)^{1\over 2}\,-\,1}
}\,\,.
\label{l_general}
\eeq
Moreover, the length of the geodesic can be obtained by integrating $ds_1$ over the worldline,
\beq
{\cal L}\,=\,2\,\int_{x_*}^{\infty}\,dx
{e^{-{\phi(x)\over 4}}\,\big[h(x)\big]^{-{1\over 4}}\,\,\sqrt{G(x)}
\over
\sqrt{1-e^{{1\over 2}\,(\phi(x))-\phi_*)}\,\,\big({h(x)\over h_*}\big)^{1\over 2}}
}\,\,.
\eeq
This integral is divergent. In order to regularize it, let us study the UV behavior of the integrand. For large $x$, the functions $h(x)$ and $G(x)$ behave as in (\ref{UVexpansion_h_phi}) and (\ref{H_G_UV}),  respectively. Thus, at leading order  for large $x$, 
\beq
\big[h(x)\big]^{-{1\over 4}}\,\,\sqrt{G(x)}\,\approx {L_0\over b}\,x^{-1}\,\,.
\eeq
In  the UV region $x\to\infty$, the integrand in ${\cal L}$ behaves approximately as
$2\,b^{-1}\,L_0\,e^{-\phi_0/4}\,x^{-1}$, which produces a logarithmic UV divergence when it is integrated. In order to tackle this divergence, let us regulate the integral by extending it up to some cutoff $x_{\max}$ and  renormalize the geodesic length by subtracting the divergent part. Accordingly, we define the renormalized geodesic length as:
\beq
{\cal L}_r\,=\,2\,\int_{x_*}^{x_{max}}\,dx
{e^{-{\phi(x)\over 4}}\,\big[h(x)\big]^{-{1\over 4}}\,\,\sqrt{G(x)}
\over
\sqrt{1-e^{{1\over 2}\,(\phi(x))-\phi_*)}\,\,\big({h(x)\over h_*}\big)^{1\over 2}}
}\,-\,{2\,L_0\,e^{-{\phi_0\over 4}}\over b}\,\log ({\cal C}\,x_{max})\,\,,
\label{renormalized_geodesic}
\eeq
where ${\cal C}$ is a constant to be fixed by  choosing a suitable normalization condition for the correlator. 

Our background interpolates between two limiting  $AdS_4$ geometries, at the UV and IR, with different radii. For an equal-time two-point function 
$\Big\langle {\cal O}(t, l)\,{\cal O}(t, 0)\Big\rangle$ the UV and IR limits should correspond to the cases in which $r_q\,l$ is small or large, respectively.  At the two endpoints  of the flow,  the theory is conformal invariant and the two-point correlator behaves as a power law in  $l$. We can use this fact to fix the normalization constant ${\cal C}$ in (\ref{renormalized_geodesic}). Actually, we will assume that 
 the field ${\cal O}$  is canonically normalized in the short-distance $r_q\,l\to 0$ limit and, therefore, the  UV limit of the two-point correlator is:
\beq
\Big\langle {\cal O}(t, l)\,{\cal O}(t, 0)\Big\rangle_{UV}\,=\,
{1\over  \big(r_q\,l/ \sqrt{\lambda} \big)^{2\Delta_{UV}}}\,\,,
\qquad\qquad
(r_q\,l\to 0)\,\,,
\label{UV_two-point}
\eeq
where the $\sqrt{\lambda}$ and $r_q$  factors have been introduced for convenience. 
In (\ref{UV_two-point}) $\Delta_{UV}$ is the conformal dimension of the operator ${\cal O}$ in the UV CFT, which for the dual of  a bulk field of mass $m$ is:
\beq
\Delta_{UV}\,=\,m\,L_0\,e^{-{\phi_0\over 4}}\,\,,
\label{DeltaUV}
\eeq
where we have taken into account that $m$ is large and that $L_0\,e^{-{\phi_0\over 4}}$ is the $AdS_4$ radius of the UV massless flavored geometry in the Einstein frame. It is shown in Appendix \ref{asymp_two-point_functions} that, indeed,  the correlators derived from (\ref{renormalized_geodesic}) display the canonical form (\ref{UV_two-point}) if the constant ${\cal C}$ is chosen appropriately (see (\ref{calC_value})).  In Appendix \ref{asymp_two-point_functions} we have also computed the first deviation from the conformal UV behavior. In this case the numerator on the right-hand side of (\ref{UV_two-point}) is not one but a function $f_{\Delta}(r_q\,l/\sqrt{\lambda})$ such that $f_{\Delta}(r_q\,l/\sqrt{\lambda}=0)=1$. We show in Appendix 
\ref{asymp_two-point_functions} that  $f_{\Delta}(r_q\,l/\sqrt{\lambda})-1\propto (r_q\,l/\sqrt{\lambda})^{2b}$ for small 
$r_q\,l$. The explicit form of the first correction to the non-conformal behavior can be computed analytically  from the mass corrections of Section \ref{UV_mass_corrections} and Appendix \ref{UV_asymptotics} (see eqs. (\ref{nearUV_corr})-(\ref{cDelta})). 
\begin{figure}[ht]
\center
\includegraphics[width=0.75\textwidth]{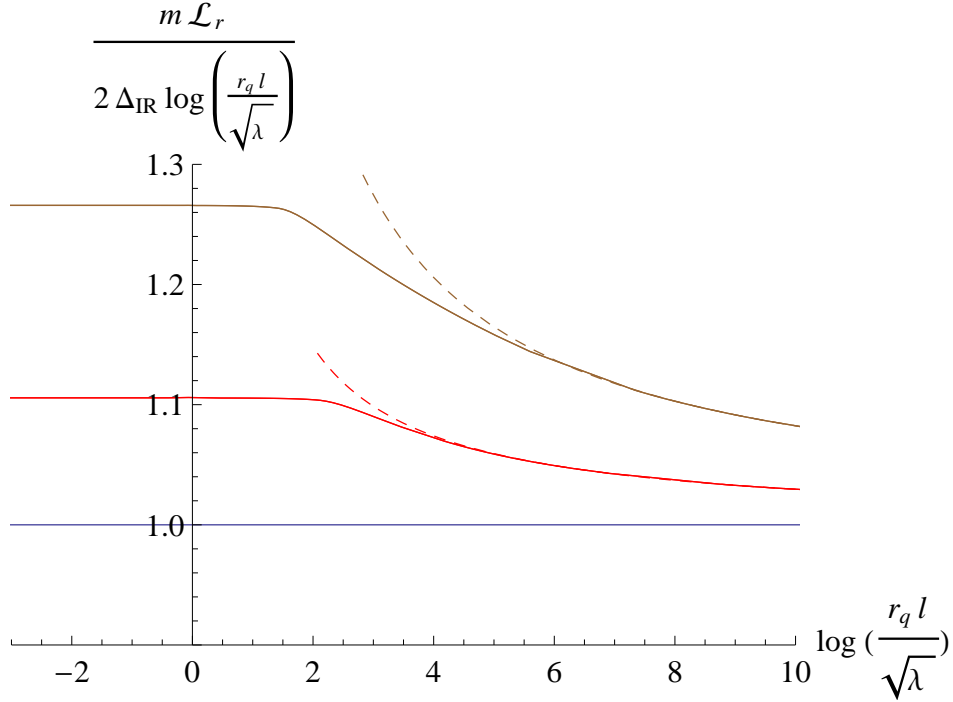}
\caption{Plot of $m\,{\cal L}_r/\big(2\Delta_{IR}\,\log (r_q\,l/\sqrt{\lambda})\big)$ versus the logarithm of 
$r_q\,l/\sqrt{\lambda}$. Notice that, according to (\ref{OO-vev}), 
$m\,{\cal L}_r=-\log \langle {\cal O}(t, l)\,{\cal O}(t, 0)\rangle$. The  three curves correspond to $\hat\epsilon=0$ (bottom, blue),  $\hat\epsilon=1$ (middle, red), and
$\hat\epsilon=9$ (top, brown). In the deep UV ($r_q\,l\to 0$) the curves approach the constant value $\Delta_{UV}/\Delta_{IR}$. The dashed curves correspond to the behavior 
(\ref{VEV_IR}), with the normalization constant ${\cal N}$ given in (\ref{Cal_N}).
}
\label{two-point_function}
\end{figure}

When the distance $r_q\,l$ is large the theory reaches a  new conformal point. Accordingly, the two-point function should behave again as a power law. Notice, however, that the conformal dimension $\Delta_{IR}$ in the IR of an operator dual to a particle of mass $m$ is different from the UV value (\ref{DeltaUV}). Indeed, in the IR the conformal dimension $\Delta_{IR}$ for an operator ${\cal O}$ of mass $m$ is the one corresponding to the unflavored ABJM theory,
\beq
\Delta_{IR}\,=\,m\,L_{ABJM}\,e^{-{\phi_{ABJM}\over 4}}\,\,,
\label{Delta_IR}
\eeq
where $L_{ABJM}$ and $\phi_{ABJM}$ are given, respectively, in (\ref{ABJM-AdSradius}) and (\ref{ABJMdilaton}). Actually, one can check that $\Delta_{UV}\ge \Delta_{IR}$ and that $\Delta_{UV}/ \Delta_{IR}\propto \hat\epsilon^{{1\over 16}}$ for large values of the deformation parameter $\hat\epsilon$. The calculation of the two-point function in the IR limit of large $r_q\,l$ is performed in detail in Appendix  
\ref{asymp_two-point_functions}, with the result:
\beq
\Big\langle {\cal O}(t, l)\,{\cal O}(t, 0)\Big\rangle_{IR}\,=\,
{{\cal N}\over  \big(r_q\,l/ \sqrt{\lambda} \big)^{2\Delta_{IR}}}\,\,,
\qquad\qquad
(r_q\,l \to \infty)\,\,,
\label{VEV_IR}
\eeq
where ${\cal N}$ is a constant whose analytic expression is written  in (\ref{Cal_N}). Notice that 
${\cal N}\not=1$ due to our choice of the constant ${\cal C}$ in (\ref{renormalized_geodesic}), which corresponds to imposing the canonical normalization (\ref{UV_two-point}) to the two-point function in the UV regime. 

The results obtained by the numerical evaluation of the integral (\ref{renormalized_geodesic}) interpolate smoothly between the conformal behaviors (\ref{UV_two-point}) and (\ref{VEV_IR}). This is shown in Fig. \ref{two-point_function}, where we plot 
$-\log \langle {\cal O}(t, l)\,{\cal O}(t, 0)\rangle/(2\Delta_{IR}\,\log(r_q\,l/\sqrt{\lambda}))$ as a function of $\log(r_q\,l/\sqrt{\lambda})$. For small values of $r_q\,l/\sqrt{\lambda}$ the curve asymptotes to the ratio  $\Delta_{UV}/\Delta_{IR}$ of conformal  dimensions, in agreement with (\ref{UV_two-point}), whereas for large  $r_q\,l/\sqrt{\lambda}$ we recover the IR behavior (\ref{VEV_IR}).

\section{Meson spectrum}
\label{mesons}
Let us now test the flow encoded in our geometry by analyzing the mass spectrum of $q\bar q$ bound states. We will loosely refer to these bound states as mesons, although our background is not confining and quarkonia would be a more appropriate name for them. To carry out our analysis we will introduce additional external quarks, with a mass $\mu_q$ not necessarily equal to the mass $m_q$ of the quarks which backreact on the geometry. To distinguish between the two types of flavors we will call valence quarks to the additional ones, whereas the unquenched $N_f$ dynamical flavors of the geometry  will be referred to as sea quarks. The ratio $\mu_q/m_q$ of the masses of the two types of quarks will be an important quantity in what follows. Indeed, $\mu_q/m_q$ is the natural parameter for the holographic renormalization group trajectory.  When  $\mu_q/m_q$  is large (small) we expect to reach a UV (IR) conformal fixed point, whereas for intermediate values of this mass ratio the theory should flow in such a way that the  screening effects produced by the  sea quarks decrease as we move towards the IR. 

Within the context of the gauge/gravity  duality, 
the valence quarks can be introduced by adding an additional flavor D6-brane, which will be treated as a probe in the backreacted geometry. The mesonic mass spectrum can be obtained from the normalizable fluctuations of the D6-brane probe. The way in which the D6-brane probe is embedded in the ten-dimensional geometry preserving the supersymmetry of the background can be determined by using kappa symmetry. For fixed values of the Minkowski and holographic coordinates, the D6-brane extends over a cycle inside the ${\mathbb C}{\mathbb P}^3$ which has two directions along the 
${\mathbb S}^4$ base and one direction along the ${\mathbb S}^2$ fiber.  In order to specify further this configuration,  let us parameterize the $SU(2)$ left invariant one-forms $\omega_i$ of the four-sphere metric (\ref{S4metric})  in terms of three angles $\hat\theta$, $\hat\varphi$ and $\hat \psi$,
\bear
\omega^1 & = & \cos\hat\psi\,d\,\hat\theta+\sin\hat\psi\,\sin\hat\theta\,d\hat\varphi\,\,, \rc
\omega^2 & = & \sin\hat\psi\,d\,\hat\theta-\cos\hat\psi\,\sin\hat\theta\,d\hat\varphi\,\,, \rc
\omega^3 & = & d\hat\psi+\cos\hat\theta \,d\hat\varphi\,\,,
\label{w123}
\eear
with $0\le \hat\theta\le \pi$, $0\le\hat\varphi<2\pi$, $0\le\hat\psi \le 4\pi$.  Then, our D6-brane probe will be  extended along the Minkowski directions and embedded in the geometry in such a way that the angles $\hat\theta$ and $\hat\varphi$ are constant and that the angle $\theta$  of the ${\mathbb S}^2$ fiber depends on the holographic variable $x$. The pullbacks (denoted by a hat) of the left-invariant $SU(2)$ one-forms (\ref{w123})  are $\hat \omega^{1}\,=\,\hat \omega^{2}\,=\,0$ and $\hat \omega^{3}\,=\,d\hat\psi$. 
The kappa symmetric configurations are those for which the function $\theta(x)$ satisfies the first order BPS equation \cite{Conde:2011sw}:
\beq
x\,{d\theta\over dx}\,=\,\cot\theta\,\,,
\eeq
which can be integrated as:
\beq
\cos\theta\,=\,{x_{*}\over x}\,\,.
\eeq
Here $x_*$ is the minimum value of the variable $x$ for the embedding, \ie, the value of $x$ for the tip of the brane. This minimum value of the coordinate $x$ for the embedding is related to the mass $\mu_q$ of the valence quarks introduced by the flavor probe. Indeed, by computing the Nambu-Goto action of a fundamental string stretched in the holographic direction between $x=0$ and $x=x_*$ we obtain $\mu_q$ as:
\beq
\mu_q\,=\,{1\over 2\pi\alpha'}\,\,\int_{0}^{x_*}\,\,
{e^{g(x)}\over x}\,dx\,\,.
\label{muq}
\eeq
In the following we will take the Regge slope $\alpha'=1$. Moreover,  to simplify the description of the embedding,  let us introduce the angular coordinate $\alpha$, defined as follows:
\beq
\xi\,=\,\tan \big({\alpha\over 2}\big)\,\,,
\eeq
and let us  define new angles  $\beta$ and $\psi$ as:
\beq
\beta\,=\,{\hat\psi\over 2}\,\,,\qquad\qquad
\psi\,=\,\varphi\,-\,{\hat\psi\over 2}\,\,,
\label{RP3-angles}
\eeq
where $\varphi$ is the angle in (\ref{cartesian_S2}). 
One can check that the ranges of the new angular variables are  $0\le\alpha <\pi$, $0\le \beta \,,\,\psi< 2\pi$. We will take the following set of worldvolume coordinates for the D6-brane:
\beq
\zeta^{a}\,=\,(x^{\mu}, x, \alpha, \beta, \psi)\,\,.
\eeq
Then, it is straightforward to verify that the  induced metric on the D6-brane worldvolume  takes the form:
\beq
ds^2_{7}\,=\,h^{-{1\over 2}}\,dx_{1,2}^2\,+\,{h^{{1\over 2}}\,e^{2g}\over x^2-x_*^2}\,
dx^2\,+\,h^{{1\over 2}}\,e^{2f}\,\Big[\,d\alpha^2\,+\,\sin^2\alpha\,d\beta^2\,\Big]\,+\,
(x^2-x_*^2)\,{h^{{1\over 2}}\,e^{2g}\over x^2}\,\big(d\psi\,+\,\cos\alpha\,d\beta\big)^2\,\,.
\label{induced_metric}
\eeq
We will restrict ourselves to study a particular set of fluctuations of the D6-brane probe, namely the  fluctuations of the  worldvolume gauge field $A_{a}$. 
The equation for these fluctuations is:
\beq
\partial_{a}\,\Big[e^{-\phi}\,\sqrt{-\det g}\,\,g^{ac}\,g^{bd}\,F_{cd}\,\Big]\,=\,0\,\,,
\label{fluct_eqs}
\eeq
where $g_{ab}$ is the induced metric (\ref{induced_metric}). More concretely,  we will study this equation for the following ansatz for $A_{a}$:
\beq
A_{\mu}\,=\,\xi_{\mu}\,e^{ik_{\nu} x^{\nu}}\,R(x)\,\,,\qquad
(\mu=0,1,2)\,\,,
\qquad\qquad
A_x\,=\,A_i\,=\,0\,\,,
\eeq
where  $\xi_{\mu}$ is a constant polarization vector and $A_i$ denote the components along the angular directions. These modes are dual to the vector mesons of the theory, with $k_{\mu}$ being  the momentum of the meson ($\eta^{\mu\nu}\,k_{\mu}\,k_{\nu}=-m^2$, with $m$ being the mass of the meson). The non-vanishing components of the field strength 
$F_{ab}$ are:
\beq
F_{\mu\nu}\,=\,i(k_{\mu}\,\xi_{\nu}\,-\,k_{\nu}\,\xi_{\mu})\,e^{ik_{\nu} x^{\nu}}\,R(x)\,\,,
\qquad\qquad
F_{x\mu}\,=\,\xi_{\mu}\,e^{ik_{\nu} x^{\nu}}\,R'(x)\,\,.
\eeq
The fluctuation equation (\ref{fluct_eqs}) is trivially satisfied when $b=i$, whereas for $b=x$ it is satisfied if the polarization is transverse:
\beq
\eta^{\mu\nu}\,k_{\mu}\,\xi_{\nu}\,=\,0\,\,.
\eeq
Moreover, by taking $b=\mu$ in (\ref{fluct_eqs}) we arrive at the following differential equation for the radial function $R$:
\beq
\partial_{x}\,\Big[\,{h^{{1\over 4}}\,e^{2f-\phi}\over x}\,(x^2-x_*^2)\,\partial_{x} R\,\Big]\,+\,
m^2\,{h^{{5\over 4}}\,e^{2f+2g-\phi}\over x}\,R\,=\,0\,\,.
\label{meson_fluct_eq}
\eeq
\begin{figure}[ht]
\center
\includegraphics[width=0.75\textwidth]{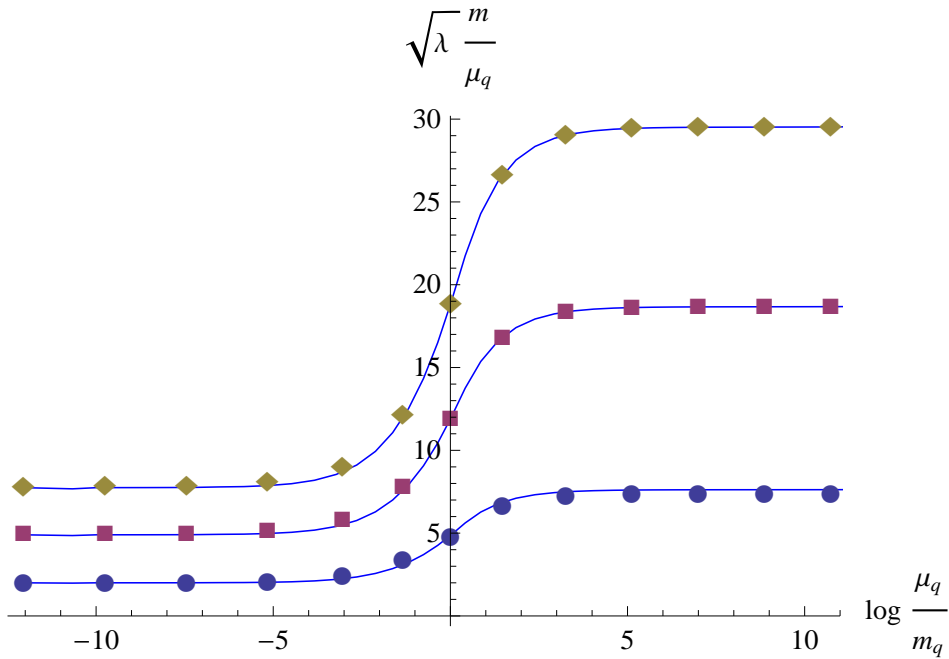}
\caption{Numerical values of the meson  masses for the first three levels  ($n=0,1,2$) as a function of the sea quark mass $m_q$ for deformation parameter $\hat\epsilon=9$. The solid curves depict  the WKB estimate (\ref{WKB_masses}).} 
\label{fig:masses}
\end{figure}

The mass levels correspond  to the   values of $m$ for which there are normalizable solutions of (\ref{meson_fluct_eq}). They can be obtained numerically by the shooting technique. One gets in this way a discrete spectrum depending on a quantization number $n$ ($n\in {\mathbb Z}$, $n\ge 0$). The numerical results for the first three levels are shown in Fig. \ref{fig:masses} as functions of the mass ratio
$\mu_q/m_q$. One notices in these results that the meson masses increase as we move from the IR ($\mu_q/m_q\to 0$) to the UV ($\mu_q/m_q\to \infty$). This non-trivial flow is due to the vacuum polarization effects of the sea quarks, which are enhanced as we move towards the UV and the sea quarks become effectively massless. This is the expected behavior of bound state masses for a theory in the Coulomb phase, since the screening effects reduce effectively the strength of the quark-antiquark force. 

One can get a very accurate description of the flow by applying the WKB approximation. The detailed calculation is presented in Appendix \ref{appendix_WKB_masses}. The WKB formula for the mass spectrum is:
\beq
m_{WKB}\,=\,{\pi\over \sqrt{2}\, \xi(x_*)}\,\sqrt{(n+1)(2n+1)}\,\,,
\qquad\qquad (n=0,1,2,\cdots)\,\,,
\label{WKB_masses}
\eeq
where  $\xi(x_*)$ is the following integral:
\beq
\xi(x_*)\,=\,\int_{x_*}^{\infty}\,dx\,{e^{g(x)}\sqrt{h(x)}\over \sqrt{x^2-x_*^2}}\,\,.
\label{WKB_xi}
\eeq
The WKB mass levels (\ref{WKB_masses}) are compared with those obtained by the shooting technique in Fig. \ref{fig:masses}. We notice from these plots that the estimate  (\ref{WKB_masses}) describes rather well the numerical results along the flow. 
Moreover, we can use the UV and IR limits of the functions $g$ and $h$ to obtain the asymptotic form of the WKB spectrum at the endpoints of the flow. This analysis is performed in detail in Appendix  \ref{appendix_WKB_masses} (see eqs. (\ref{m_WKB_UV}) and (\ref{m_WKB_IR})). As expected, in  the deep IR the mass levels coincide with those of the unflavored ABJM model. In this latter model the mass spectrum of vector mesons can be computed analytically since the fluctuation equation can be solved in terms of hypergeometric functions \cite{Hikida:2009tp,Jensen:2010vx}. When $\mu_q/m_q\to\infty$ the meson masses coincide with those obtained for the massless flavored model of \cite{Conde:2011sw}.

We can use the WKB formulas  (\ref{m_WKB_UV}) and (\ref{m_WKB_IR})  for the 
spectrum at the endpoints of the renormalization group flow to estimate the 
variation generated in the meson masses by changing the sea quark mass $m_q$ and switching on and off gradually the screening effects. It is interesting to point out that, within the WKB approximation, the ratio of these masses only depends on the number of flavors, and is given by:
\beq
{m_{_{WKB}}^{^{(UV)}}\over m_{_{WKB}}^{^{(IR)}}}\,=\,{\sqrt{\pi}\over 2}\,\,
{\Gamma\Big({b+1\over 2b}\Big)
\over 
\Gamma\Big({2b+1\over 2b}\Big)}\,\,{1\over \sigma}\,\,,
\label{UV-IR-mass_ratio}
\eeq
where $\sigma$ is the screening function defined in (\ref{screening-sigma}). As expected, when $N_f=0$ the right-hand side of (\ref{UV-IR-mass_ratio}) is equal to one, 
\ie,  there is no variation of the masses along the flow. On the contrary, when $N_f>0$ the UV/IR mass ratio in (\ref{UV-IR-mass_ratio}) is always greater than one, which means that the masses grow as we move towards the UV and the screening effects become more important. In Appendix \ref{appendix_WKB_masses} we have expanded  
(\ref{UV-IR-mass_ratio}) for low values of the deformation parameter $\hat\epsilon$ (see (\ref{UV/IR_mass_ratio_lowNf})). Moreover, for large  $\hat\epsilon$ the UV/IR mass ratio grows as 
$\sqrt{\hat\epsilon}$ (see (\ref{UV/IR_mass_ratio_largeNf}) for the explicit formula).

\section{Summary and conclusions}
\label{conclu}

In this paper we obtained a  holographic dual to Chern-Simons matter theory with unquenched flavor in the strongly-coupled Veneziano limit. The flavor degrees of freedom were added by a set of D6-branes smeared along the internal directions, which backreacted on the geometry by squashing it, while preserving ${\cal N}=1$ supersymmetry. We considered massive flavors and found a non-trivial holographic renormalization group flow connecting two scale-invariant fixed points: the unflavored ABJM theory at the IR and the massless flavored model at the UV. 

The quark mass $m_q$ played an important role as a control parameter of the solution. By increasing $m_q$ our solutions became closer to the unflavored ABJM model and we smoothly connected  the unquenched flavored model to the ABJM theory without fundamentals. After this soft introduction of flavor no pathological behavior was found. Indeed, our backgrounds had good IR and UV behaviors, contrary to what happens to other models with unquenched flavor \cite{Nunez:2010sf}. This made the ABJM model especially adequate to analyze the effects of unquenched fundamental matter in a holographic setup. 

We analyzed different flavor effects in our model. In general, the screening effects due to loops of fundamentals were controlled by the relative value of the quark mass $m_q$ with respect to the characteristic length scale $l$ of the observable. If $m_q\,l$ was small, which corresponds to the UV regime, the flavor effects were important, whereas they were suppressed if $m_q\,l$ is large, \ie, at the IR. Among the different observables that we analyzed, the holographic entanglement entropy for a disk was specially appropriate since it counts precisely the effective number of degrees of freedom which are relevant at the length scale  given by the radius of the disk. By using the refined entanglement entropy ${\cal F}$ introduced in \cite{Liu:2012eea}, we explicitly obtained the running of ${\cal F}$ and  verified the reduction of the number of degrees at the IR that was mentioned above. The other observables studied also supported this picture.

We end this paper with a short discussion on the outlook. We are convinced that our model could serve as a starting point to gain new insights on the effects of unquenched flavor in other holographic setups.  One possible generalization could be the construction of a black hole for the unquenched massive flavor. Such a background could serve to study the meson melting phase transition which occurs when the tip of the brane approaches the horizon. This system was studied in \cite{Jokela:2012dw}, in the case in which  the massive flavors are quenched and the corresponding flavor brane is a probe.  Another possibility would be trying to find a gravity dual of a theory in which the sum of the two Chern-Simons levels is non-vanishing. According to \cite{Gaiotto:2009mv} we should find a flavored solution of type  IIA supergravity with non-zero Romans mass.

Our program is  also to converge toward increasingly realistic holographic condensed matter models capable of testable predictions. To make contact with any condensed matter system, one is forced to consider non-vanishing components of the gauge field in the background or at the probe level. A natural flow of ideas taking one to land in  the former case typically requires a deep understanding of the probe brane dynamics with worldvolume gauge fields turned on. As an initial step in this direction, we have started exploring what physical phenomena we will encompass by turning on a charge density, magnetic field, and internal flux on the worldvolume of an additional probe D6-brane. The variety of different phenomena seems incredibly rich, much in parallel with recent works on the D3-D7' system \cite{d3d7} and the closely related D2-D8' system \cite{d2d8}. Our findings will be reported elsewhere.

\section*{Acknowledgments}

We are grateful to Daniel Are\'an, Manuel Asorey,  Jose Ignacio Latorre, David Mateos,   Carlos N\'u\~nez, and \'Angel Paredes   for useful discussions. 
The  works of Y.~B., N.~J., 
 and A.~V.~R. are funded in part by the Spanish grant 
FPA2011-22594,  by Xunta de Galicia (Conseller{\'i}a de Educaci\'on, grant 
INCITE09 206 121 PR and grant PGIDIT10PXIB206075PR),  by the 
Consolider-Ingenio 2010 Programme CPAN (CSD2007-00042), and by FEDER.
The work of E.C. is partially supported by IISN - Belgium (conventions 4.4511.06 and 4.4514.08), by the ``Communaut\'e Fran\c{c}aise de Belgique" through the ARC program and by the ERC through the ``SyDuGraM" Advanced Grant.  Y. B. is supported by the Spanish FPU fellowship FPU12/00481.
N.~J. is supported also through the Juan de la Cierva program. 
N.~J. and A.~V.~R. wish to thank Centro de Ciencias de Benasque Pedro Pascual and N.~J. the Kavli IPMU for warm hospitalities while this work was in progress.

\appendix
\vskip 1cm
\renewcommand{\theequation}{\rm{A}.\arabic{equation}}
\setcounter{equation}{0}
\medskip

\section{BPS equations}
\label{BPS}

In this Appendix we will derive the master equation (\ref{master_eq_W}), as well as the equations  that allow to construct the metric and dilaton from the master function $W(x)$ (\ie, (\ref{g-f-W}), (\ref{warp_factor_x}), and (\ref{dilaton_x})). 

Let us begin by writing the BPS equations that guarantee the preservation of ${\cal N}=1$ SUSY. They can be written in terms of the function $\Lambda$ introduced in \cite{Conde:2011sw}, which is  defined as the following combination  of the dilaton and the warp factor:
\beq
e^{\Lambda}\equiv e^{\phi}\,h^{-{1\over 4}}\,\,.
\label{Lambda-def}
\eeq
Then, it was proved in \cite{Conde:2011sw} that $\Lambda(r)$, $f(r)$, and $g(r)$ are solutions to the following system of first-order differential equations:
\bear
&&{d\Lambda\over dr}\,=\,k\,\eta\,e^{\Lambda-2f}\,-\,
{k\over 2}\,e^{\Lambda-2g}\,\,,\rc\rc
&&{d f\over dr}\,=\,{k\,\eta\over 4}\,e^{\Lambda-2f}\,-\,
{k\over 4}\,e^{\Lambda-2g}\,+\,e^{-2f+g}\,\,,\rc\rc
&&{d g\over dr}\,=\,{k\,\eta\over 2}\,e^{\Lambda-2f}\,+\,e^{-g}\,-\,e^{-2f+g}\,\,.
\label{3-system-flavored}
\eear
Moreover, the warp factor $h(r)$ can be recovered from $\Lambda$, $f$, and $g$ through:
\beq
h(r)\,=\,\,e^{-\Lambda(r)}\,\,\Big[\,\alpha\,-\,3\pi^2N\,
\int^r\,\,
e^{2\Lambda(z)-4f(z)-2g(z)}\,\,dz\,\,\Big]\,\,,
\label{warp-integral}
\eeq
where $\alpha$ is an integration constant. Given $h$ and $\Lambda$, the dilaton $\phi$ is obtained from (\ref{Lambda-def}).  The function $K$ of the RR four-form can be related to the other functions of the background by using (\ref{K-N}).  Alternatively, $K$ can be obtained from the BPS system as:
\beq
K\,=\,{d\over dr}\,\,\Big(\,e^{-\phi}\,\,h^{-{3\over 4}}\,\Big)\,\,.
\label{K-phi-h}
\eeq
In terms of the $x$ variable defined in (\ref{r-x-diff-eq}), the BPS system (\ref{3-system-flavored}) becomes:
\bear
&&x\,{d\Lambda\over dx}=k\,\eta\,e^{\Lambda-2f+g}-\frac{k}{2}\,e^{\Lambda-g}\,,\rc\rc
&&x\,{df\over dx}=\frac{k}{4}\,\eta\,e^{\Lambda-2f+g}-\frac{k}{4}\,e^{\Lambda-g}+e^{-2f+2g}\,,\rc\rc
&&x\,{dg\over dx}=\frac{k}{2}\,\eta\,e^{\Lambda-2f+g}+1-e^{-2f+2g}\,.
\label{BPS-system-x}
\eear
In order to reduce this system, let us define as in \cite{Conde:2011sw} the functions $\Sigma(x)$ and $\Delta(x)$,
\beq
\Sigma\,\equiv\, \Lambda-f\,\,,
\qquad\qquad
\Delta\,\equiv\, f-g\,\,.
\eeq
Then, one can easily show that $\Sigma(x)$ and $\Delta(x)$ satisfy the system:
\bear
&&x\,{d\Sigma\over dx}={k\over 4}\,e^{\Sigma}\left(3\eta\,e^{-\Delta}-e^{\Delta}\right)-e^{-2\Delta}\,,\rc\rc
&&x\,{d\Delta\over dx}=-{k\over 4}\,e^{\Sigma}\left(\eta\,e^{-\Delta}+e^{\Delta}\right)-1+2e^{-2\Delta}\,,
\label{system-Sigma-Delta}
\eear
whereas $g$ can be obtained from $\Sigma$ and $\Delta$ by integrating the equation:
\begin{equation}
x\,{d g\over dx}={k\over 2}\,\eta\,e^{\Sigma-\Delta}+1-e^{-2\Delta}\,.
\label{eqn:rec.g}
\end{equation}

Let us next define the master function  $W(x)$ as in (\ref{W_definition}). One immediately verifies that, in terms of the 
functions $\Delta$ and $\Sigma$,  this definition is equivalent to
\beq
W(x)= {4\over k}\,e^{\Delta-\Sigma}\,x\,\,.
\label{W-def}
\eeq
By computing the derivative of (\ref{W-def}) and using the BPS system (\ref{system-Sigma-Delta}), one can easily prove that:
\beq
{dW\over dx}\,=\,{12\over k}\,e^{-\Sigma-\Delta}\,-\,4\eta\,\,.
\label{W-prime}
\eeq
From (\ref{W-prime}) one immediately finds:
\beq
e^{\Sigma+\Delta}\,=\,{12\over k}\,{1\over W'+4\eta}\,\,,
\label{e-Sigma+Delta}
\eeq
where the prime denotes derivative with respect to $x$. Moreover, from the BPS system we can calculate the derivative of $\Sigma+\Delta$ and write the result as:
\beq
x\,{d\over dx}\,\Big(e^{\Sigma+\Delta}\Big)\,=\,
{2x\,\eta\, e^{\Sigma+\Delta}\over W}\,-\,
\Big[{k\over 2}\,e^{\Sigma+\Delta}\,+\,1\Big]\,e^{\Sigma+\Delta}\,+\,{4\over k}\,{x\over W}\,\,.
\label{d-Sigma+Delta}
\eeq
Plugging (\ref{e-Sigma+Delta}) into (\ref{d-Sigma+Delta}), we arrive at the following second-order equation for $W(x)$:
\beq
x\,{d\over dx}\,\Bigg(
{1\over W'+4\eta}\Bigg)\,+\,{W'+4\eta+6\over (W'+4\eta)^2}\,-\,
{x\over 3}\,{W'+10\eta\over W(W'+4\eta)}
\,=\,0\,\,,
\eeq
which can be straightforwardly shown to be equivalent to the master equation (\ref{master_eq_W}). 

Let us see now how one can reconstruct  the full solution from the knowledge of the function $W(x)$. First of all, we notice that from the expression of $W$ in (\ref{W-def}), we get:
\beq
e^{\Sigma-\Delta}\,=\,{4\over k}\,{x\over W}\,\,.
\eeq
By combining this expression with  (\ref{e-Sigma+Delta}) we obtain $\Sigma$ and $\Delta$:
\beq
e^{2\Sigma}\,=\,{48\over k^2}\,\,{x\over W(W'+4\eta)}\,\,,
\qquad\qquad
e^{2\Delta}\,=\,{3 W\over x(W'+4\eta)}\,\,.
\label{Sigma-Delta-q}
\eeq
By noticing that $e^{2\Delta}=q$ we arrive at the representation  of the squashing function $q$ written in 
(\ref{q_W_Wprime}). Moreover, by using this result in (\ref{eqn:rec.g}), we obtain the differential equation satisfied by $g$:
\beq
{dg\over dx}\,=\,{2\eta\over 3W}\,-\,{W'\over 3W}\,+\,{1\over x}\,\,,
\label{g_diff_eq_in_x}
\eeq
which allows to obtain $g(x)$ once $W(x)$ is known. The result of this integration is just the expression written in (\ref{g-f-W}). Moreover, taking into account the expression of the squashing factor $q$  we get precisely the expression of $f$ written in (\ref{g-f-W}).

Let us now compute $\Lambda$ by using $\Lambda=\Sigma+\Delta+g$ and (\ref{e-Sigma+Delta}). We get:
\beq
e^{\Lambda}\,=\,{12\over k}\,{e^{g}\over W'+4\eta}\,\,,
\eeq
and, after using (\ref{g-f-W}), we arrive at:
\beq
e^{\Lambda}\,=\,{12\over k}\,
{x\over W^{{1\over 3}} (W'+4\eta)}\,\exp\Big[{2\over 3}\int^x\,{\eta(\xi)d\xi\over W(\xi)}\Big]\,\,.
\label{Lambda-W}
\eeq
By using this result and (\ref{g-f-W}) in (\ref{warp-integral}), we get that the warp factor can be written as in (\ref{warp_factor_x}). The expression (\ref{dilaton_x}) for the dilaton is just a consequence of the definition  of $\Lambda$ in (\ref{Lambda-def}) and of (\ref{Lambda-W}).

\subsection{Equations of motion}

Let us now verify that the first-order BPS system (\ref{3-system-flavored}) implies the second-order equations of motion for the different fields. Let us work in Einstein frame and write the total action as:
\beq
S\,=\,S_{IIA}\,+\,S_{{\rm sources}}\,\,,
\label{total-action}
\eeq
where the action of type IIA supergravity is given by:
\beq
S_{IIA}\,=\,{1\over 2\kappa_{10}^2}\,\,\Bigg[\,
\int \sqrt{-g}\,\Big(R\,-\,{1\over 2}\,\partial_{\mu}\phi\,\partial^{\mu}\,\phi\,\Big)\,-\,
{1\over 2}\,\int\,\Big[\,e^{{3\phi\over 2}}\,\,{}^*F_2\wedge F_2+
e^{{\phi\over 2}}\,\,{}^*F_4\wedge F_4\Big]\Bigg]\,\,,
\label{IIA-action}
\eeq
and the source contribution is the DBI+WZ action for the set of smeared D6-branes. Let us write this last action as in \cite{Conde:2011sw}. First of all, we introduce a charge distribution three-form $\Omega$. Then, the DBI+WZ action 
is given by:
\beq
S_{{\rm sources}}\,=\,-T_{D_6}\,\int\,\Big(\,e^{{3\phi\over 4}}\,{\cal K}\,-\,C_7\,
\Big)\,\wedge\,\Omega\,\,,
\label{source-action}
\eeq
where the DBI term has been written in terms of the so-called calibration form (denoted by ${\cal  K}$), whose pullback to the worldvolume is equal to the induced volume form  for the supersymmetric embeddings. The expression of  ${\cal  K}$ has been written in \cite{Conde:2011sw}. Let us reproduce it here for completeness:
\beq
{\cal K}\,=\,-e^{012}\,\wedge\big(\,
e^{3458}\,-\,e^{3469}\,+\, e^{3579}\,+\,e^{3678}\,+\,e^{4567}\,+\,e^{4789}\,+\,e^{5689}
\,\big)\,\,,
\label{cal-K-explicit}
\eeq
where the $e^i$'s are the one-forms of the basis corresponding to the forms (\ref{Es}) and (\ref{calS}) (see \cite{Conde:2011sw} for further details). Notice that the equation of motion for $C_7$ derived from (\ref{total-action}) is just $dF_2=2\pi\,\Omega$. Therefore, the $\Omega$ for our ansatz can be read from the right-hand side of (\ref{massive-Omega}).

The Maxwell equations for the forms $F_2$ and $F_4$ derived from (\ref{total-action}) are:
\beq
d\,\Big(\,e^{{3\phi\over 2}}\,\,{}^*F_2\,\Big)\,=\,0\,\,,\qquad\qquad
d\,\Big(\,e^{{\phi\over 2}}\,\,{}^*F_4\,\Big)\,=\,0\,\,,
\label{Maxwell_F2_F4}
\eeq
while the equation for the dilaton is:
\beq
d\,{}^*d\phi\,=\,{3\over 4}\,\,e^{{3\phi\over 2}}\,\,{}^*F_2\wedge F_2+{1\over 4}\,
e^{{\phi\over 2}}\,\,{}^*F_4\wedge F_4\,+\,
{3\over 2}\,\kappa_{10}^2\,T_{D_6}\,\,e^{{3\phi\over 4}}\,{\cal K}\wedge\,\Omega\,\,.
\label{dilaton_eom}
\eeq
One can verify that, for our ansatz, (\ref{Maxwell_F2_F4}) and (\ref{dilaton_eom}) are a consequence of the BPS equations (\ref{3-system-flavored}). To carry out this verification we need to know the radial derivatives of $h(r)$ and $\phi(r)$ (which are not written in (\ref{3-system-flavored})). The derivative of $h$ can be related to the derivative of $\Lambda(r)$,
\beq
{dh\over dr}\,=\,-h\,{d\Lambda\over dr}\,-\,3\pi^2\,N\,e^{\Lambda\,-\,4f\,-\,2g}\,\,.
\eeq
The radial derivative of the dilaton can be put in terms of the derivative of $\Lambda$ and $h$ by using (\ref{Lambda-def}):
\beq
{d\phi\over dr}\,=\,{d\Lambda\over dr}\,+\,{1\over 4h}\,{dh\over dr}\,\,.
\eeq
It remains to check Einstein equations, which read:
\bear
&&R_{\mu\nu}\,-\,{1\over 2}\,g_{\mu\nu}\,R\,=\,
{1\over 2}\,\partial_{\mu}\phi\,\partial_{\nu}\phi\,-\,{1\over 4}\,g_{\mu\nu}\,
\partial_{\rho}\phi\,\partial^{\rho}\phi\,+\,
{1\over 4}\,e^{{3\phi\over 2}}\,\Big[\,2F_{\mu\rho}^{(2)}\,F_{\nu}^{(2)\,\,\rho}\,-\,{1\over 2}\,g_{\mu\nu}\,F_2^2\,\Big]\rc\rc
&&\qquad\qquad\qquad\qquad
+{1\over 48}\,
e^{{\phi\over 2}}\,\Big[\,4F_{\mu\rho\sigma\lambda}^{(4)}\,F_{\nu}^{(4)\,\,\rho\sigma\lambda}\,-\,
{1\over 2}\,g_{\mu\nu}\,F_4^2\,\Big]\,+\,T_{\mu\nu}^{{\rm sources}}\,\,,
\label{Einstein-eq}
\eear
where $T_{\mu\nu}^{{\rm sources}}$ is the stress-energy tensor for the flavor branes, which is defined as:
\beq
T_{\mu\nu}^{{\rm sources}}\,=\,-{2\kappa_{10}^2\over \sqrt{-g}}\,\,
{\delta S_{{\rm sources}}\over \delta g^{\mu\nu}}\,\,.
\label{Tmunu_def}
\eeq
In order to write the explicit expression for $T_{\mu\nu}^{{\rm sources}}$ derived from the definition (\ref{Tmunu_def}), let us introduce the following operation for any two $p$-forms $\omega_{(p)}$ and 
$\lambda_{(p)}$:
\beq
	\omega_{p} \lrcorner \lambda_{(p)} = \frac{1}{p!} \omega^{\mu_1 ... \mu_p}
	 \lambda_{\mu_1 ... \mu_p}\,\,.
\eeq
Then, by computing explicitly the derivative of the action (\ref{source-action}) with respect to the metric,  one can check that:
\beq
T_{\mu\nu}^{{\rm sources}}\,=\,\kappa_{10}^2\,T_{D_6}\,\,e^{{3\phi\over 4}}\,
\Big[\,\,g_{\mu\nu}\,
{}^*{\cal K} \lrcorner\,\Omega\,-\,
{1\over 2}\,\Omega_{\mu}^{\,\,\,\rho\sigma}\,\big({}^*{\cal K}\big)_{\nu\rho\sigma}
\Big]\,\,.
\eeq
It is now straightforward to compute explicitly the different components of this tensor. Written in  flat components in the basis in which the calibration form has the form (\ref{cal-K-explicit}), we get:\footnote{As compared to the  case studied in \cite{Conde:2011sw}, now we have terms proportional to $d\eta/dr$ that were absent for massless flavors.  
}
\begin{align}
&T_{00}=-T_{11}=-T_{22}=k\Big(\,\eta-1+\frac{e^g}{2}\,{d\eta\over d r}\,\Big)\,e^{-2f-g+{3\phi\over 2}}\,\,h^{-{3\over 4}}\,\,,\rc
&T_{33}=k(\eta-1)\,e^{-2f-g+{3\phi\over 2}}\,\,h^{-{3\over 4}}\,\,,\rc
&T_{ab}=-{k\over 2}\,\Big(\,\eta-1+\frac{e^g}{2}\,{d\eta\over d r}\,\Big)\,e^{-2f-g+{3\phi\over 2}}\,\,h^{-{3\over 4}}\delta_{ab}\,\,,
\qquad\qquad (a,b=4,\ldots, 7)\,,\rc
&T_{88}=T_{99}=-{k\over 2}\,\Big(\,\eta-1+e^g\,{d\eta\over d r}\,\Big)\,e^{-2f-g+{3\phi\over 2}}\,\,h^{-{3\over 4}}\,.
\end{align}
By using these values one can verify that,  the Einstein equations (\ref{Einstein-eq}) are indeed satisfied as a consequence of the first-order system (\ref{3-system-flavored}). Notice that, for the massive flavored background of Section \ref{massive_flavor},  $d\eta/dr $ (and, therefore, $T_{\mu\nu}$) has a finite discontinuity at $r=r_q$. It follows from (\ref{Einstein-eq}) that the Ricci tensor $R_{\mu\nu}$  has also a finite jump at this point.

\vskip 1cm
\renewcommand{\theequation}{\rm{B}.\arabic{equation}}
\setcounter{equation}{0}
\medskip

\section{Mass corrections in the UV }
\label{UV_asymptotics}

In this appendix we show how to obtain the first corrections to the conformal behavior of the metric and the dilaton from the UV asymptotic expansion of the master function $W(x)$ written in (\ref{W_asymp_UV}). First we notice that the ratio $\eta_0/ A_0$ can be written in terms of $b$ as:
\beq
{\eta_0\over A_0}\,=\,{3\over 2b}\,-\,1\,\,,
\label{eta0/A0}
\eeq
leading to  a useful identity:
\beq
x^{{1\over b}\,-\,{2\over 3}}\,\,\exp\Big[\,-{2\over 3}\,\int_{1}^{x}\,{\eta_0\over A_0}\,
{d\xi\over \xi}\,\Big]\,=\,1\,\,.
\eeq
Inserting the unit written in this way in the integral appearing in the expression of $e^{g}$ in (\ref{background_functions_xge1}), we get:
\beq
\exp\Big[{2\over 3}\int^x_1\,{\eta(\xi)d\xi\over W(\xi)}\Big]\,=\,x^{{1\over b}\,-\,{2\over 3}}\,\,
J\,\,{\cal G}(x)\,\,,
\eeq
where $J$  is the following constant integral (depending on $\hat\epsilon$):
\beq
J\,\equiv\,\exp\Bigg[{2\over 3}\int^{\infty}_1\,\Big[{\eta(\xi)\over W(\xi)}\,-\,
{\eta_0\over A_0\,\xi}\Big]
d\xi\,\Bigg]\,\,,
\eeq
and ${\cal G}(x)$ is the function
\beq
{\cal G}(x)\equiv 
\exp\Bigg[{2\over 3}\int^{x}_{\infty}\,\Big[{\eta(\xi)\over W(\xi)}\,-\,
{\eta_0\over A_0\,\xi}\Big]
d\xi\,\Bigg]\,\,.
\eeq
Using these results, we can immediately write:
\beq
e^{g}\,=\, r_q\,\Big[{(\hat \gamma+1)^2\over 2\,\hat \gamma}\Big]^{{1\over 3}}\,\,
J\,x^{{1\over b}}\,\,
\Big[{x\over W(x)}\Big]^{{1\over 3}}\,\,{\cal G}(x)
\,\,.
\eeq
The function ${\cal G}(x)$ can be easily expanded for large $x$. At first non-trivial order we get:
\beq
{\cal G}(x)\,=\, 1\,+\,{1\over 3 A_0}\,\Big[\,\eta_0\,-\,1\,+A_2\,{\eta_0\over A_0}\,\Big]\,\,
{1\over x^2}\,+\,\cdots\,\,.
\eeq
Using that, for large $x$,
\beq
\Big[{x\over W(x)}\Big]^{{1\over 3}}\,=\,
{1\over A_0^{{1\over 3}}}\,
\Big[\,1\,-\,{A_2\over 3\,A_0}\,{1\over x^2}\,+\,\cdots\Big]\,\,,
\eeq
we easily arrive at the expansion of $e^g$ written in (\ref{g_f_UVexpansions_x}). 
Let us next  find the large $x$ expansion for $f$, which can be obtained by using  the expansion of $g$ and $q$ in the relation $e^{f}=\sqrt{q}\,e^{g}$. We get:
\beq
e^{f}\,=\, \sqrt{q_0}\,\,
{\kappa\,\, r_q\over b}\,\,
J\,x^{{1\over b}}\,\,
\Big[\,1\,+\,{f_2\over x^2}\,+\,\cdots\Big]\,\,,
\eeq
where the coefficient $f_2$ is given by:
\beq
f_2\,=\,g_2\,+\,{1\over 2}\,{q_2\over q_0}\,=\,g_2\,+\,{2-b\over 2b}\,q_2\,\,.
\label{f2_g2_q2}
\eeq
It is straightforward to demonstrate that (\ref{f2_g2_q2}) coincides with the value of $f_2$ written in (\ref{g2_f2}).

Let us now analyze the UV expansion of $h$. First, we evaluate the integral appearing in 
(\ref{warp_factor_x}):
\beq
\int_{x}^{\infty}\,
{\xi\,e^{-3 g(\xi)}\over W(\xi)^2}
\,d\xi\,=\,
{2\hat \gamma\,b\over 3\,(\hat \gamma+1)^2\,A_0\,(Jr_q)^3}\,\,
x^{-{3\over b}}\,
\Big[\,1\,-{3\over 3+2b}\,\Big(3g_2+2 {A_2\over A_0}\Big)\,{1\over x^2}\,+\,\cdots
\Big]\,\,.
\eeq
By combining this result with the expansions (\ref{g_f_UVexpansions_x}) and (\ref{Wprime-eta-expansion}) we get:
\beq
h(x)\,=\,L_0^4\,
\Big[{2\,\hat \gamma\,A_0\over (\hat \gamma+1)^2 }\Big]^{{4\over 3}}\,\,
{x^{-{4\over b}}\over (J\,b\,r_q)^4}\,\Big[1\,+\,{h_2\over x^2}\,+\,\cdots\Big]\,\,,
\label{UVexpansion_h_x}
\eeq
where $L_0$ is the UV AdS  radius (\ref{L0_explicit}) and the coefficient $h_2$ is:
\beq
h_2\,=\,-{2(6+b)\over 3+2b}\,g_2\,-\,\Big({6\over 3+2b}\,+\,{b\over 3(2-b)}\Big)\,{A_2\over A_0}\,-\,{2\over 3}\,{3-2b\over 2-b}\,{\eta_0-1\over \eta_0}\,\,.
\label{h2_first}
\eeq
One can readily check that the prefactor in (\ref{UVexpansion_h_x}) coincides with the one in (\ref{UVexpansion_h_phi}) and that the coefficient $h_2$ written in  (\ref{h2_first}) is the same as the one in (\ref{h_2_phi_2}).

Let us now obtain the UV expansion of the function $\Lambda(x)$, defined in (\ref{Lambda-def}), which can be written as:
\beq
e^{\Lambda(x)}\,=\,{12\,r_q\,J\over k}\,\,
\Big[{(\hat \gamma+1)^2\over 2\,\hat \gamma}\Big]^{{1\over 3}}\,\,
{x^{{1\over b}}\over W'+4\eta}
\,\,\Big[{x\over W(x)}\Big]^{{1\over 3}}\,\,{\cal G}(x)\,\,.
\eeq
The UV expansion of this expression is:
\beq
e^{\Lambda}\,=\, {12\,r_q\,J\over k\,(A_0+4\eta_0)}\,\,
\Big[{(\hat \gamma+1)^2\over 2\,\hat \gamma\,A_0}\Big]^{{1\over 3}}\,\,
x^{{1\over b}}\,\,
\Big(1\,+\,{\Lambda_2\over x^2}\,+\,\cdots\Big)\,\,,
\eeq
where the coefficient $\Lambda_2$ can be written in terms of $g_2$ as:
\beq
\Lambda_2\,=\,g_2\,+\,{A_2+4(\eta_0-1)\over A_0+4\eta_0}\,=\,
g_2\,+\,{b\over 3(2-b)}\,\Big[{2(3-2b)\over b}\,{\eta_0-1\over \eta_0}\,+\,
{A_2\over A_0}\,\Big]\,\,.
\eeq
By using the value  of $g_2$ in (\ref{g2_f2}), we can find a more explicit expression for  $\Lambda_2$:
\beq
\Lambda_2\,=\,{3-2b\over 3}\,\Big[\,{1\over 2b}\,+\,{2\over 2-b}\,\Big]\,{\eta_0-1\over \eta_0}\,+\,{1\over 3}\Big[\,{3-4b\over 2b}\,+\,{b\over 2-b}\Big]\,\,{A_2\over A_0}\,\,.
\eeq
Let us next obtain the expansion of $e^{\phi}$ by using $e^{\phi}=e^{\Lambda}\,h^{{1\over 4}}$.  We get:
\beq
e^{\phi}\,=\, e^{\phi_0}\,\Big(1\,+\,{\phi_2\over x^2}\,+\,\cdots\Big)\,\,,
\eeq
where the prefactor is
\beq
e^{\phi_0}\,=\,{12\,L_0\over k(A_0+4\eta_0)\,b}\,\,,
\eeq
and can be shown to be the same as the asymptotic value of the dilaton written in (\ref{dilaton-AdS_asymp}). The coefficient $\phi_2$ is:
\beq
\phi_2\,=\,\Lambda_2\,+\,{1\over 4}\,h_2\,\,,
\eeq
which has been written  more explicitly in (\ref{h_2_phi_2}).

Finally,  let us write these results in terms of the $r$ variable. At  next-to-leading order, the relation between the coordinates $x$ and $r$ is obtained by integrating the differential equation:
\beq
{dr\over dx}\,=\,{e^{g}\over x}\,
=\,{\kappa\,r_q\over b}\,\big[\,
x^{{1\over b}-1}\,+\,g_2\,x^{{1\over b}-3}\,+\cdots\big]\,\,.
\eeq
We get:
\beq
r\, =\,\kappa\,r_q\,\big[\,x^{{1\over b}}\,-\,{g_2\over 2b-1}\,
x^{{1\over b}-2}\,+\cdots\big]\,\,.
\label{r-x_relation}
\eeq
This relation can be inverted as:
\beq
x= \Big({r\over \kappa\,r_q}\Big)^{b}\,\Big[\,
1\,+\,{b\over 2b-1}\,g_2\, \Big({ \kappa\,r_q\over r}\Big)^{2b}\,+\,\cdots
\,\Big]\,\,.
\label{x-r_relation}
\eeq
By plugging the expansion (\ref{x-r_relation}) into (\ref{g_f_UVexpansions_x}) and (\ref{UVexpansion_h_phi}) one easily arrives at (\ref{All_UV_expansions_r}) and (\ref{tilde_UV_coeff}).

\vskip 1cm
\renewcommand{\theequation}{\rm{C}.\arabic{equation}}
\setcounter{equation}{0}
\medskip

\section{More on the entanglement entropy}
\label{entanglement_appendix}

Let us study analytically the entanglement entropy and the ${\cal F}$ function near the UV fixed point at $R=0$. We shall represent $S(R)$ in terms of a local functional 
${\cal L}={\cal L}(H(x), G(x), \rho(x))$ as:
\beq
S(R)\,=\,\int_{x_*}^{\infty}\,{\cal L}\,dx\,\,.
\label{entropy_calL}
\eeq
We will use this representation to compute the first-order variation of $S$ when the background functions $H$ and $G$ and the embedding function $\rho$ are varied around their UV values:
\beq
H(x)\,=\,H_{UV}(x)+\delta H(x)\,\,,
\qquad
G(x)\,=\,G_{UV}(x)+\delta G(x)\,\,,
\qquad
\rho(x)\,=\,\rho_{UV}(x)+\delta \rho(x)\,\,,
\label{H_G_rho_UVexpansion}
\eeq
where $H_{UV}=H_{\infty}\,x^{{4\over b}}$, $G_{UV}=G_{\infty}\,x^{-2-{2\over b}}$ and $\rho_{UV}(x)$ is the function written in (\ref{rho_exact}). At first order the corrections  $\delta H(x)$ and $\delta G(x)$ can be parameterized as:
\beq
\delta H(x)\,=\,H_{\infty}\,H_2\,x^{{4\over b}-2}\,\,,
\qquad
\delta G(x)\,=\,G_{\infty}\,G_2\,x^{-4-{2\over b}}\,\,.
\label{deltaHG_UV}
\eeq
The constants $H_2$ and $G_2$  in (\ref{deltaHG_UV}) can be related to the ones characterizing the behavior of the background at the UV:
\beq
H_2\,=\,2\,(h_2+4f_2+2g_2-2\phi_2)\,\,,
\qquad\qquad
G_2\,=\,h_2\,+\,2 g_2\,\,.
\eeq
It is useful  to define the following combination of $H_2$ and $G_2$:
\beq
{\cal H}_2\,=\, {H_2\over 2}+{G_2\over 2b+1}
\label{calH_2}\,\,.
\eeq
The perturbation of the profile $\delta \rho(x)$ is the  correction, at first-order,  of the solution of (\ref{Euler_Lagrange_disk}) which satisfies $\delta \rho(x\to\infty)=0$. 
By computing the first variation of  (\ref{Euler_Lagrange_disk})  we find that  $\delta \rho(x)$ is solution to the following second-order inhomogeneous  differential equation:
\bear
&&{d\over dx}\,\Bigg[{\sqrt{H_{UV}}\,\rho_{UV}\,\rho\,'_{UV}\over \sqrt{(\rho\,'_{UV})^2+G_{UV}}}\,\Big({\delta\rho'\over \rho\,'_{UV}}\,+\,{\delta\rho\over \rho_{UV}}\,+\,
{\delta H\over 2 H_{UV}}\,-\,{1\over 2}\,\,
{2\,\rho\,'_{UV}\,\delta\rho\,'\,+\,\delta G\over (\rho\,'_{UV})^2+G_{UV}}\Big)\Bigg]
\rc\rc\rc
&&\qquad\qquad
-\sqrt{H_{UV}}\,\sqrt{(\rho\,'_{UV})^2+G_{UV}}\,\Big({\delta H\over 2\,H_{UV}}\,+\,
{1\over 2}\,\,
{2\,\rho\,'_{UV}\,\delta\rho\,'\,+\,\delta G\over (\rho\,'_{UV})^2+G_{UV}}\Big)\,=\,0\,\,.
\label{ODE_deltarho}
\eear
More explicitly,  after using the UV values of $H$, $G$, and $\rho$, the differential  equation (\ref{ODE_deltarho}) can be written as:
\bear
&&{d\over dx}\,\Bigg[{2\,x^{{3\over b}+1}\,\rho^4_{UV}\over R^2\,G_{\infty}}\,\,
\delta\rho'\,+\,2b\,x^{{1\over b}}\,\delta\rho\,+\,b\,H_2\,x^{{1\over b}-2}\,\rho_{UV}\,-\,
{b\,G_2\over R^2}\,x^{{1\over b}-2}\,\rho^3_{UV}\Bigg]\rc\rc
&&\qquad\qquad\qquad\qquad
-2b\,x^{{1\over b}}\,\delta\rho'\,-\,{R^2\,H_2\,x^{{1\over b}-3}\over \rho_{UV}}\,-\,
G_2\,x^{{1\over b}-3}\, \rho_{UV}\,=\,0\,\,.
\label{ODE_deltarho_explicit}
\eear

Let us next calculate the first-order variation of the entanglement entropy. From (\ref{entropy_calL}) we get:
\beq
\delta S\,=\,\int_{x_*}^{\infty}\,dx\,\Bigg[
{\partial {\cal L}\over \partial H}\Big|_{UV}\delta H\,+\,
{\partial {\cal L}\over \partial G}\Big|_{UV}\delta G\Bigg]\,+\,
\Pi_{UV}(x)\,\delta\rho (x)\Bigg|^{x=\infty}_{x=x_*}\,\,,
\label{deltaS_UV}
\eeq
where $\Pi_{UV}(x)$ is defined as:
\beq
\Pi_{UV}(x)\equiv 
{\partial {\cal L}\over \partial \rho{\,'}}\Big|_{UV}\,=\,{2\over 3\pi^2}\,
\sqrt{H_{UV}(x)}\,\,{\rho_{UV}\,\rho\,'_{UV}\over 
\sqrt{(\rho\,'_{UV})^2+G_{UV}(x)}}\,\,.
\eeq
Using  the explicit form of $\rho_{UV}$ we get:
\beq
\Pi_{UV}(x)\,=\,{2b\over 3\pi^2}\,
\sqrt{H_{\infty}\,G_{\infty}}\,\,x^{{1\over b}}\,
\sqrt{1\,-\,{b^2\,G_{\infty}\over R^2}\,x^{-{2\over b}}}\,\,,
\eeq
and it follows that:
\beq
\Pi_{UV}(x=x_*)=0\,\,.
\eeq
Therefore, the lower limit contribution to the last term in (\ref{deltaS_UV}) vanishes. Let us now study the contribution to this term of the $x\to\infty$ limit. We will now check
that $\delta\rho(x)$ decreases as $x\to\infty$ in such a way that:
\beq
\lim_{x\to\infty}\,\Pi_{UV}(x)\,\,\delta\rho(x)\,=\,0\,\,.
\label{Pi_rho_xinfty}
\eeq
Thus, the upper limit of the last term in (\ref{deltaS_UV}) also vanishes. To prove (\ref{Pi_rho_xinfty}) we have to integrate the differential equation (\ref{ODE_deltarho_explicit}) and extract the large $x$ behavior of the solution. Amazingly, (\ref{ODE_deltarho_explicit}) can be integrated analytically. Its general solution can be written as the sum of two terms:
\beq
\delta\rho\,=\,\delta\rho_P\,+\delta\rho_G\,\,,
\eeq
where $\delta\rho_P$ is a particular solution of the equation and $\delta\rho_G$ is a general solution of the homogeneous part of (\ref{ODE_deltarho_explicit}). We have found a particular solution  $\delta\rho_P$, which can be written in terms of hypergeometric functions and is given by:
\bear
&&\delta\rho_P={R^{2b+2}{\cal H}_2\over 2b(b-1)(2b-1)G_{\infty}\,\rho_{UV}}\,\times\rc\rc
&&\qquad\qquad
\times
\Bigg[\Big(Rx^{{1\over b}}-b\,\sqrt{G_{\infty}}\Big)^{2-2b}
{}_2F_1\Big(2b-2, 2b-2;2b-1;{b\,\sqrt{G_{\infty}}\over b\sqrt{G_{\infty}}-R\,x^{{1\over b}}}\Big)\qquad\qquad
\rc\rc
&&\qquad\qquad
+\Big(R\,x^{{1\over b}}+b\,\sqrt{G_{\infty}}\Big)^{2-2b}\,
{}_2F_1\Big(2b-2, 2b-2;2b-1;{b\sqrt{G_{\infty}}\over b\sqrt{G_{\infty}}+R\,x^{{1\over b}}}\Big)\Bigg]
\rc\rc
&&\qquad\qquad
-\Big[{b^2\,G_2\,G_{\infty}\over 2\,(2b+1)}\,x^{-{2\over b}}\,+\,{R^2\,{\cal H}_2\over 2b-1}\,+\,{R^4\,{\cal H}_2\over b\,(b-1)\,(2b-1)\,G_{\infty}}\,x^{{2\over b}}\,\Big]\,
{1\over x^2\,\rho_{UV}}\,\,.
\label{delta_rho_particular}
\eear
We are only interested in the behavior of $\delta\rho_P$ for large $x$. It is straightforward to prove that, for large $x$,  $\delta\rho_P$ can be approximated as:
\beq
\delta\rho_P\approx {b^2\,G_{\infty}\,\big[bH_2\,-\,(b-1)\,G_2\big]\over 2\,(b+1)\,(2b-1)\,R}\,\,x^{-2-{2\over b}}\,\,.
\eeq
Thus, as $\Pi_{UV}\propto x^{{1\over b}}$ for large $x$,
\beq
\lim_{x\to\infty}\,\Pi_{UV}(x)\,\,\delta\rho_P(x)\,=\,0\,\,.
\eeq
The general solution of the homogeneous differential equation which vanishes as $x\to\infty$ is:
\beq
\delta\rho_G\,=\,{C\over x^{{1\over b}}\,\rho_{UV}}
\Big[{R\,x^{{1\over b}}\over 2b\,\sqrt{G_{\infty}}}\,
\log\,{R\,x^{{1\over b}}\,+\,b\,\sqrt{G_{\infty}}\over R\,x^{{1\over b}}\,-\,b\,\sqrt{G_{\infty}}}\,\,-\,1
\Big]\,\,,
\label{delta_rho_general}
\eeq
where $C$ is an arbitrary constant. For large $x$ the function in (\ref{delta_rho_general}) behaves as:
\beq
\delta\rho_G\,\approx  {C\,b^2\,G_{\infty}\over 3\,R^3}\,x^{-{3\over b}}\,\,.
\eeq
Therefore,
\beq
\lim_{x\to\infty}\,\Pi_{UV}(x)\,\,\delta\rho_G(x)\,=\,0\,\,.
\eeq
Then, eq. (\ref{Pi_rho_xinfty}) holds  and the last term in (\ref{deltaS_UV}) does not contribute to $\delta S$, as claimed above. Let us now calculate the other two contributions. First of all, the term due to the variation of $H$ is given by:
\bear
&&\int_{x_*}^{\infty}\,dx\,
{\partial {\cal L}\over \partial H}\Big|_{UV}\delta H\,=\,{1\over 3\pi^2}\,
\int_{x_*}^{\infty}\,dx\,\,
{\rho_{UV}\sqrt{(\rho\,'_{UV})^2+G_{UV}(x)}\over
\sqrt{H_{UV}(x)}}\,\,\delta H\rc\rc
&&\qquad\qquad\qquad\qquad
={1\over 3\pi^2}\,{bH_2\over 2b-1}\,\sqrt{H_{\infty}G_{\infty}}\,\,R\,x_*^{{1\over b}-2}\,\,,
\eear
which, after using (\ref{xstar})  to eliminate $x_*$, becomes:
\beq
\int_{x_*}^{\infty}\,dx\,
{\partial {\cal L}\over \partial H}\Big|_{UV}\delta H\,=\,{1\over 3\pi^2}\,
{H_2\,b^{2-2b}\over 2b-1}\,\sqrt{H_{\infty}}\,G_{\infty}^{1-b}\,R^{2b}\,\,.
\eeq
Similarly,
\bear
&&\int_{x_*}^{\infty}\,dx\,
{\partial {\cal L}\over \partial G}\Big|_{UV}\delta G\,=\,
{1\over 3\pi^2}\,\int_{x_*}^{\infty}\,dx\,
{\sqrt{H_{UV}(x)}\,\,\rho_{UV}\over
\sqrt{(\rho\,'_{UV})^2+G_{UV}(x)}}\,\,\delta G\rc\rc
&&\qquad\qquad\qquad\qquad
=\,
{2\over 3\pi^2}\,{b^2\over 4b^2-1}\,
{G_2\,\sqrt{H_{\infty}}\,G_{\infty}\over x_{*}^2}\,\,,
\eear
which, after eliminating $x_*$, gives:
\beq
\int_{x_*}^{\infty}\,dx\,
{\partial {\cal L}\over \partial G}\Big|_{UV}\delta G\,=\,
{2\over 3\pi^2}\,{G_2\,b^{2-2b}\over 4b^2-1}\,\sqrt{H_{\infty}}\,\,G_{\infty}^{1-b}\,\,R^{2b}\,\,.
\eeq
Thus,  $\delta S$ can be written as:
\beq
\delta S\,=\,{2\over 3\pi^2}\,{b^{2-2b}\over 2b-1}\,\,
\sqrt{H_{\infty}}\,\,G_{\infty}^{1-b}\,\,{\cal H}_2\,
R^{2b}\,\,,
\eeq
where ${\cal H}_2$ is the constant defined in (\ref{calH_2}). Using the definition (\ref{calF_definition}), it follows that the change in the function ${\cal F}$ is:
\beq
\delta {\cal F}\,=\,{2\over 3\pi^2}\,b^{2-2b}\,\sqrt{H_{\infty}}\,\,G_{\infty}^{1-b}
\,\,{\cal H}_2\,R^{2b}\,\,.
\eeq
Taking into account (\ref{HG_infinity})  and  (\ref{bHG_F}) this expression can be written as:
\beq
\delta {\cal F}\,=\,F_{UV} ({\mathbb S}^3)
\,\,{\cal H}_2\,
\Big({\kappa\over L_0^2}\Big)^{2b}\,(r_q\,R)^{2b}\,=\,F_{UV} ({\mathbb S}^3)\,
\,\,{\cal H}_2\,
x_*^{-2}\,\,.
\label{deltaF_R2b}
\eeq
Therefore, we can write  the  ${\cal F}$ function  near the UV fixed point as:
\beq
{\cal F}\,=\,F_{UV} ({\mathbb S}^3)\,+\,c_{UV}\,(r_q\,R)^{2b}\,\,,
\label{calF_nearUV_b}
\eeq
where the constant coefficient $c_{UV}$ can be read from (\ref{deltaF_R2b}),
\beq
c_{UV}\,=\,\Big({\kappa\over L_0^2}\Big)^{2b}\,\, F_{UV} ({\mathbb S}^3)
{\cal H}_2\,\,.
\label{c_UV}
\eeq

Eq. (\ref{calF_nearUV_b}) coincides with (\ref{calF_nearUV}) when the former is written in terms of the dimension  $\Delta_{UV}=3-b$ of the quark-antiquark bilinear operator in the massless flavored theory. Moreover, as $F_{UV} ({\mathbb S}^3)$ is always positive,  the sign of $c_{UV}$ depends on the sign of the coefficient ${\cal H}_2$. We have checked that ${\cal H}_2$ is negative for all values of the deformation parameter $\hat\epsilon$, in agreement with the expectation that ${\cal F}$  is maximized at the UV fixed point.

\subsection{Entanglement entropy on the strip}

Let us  now evaluate the entanglement entropy in   the case in which  the region $A$ is the strip $-{l\over 2}\,\le x^1\,\le +{l\over 2}$ of width $l$ in the $(x^1,x^2)$-plane (see Fig. \ref{Strip_surface}). In this case we consider a constant time surface $\Sigma$, whose embedding in the ten-dimensional space is determined by a function $x=x(x^1)$.  The induced metric on $\Sigma$ is,
\beq
ds^2_{8}\,=\,h^{-{1\over 2}}\,\Big[\,\Big(1+G(x)\,x'^{\,2}\,\Big)\, (dx^1)^2\,+\,
(dx^2)^2\,\Big]+h^{{1\over 2}}\,\Big[\,
e^{2f}\,ds_{{\mathbb S}^4}^2+
e^{2g}\,\Big(\,\big(E^1\big)^2\,+\,\big(E^2\big)^2\Big)\,\Big]\,\,,
\eeq
where $x'$ denotes the derivative of the holographic coordinate $x$ with respect to the cartesian coordinate $x^1$ and   the function $G(x)$ has been defined in (\ref{G_def}).
The entropy  functional  for the strip of width $l$ is given by:
\beq
S(l)\,=\,{V_6\,L_2\over 4 \,G_{10}}\,
\int_{-{l\over 2}}^{+{l\over 2}}\,dx^1\,\sqrt{H(x)}\,\sqrt{1+G(x)\,x'^{\,2}}\,\equiv\,
\int_{-{l\over 2}}^{+{l\over 2}}\,dx^1\,{\cal L}_{strip}
\,\,,
\eeq
where   $V_6$ is the volume of the internal manifold, whose value was given after (\ref{entropy_total_disk}),  $H(x)$ is the function defined in (\ref{H_def}), and $L_2=\int dx^2$ is the length of the strip.  As the integrand in $S(l)$ does not depend on the coordinate $x^1$, the corresponding  Euler-Lagrange equation admits the following first integral:\begin{figure}[ht]
\center
\includegraphics[width=0.35\textwidth]{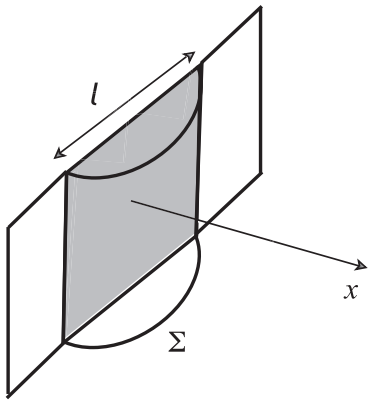}
\caption{The surface $\Sigma$ ends on a strip of length $l$ at the boundary.  } 
\label{Strip_surface}
\end{figure}

\beq
x'\,{{\cal L}_{strip}\over \partial x'}\,-\,{\cal L}_{strip}\,=\,{\rm constant}\,\,,
\eeq
or, more explicitly,
\beq
{\sqrt{H(x)}\over \sqrt{1\,+\,G(x)\,x'^{\,2}}}\,=\,\sqrt{H_*}\,\,,
\eeq
where $H_{*}=H(x=x_*)$ and $x_*$  is the holographic coordinate of the turning point. It follows from this last expression that $x'$ is given by:
\beq
x'\,=\,\pm {1\over \sqrt{G(x)}}\,\sqrt{{H(x)\over H_*}\,-\,1}\,\,.
\eeq
Therefore,  the width of the strip is given by the integral:
\beq
l\,=\,2 \sqrt{H_*}\,\int_{x_*}^{\infty}\,dx\,{\sqrt{G(x)}\over \sqrt{H(x)-H_*}}\,\,,
\label{width_strip}
\eeq
and the entropy $S(l)$ is:
\beq
S(l)\,=\,{V_6\,L_2\over 2 G_{10}}\,
\int_{x_*}^{\infty}\,dx\,{\sqrt{G(x)}\,H(x)\over \sqrt{H(x)-H_*}}\,\,.
\label{S_strip_total}
\eeq
The integral (\ref{S_strip_total}) for $S(l)$ is divergent  in the UV. Indeed, from the  behavior of the functions $H(x)$ and $G(x)$  at large $x$ (eqs. (\ref{H_G_UV}) and (\ref{HG_infinity})), it follows that, when $x\to\infty$,  the integrand in (\ref{S_strip_total})  behaves as: 
\beq
{\sqrt{G(x)}\,H(x)\over \sqrt{H(x)-H_*}}\,
\approx\,\sqrt{H_{\infty}\,G_{\infty}}\,\,x^{{1\over b}\,-\,1}\,\,,
\eeq
and $S(l)$ therefore diverges as $x_{\max}^{{1\over b}}$, where $ x_{\max}$ is the upper limit of the integral, which can be regarded as a UV cutoff. More explicitly, one can rearrange the integral in (\ref{S_strip_total}) as:
\bear
&&\int_{x_*}^{x_{\max}}\,dx\,{\sqrt{G(x)}\,H(x)\over \sqrt{H(x)-H_*}}\,=\,
\int_{x_*}^{x_{\max}}\,dx\,\Big[{\sqrt{G(x)}\,H(x)\over \sqrt{H(x)-H_*}}\,-\,
\sqrt{H_{\infty}\,G_{\infty}}\,\,x^{{1\over b}\,-\,1}\,\Big]\rc\rc
&&\qquad\qquad\qquad\qquad\qquad\qquad\qquad\qquad\qquad
+b\,\sqrt{H_{\infty}\,G_{\infty}}\,\big[\,x_{max}^{{1\over b}}\,-\,x_{*}^{{1\over b}}\,\big]\,\,,
\eear
where the divergent term is explicitly shown. We now define the rescaled functions $\hat G(x)$ and $\hat H(x)$ as:
\beq
\hat G(x)\equiv {G(x)\over G_{\infty}}\,\,,
\qquad\qquad
\hat H(x)\equiv {H(x)\over H_{\infty}}\,\,,
\qquad\qquad
\hat H_{*}\equiv {H_{*}\over H_{\infty}}
\,\,.
\eeq
Then, the finite term in $S(l)$ can be written as:
\beq
S_{finite}(l)\,=\,{V_6\,L_2\over 2 G_{10}}\,
\sqrt{H_{\infty}\,G_{\infty}}\Bigg[
\int_{x_*}^{\infty}\,dx\,\Bigg({\sqrt{\hat G(x)}\,\hat H(x)\over \sqrt{\hat H(x)-\hat H_*}}\,-\,
x^{{1\over b}-1}\,\Bigg)\,-\,b\,x_{*}^{{1\over b}}\,\Bigg]\,\,,
\label{Sfinite_strip}
\eeq
whereas the divergent term takes the form:
\beq
S_{div}\,=\,{V_6\,L_2\over 2 G_{10}}\,b\,
\sqrt{H_{\infty}\,G_{\infty}}\,\,x_{max}^{{1\over b}}\,\,.
\eeq
\begin{figure}[ht]
\center
\includegraphics[width=0.75\textwidth]{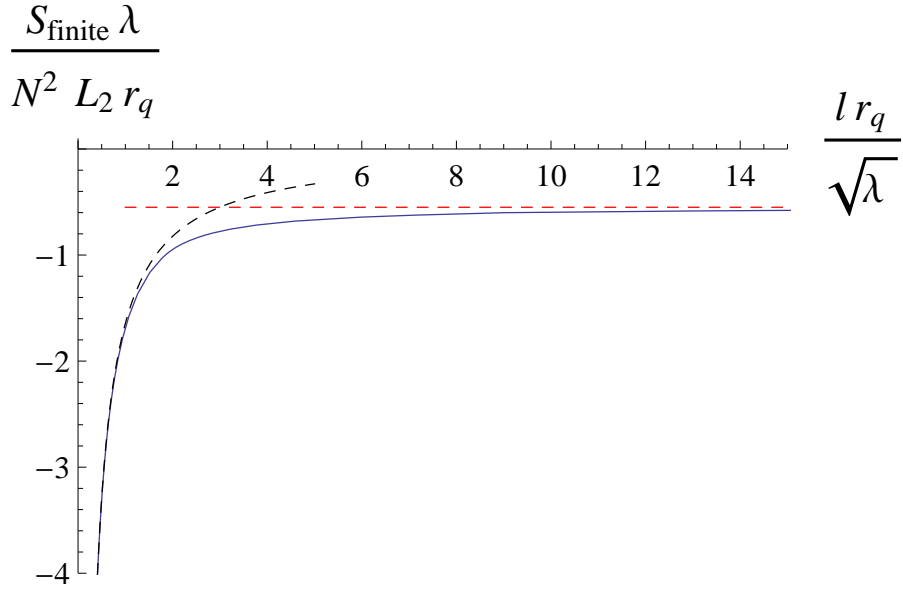}
\caption{Plot of the entanglement entropy versus the width of the strip. 
The solid curve corresponds to the numerical results for $\hat\epsilon=9$, while the 
the black dashed curve on the left  is the analytic UV result (\ref{Sfinite_strip_UV}) and the dashed red line  on the right corresponds to the IR value (\ref{S_infty}).} 
\label{Strip_entanglement}
\end{figure}

We have evaluated numerically the right-hand side of (\ref{Sfinite_strip}) as a function of the strip width $l$.  The result is displayed in Fig. \ref{Strip_entanglement}. One notices from these results that $S_{finite}(l)/L_2$ is negative and diverges as $-c/l$ for small $l$, where $c$ is a constant. Actually, for small $l$ the surface $\Sigma$ is in the UV region of the geometry and one can evaluate the entropy analytically in this limit. This is the subject of the next subsection, where we show that the constant $c$ is proportional to the free energy on the three-sphere of the massless flavored theory (see eq. (\ref{Sfinite_strip_UV}) below). The analytical UV calculation is compared to the numerical results in Fig. \ref{Strip_entanglement}.

\subsubsection{UV limit}

In order to study the UV limit it is convenient to change variables in the integrals (\ref{width_strip}) and (\ref{Sfinite_strip}). Let us introduce a new variable $z$, related to $x$ as $x=x_*\,z$. Then, $l$ can be represented as:
\beq
l\,=\,2\, x_{*}^{{2\over b}+1}\,\sqrt{G_{\infty}}\,\int_{1}^{\infty}\,dz\,{\sqrt{\hat G(x_*\,z)}\over \sqrt{\hat H(x_*\,z)-x_{*}^{4/ b}}}\,\,,
\eeq
while $S_{finite}(l)$ is given by:
\beq
S_{finite}(l)\,=\,-{V_6\,L_2\over 2 \,G_{10}}\,
\sqrt{H_{\infty}\,G_{\infty}}\,\,x_{*}^{{1\over b}}\,
\Bigg[b-\int_{1}^{\infty}\,dz\Bigg(x_{*}^{1-{1\over b}}\,
{\sqrt{\hat G(x_*\,z)}\,\hat H(x_*\,z)\over \sqrt{\hat H(x_*\,z)-x_{*}^{4/ b}}}\,-\,
z^{{1\over b}-1}\Bigg)\Bigg]\,\,.
\eeq
Let us now obtain the expressions of $l$ and $S_{finite}(l)$ in the limit in which $x_{*}\to\infty$. In this case the argument of the functions $\hat G$ and $\hat H$ in the integrals is always large and one can take $\hat G(x)\approx x^{-2-{2\over b}}$, $\hat H(x)\approx x^{{4\over b}}$. We get:
\beq
l\approx 2\,x_{*}^{-{1\over b}}\,\sqrt{G_{\infty}}\,\,I_1\,\,,
\label{l_UV_approx}
\eeq
where $I_1$ is the following integral:
\beq
I_1\,=\,\int_{1}^{\infty}\,{dz\over z^{1+{1\over b}}\,\sqrt{z^{{4\over b}}\,-\,1}}\,=\,
{b\,\sqrt{2}\,\pi^{{3\over 2}}\over  \big[\Gamma\big({1\over 4}\big)\big]^2}\,\,.
\label{I1}
\eeq
Using (\ref{I1}) in (\ref{l_UV_approx}), we get:
\beq
l\approx {b\,\sqrt{G_{\infty}}\over x_{*}^{{1\over b}}}\,\,
{2\,\sqrt{2}\,\pi^{{3\over 2}}\over \big[\Gamma\big({1\over 4}\big)\big]^2 }\,\,.
\eeq
Similarly, the finite part of the entropy is:
\beq
S_{finite}(l)\,\approx\,-{V_6\,L_2\over 2\, G_{10}}\,
\sqrt{H_{\infty}\,G_{\infty}}\,\,x_{*}^{{1\over b}}\,
\big(\,b\,-\,I_2\,\big)\,\,,
\label{S_finite_strip_I2}
\eeq
where $I_2$ is the integral:
\beq
I_2\,=\,\int_{1}^{\infty}\,z^{{1\over b}-1}\,\Bigg[
{z^{{2\over b}}\over \sqrt{z^{{4\over b}}\,-\,1}}\,-\,1\,\Bigg]\,dz\,=\,
b\,\Bigg[1\,-\,{\sqrt{2}\,\pi^{{3\over 2}}\over \big[\Gamma\big({1\over 4}\big)\big]^2}
\Bigg]\,\,.
\label{I2}
\eeq
Plugging (\ref{I2}) into (\ref{S_finite_strip_I2}), we arrive at:
 \beq
S_{finite}(l)\,\approx\,-{V_6\,L_2\over 2\, G_{10}}\,b\,
\sqrt{H_{\infty}\,G_{\infty}}\,\,x_{*}^{{1\over b}}\,
{\sqrt{2}\, \pi^{{3\over 2}}\over \big[\Gamma\big({1\over 4}\big)\big]^2 }\,\,.
\eeq
Eliminating $x_*$ in favor of $l$, we get:
\beq
S_{finite}(l)\,\approx\,-{2\pi^3\,V_6\,L_2\over G_{10}}\,\,
{b^2\,G_{\infty}\,\sqrt{H_{\infty}}\over 
\big[\Gamma\big({1\over 4}\big)\big]^4}\,\,
{1\over l}\,\,.
\eeq
However, from the definition of $G_{\infty}$ and  $H_{\infty}$ in (\ref{HG_infinity})  it follows that:
\beq
b^2\,G_{\infty}\,\sqrt{H_{\infty}}\,=\,
{L_0^8\,q_0^2\,e^{-2\phi_0}\over b^6}\,=\,{3\pi^2\over 2}
F_{UV} ({\mathbb S}^3)\,\,,
\label{GH-F-UV}
\eeq
where $F_{UV} ({\mathbb S}^3)$ is the free energy of the massless flavored theory on the three sphere (see (\ref{F_UV})). It follows that:
\beq
{S_{finite}(l)\over L_2}\,\approx\,-{4\pi^2\,F_{UV} ({\mathbb S}^3)\over
\big[\Gamma\big({1\over 4}\big)\big]^4}\,\,{1\over l}\,\,,
\label{Sfinite_strip_UV}
\eeq
which is the result we were looking for. From the comparison of Fig.  \ref{Strip_entanglement} to the numerical results we conclude that (\ref{Sfinite_strip_UV}) is a good description of the entropy in the small $l$ region. Notice that the coefficient of $1/l$ on the right-hand side of (\ref{Sfinite_strip_UV}) is determined by the free energy of the massless flavored theory. This is expected on general grounds in the UV region, since the masses of the quarks can be neglected in the high-energy regime.

\subsubsection{IR limit}
\label{Strip_entanglement_S_IR}

Let us now evaluate $S_{finite}(l)$ in the regime in which $l$ is large. In this case the surface $\Sigma$ penetrates deeply in the bulk and its tip is near the origin (\ie,  $x_*$ is small). We will proceed    as in Section \ref{IR_limit_disk_entanglement} and split   the interval $[x_*,\infty]$ in the integrals (\ref{width_strip}) and (\ref{Sfinite_strip}) as $[x_*,\infty]=[x_*,x_a]\cup [x_a, \infty]$, where $x_a<1$ is considered  small enough so that one can use  the unflavored background functions in the interval $[x_*,x_a]$. Moreover, when 
$x\in [x_a, \infty]$ with $x_a\gg x_*$,  $H_*\propto x^4_{*}\ll H(x)\propto x^4$ and one can neglect the terms containing $H_*$ in the integrals. Then, the strip width $l$ can be approximated as:
\beq
l\,\approx {2c\over \gamma}\,\,{L^2_{ABJM}\over x_*}\,\,
\int_{1}^{{x_a\over x_*}}\,\,{dz\over z^2\,\sqrt{z^4-1}}\,+\,
2\sqrt{H_*}\,\int_{x_a}^{\infty}\,\,dx\,{\sqrt{G(x)}\over \sqrt{H(x)}}\,\,.
\label{strip_width_IR}
\eeq
When, $x_a\gg x_*$  and $x_*\to 0$  we can extend the first integral in (\ref{strip_width_IR}) up to $\infty$ and  neglect the contribution of the second integral (which is proportional to  $\sqrt{H_*}\propto x_*^2$). Then, $l$ can be approximately written as:
\beq
l\,\approx {c\over \gamma}\,L^2_{ABJM}\,{2\sqrt{2}\,\pi^{{3\over 2}}\over 
\big[\Gamma\big({1\over 4}\big)\big]^2}\,\,{1\over x_*}\,\,,
\label{strip_width_IR_simp}
\eeq
and, as expected, a small value of $x_*$ leads to a large value of $l$. Similarly, we can perform the same type of manipulations in the expression (\ref{Sfinite_strip}) of $S_{finite}$. We find:
\bear
&&{S_{finite}(l)\over L_2}\,\approx\,{2\over 3\pi^3}\,\,{\gamma\over c}\,\,
L^6_{ABJM}\,e^{-2\phi_{ABJM}}\,\,x_*\,\int_1^{{x_a\over x_*}}\,\,
{z^2\,dz\over \sqrt{z^4-1}}\rc\rc
&&\qquad\qquad\,\,+
{2\over 3\pi^3}\,\sqrt{G_{\infty}\,H_{\infty}}\,\Bigg[
\int_{x_a}^{\infty}\,\Big(\sqrt{\hat G(x)\,\hat H(x)}\,-\,x^{{1\over b}-1}\Big)\,dx\,-\,
b\,x_a^{{1\over b}}\,\Bigg]\,\,.\qquad
\label{strip_entropy_IR}
\eear
Performing the first integral in (\ref{strip_entropy_IR}) in the limit $x_*\to 0$ for fixed $x_a$, we arrive at:
\beq
{S_{finite}(l)\over L_2}\,\approx\,-
{2\sqrt{2}\over 3\pi^{{3\over 2}}\,\big[\Gamma\big({1\over 4}\big)\big]^2}\,\,
{\gamma\over c}\,\,L^6_{ABJM}\,e^{-2\phi_{ABJM}}\,x_*\,+\,{\cal S}_{\infty}\,\,,
\label{strip_entropy_IR_simp}
\eeq
where ${\cal S}_{\infty}$ is the constant:
\beq
{\cal S}_{\infty}={2\over 3\pi^3}\sqrt{G_{\infty}\,H_{\infty}}\Big[
\int_{x_a}^{\infty}\Big(\sqrt{\hat G(x)\hat H(x)}-x^{{1\over b}-1}\Big)dx-
bx_a^{{1\over b}}\Big]+{2\over 3\pi^3}\,{\gamma\over c}\,
L^6_{ABJM}e^{-2\phi_{ABJM}}x_a\,\,.
\label{calS_infty}
\eeq
We can now eliminate in (\ref{strip_entropy_IR_simp}) the turning point $x_*$ in favor of $l$ by using (\ref{strip_width_IR_simp}). We find:
\beq
{S_{finite}(l)\over L_2}\,\approx\,-
{4\pi^2\,F_{IR} ({\mathbb S}^3)\over
\big[\Gamma\big({1\over 4}\big)\big]^4}\,\,{1\over l}\,+\,{\cal S}_{\infty}\,\,,
\label{strip_entropy_IR_final}
\eeq
where we have written the result in terms of   
$F_{IR} ({\mathbb S}^3)$. The first term in (\ref{strip_entropy_IR_final}) is just the  strip entanglement  entropy for the unflavored theory. The constant ${\cal S}_{\infty}$ represents the asymptotic value of $S_{finite}(l)/ L_2$  as $l\to\infty$. One can approximate this constant by taking $x_a\to 0$ in (\ref{calS_infty}). After using (\ref{bHG_F}) to relate $G_{\infty}$ and $H_{\infty}$ to  
$F_{UV} ({\mathbb S}^3)$, we arrive at:
\beq
{\cal S}_{\infty}\,\approx\,r_q\,{\kappa\,F_{UV} ({\mathbb S}^3)\over \pi b\,L_0^2}\,\,
\int_{0}^{\infty}\Big(\sqrt{\hat G(x)\hat H(x)}-x^{{1\over b}-1}\Big)\,dx\,\,.
\label{S_infty}
\eeq

\vskip 1cm
\renewcommand{\theequation}{\rm{D}.\arabic{equation}}
\setcounter{equation}{0}
\medskip

\section{Asymptotic quark-antiquark potential}
\label{asymp_qq_potential}

The purpose of this appendix is to obtain the analytic expressions for the $q\bar q$ potential energy in the UV and IR limit. We will start by calculating the leading and subleading UV potential. 

\subsection{UV potential}

Let us find the approximate value of the $q\bar q$ potential in the case in which the distance $d$ is small and the hanging string only explores the UV of the geometry. This is equivalent to considering the limit in which the turning point $x_*$ is large. It is then more convenient to perform a change of variables in the integrals (\ref{d_x-integral}) and (\ref{E_x-integral}) and write $d$ and $E_{q\bar q}$ as:
\bear
d&=&2\,\int_{1}^{\infty}\,
{e^{g(x_*\,z)}\,h(x_*\,z)\over z\,\sqrt{h_*-h(x_*\,z)}}\,\,dz\,\,,\rc\rc
E_{q\bar q}&=&{1\over \pi}\,\int_{1}^{\infty}\,\,
{e^{g(x_*\,z)}\over z}\,\,\Big[\,{\sqrt{h_*}\over \sqrt{h_*-h(x_*\,z)}}\,-\,1\Big]
dz\,-\,{r_*\over \pi}\,\,.
\label{d_E_z-integral}
\eear
We want to compute the leading value of $E_{q\bar q}$, as well as the first deviation from the conformal behavior. For this reason we will make use of the asymptotic expressions of the different functions of the metric derived in Section \ref{UV_mass_corrections}.

Let us begin by computing the integrals in (\ref{d_E_z-integral}) in a power series expansion for large $x_*$. From (\ref{g_f_UVexpansions_x}) and (\ref{UVexpansion_h_phi}) we get:
\beq
e^{g(x_*\,z)}\,\,h(x_*\,z)\,=\,
{L_0^4\over b\,\kappa^3\,r_q^3}\,\,x^{-{3\over b}}_*\,z^{-{3\over b}}\,
\Big(1\,+\,{h_2+g_2\over x_*^2\,z^2}\,+\,\cdots\,\Big)\,\,,
\eeq
where  $g_2$ and $h_2$ are given in (\ref{g2_f2}) and (\ref{h_2_phi_2}) and $\kappa$ has been defined in (\ref{kappa_def}). Then, it follows that:
\beq
{1\over \sqrt{h_*-h(x_*\,z)}}\,=\,{\kappa^2\,r_q^2\over L_0^2}\,\,
x^{{2\over b}}_*\,\,{z^{{2\over b}}\over \sqrt{z^{{4\over b}}-1}}\,\,
\Big(1\,-\,{h_2\over 2\,x_*^2}\,{z^{{4\over b}+2}-1\over z^2\,
(z^{{4\over b}}-1)}
\,+\,\cdots\,\Big)\,.
\eeq
Using these results we obtain the following expansion of the $q\bar q$ separation $d$:
\beq
d\,=\,{2\,L_0^2\over b\,\kappa\,r_q\,x^{{1\over b}}_*}\,\,
\Big[\,I_1\,+\,{1\over x_*^2}\,\,\Big(\,(h_2+g_2)\,I_3\,-\,{h_2\over 2}\,I_4\,\Big)
+\,\cdots \Big]\,\,,
\eeq
where $I_1$ is the integral (\ref{I1}),  and $I_3$ and $I_4$ are:
\bear
&&I_3\,=\,\int_{1}^{\infty}\,{dz\over z^{3+{1\over b}}\,\sqrt{z^{{4\over b}}\,-\,1}}\,=\,
{b\sqrt{\pi}\over 4}\,\,G(b)\,\,,\rc\rc
&&I_4\,=\,\int_{1}^{\infty}\,
{z^{{4\over b}+2}\,-\,1\over  z^{3+{1\over b}}\,(z^{{4\over b}}\,-\,1)^{{3\over 2}}}\,\,dz\,=\,
{b\sqrt{\pi}\over 2}\,\Bigg[\,{3+2b\over 4}\,\,G(b)\,-\,
{\sqrt{2}\,\,\pi\over  \big[\Gamma\big({1\over 4}\big)\big]^2}\,
\Bigg]\,\,,
\label{I3_I4}
\eear
where $G(b)$ is the following ratio of Euler Gamma functions:
\beq
G(b)\equiv
{\Gamma\Big({3+2b\over 4}\Big)
\over 
\Gamma\Big({5+2b\over 4}\Big)}\,\,.
\eeq
Using these results we arrive at:
\beq
d\,\approx\,{2\,L_0^2\over \kappa\, r_q\,x_*^{{1\over b}}}\,
\Bigg[\,{\sqrt{2}\,\pi^{{3\over 2}}\over \big[\Gamma\big({1\over 4}\big)\big]^2}\,+\,
{\sqrt{\pi}\over 4\,x_*^2}\,\Big(g_2+(1-2b)\,{h_2\over 4}\Big)\,G(b)\,+\,
{\sqrt{2}\,\pi^{{3\over 2}}\over 4\, x_*^2}\,
{h_2\over \big[\Gamma\big({1\over 4}\big)\big]^2}\,\Bigg]\,\,.
\label{d-x_*}
\eeq

Let us now compute the $q\bar q$ energy at leading and next-to-leading order in the UV (large $x_*$ or small $d$). First, we need the expansion:
\beq
{\sqrt{h_*}\over \sqrt{h_*-h(x_*\,z)}}\,=\,
{z^{{2\over b}}\over \sqrt{z^{{4\over b}}-1}}\,+\,
{h_2\over 2\,x_*^2}\,\,{z^{{2\over b}-2}\,(1-z^2)\over (z^{{4\over b}}\,-\,1)^{{3\over 2}}}
\,+\,\cdots\,\,.
\eeq 
Then, it is easy to verify that $E_{q\bar q}$ can be expanded as:
\beq
E_{q\bar q}\approx
{\kappa\, r_q\over \pi\,b}\,x_*^{{1\over b}}\Bigg[\,
I_2\,+\,{1\over x_*^2}\,\Big(\,{h_2\over 2}\,I_5\,+\,g_2\,I_6\,\Big)\,\Bigg]\,-\,
{r_*\over \pi}\,\,,
\eeq
where  $I_2$ is the integral (\ref{I2})  and $I_5$ and $I_6$ are:
\bear
&&I_5\,=\,\int_{1}^{\infty}\,{z^{{3\over b}-3}(1-z^2)\over (z^{{4\over b}}-1)^{{3\over 2}}}
\,dz\,=\,{b\sqrt{\pi}\over 2}\,\,\Bigg[
{\sqrt{2}\,\pi\over \big[\Gamma\big({1\over 4}\big)\big]^2}\,-\,{2b+1\over 4}\,\,G(b)\,\Bigg]
\,\,,\rc\rc
&&I_6\,=\,\int_{1}^{\infty}\,z^{{1\over b}-3}\,\Bigg[
{z^{{2\over b}}\over \sqrt{z^{{4\over b}}\,-\,1}}\,-\,1\,\Bigg]\,dz\,=\,{b\over 2b-1}\,
\Big[\,{(2b+1)\sqrt{\pi}\over 4}\,\,G(b)\,-\,1\,\Big]\,\,.
\eear
To compute $E_{q\bar q}$ we also need $r_*$ as a function of $x_*$. It follows from (\ref{r-x_relation}) that:
\beq
r_*\,=\,\kappa\,r_q\,x_*^{{1\over b}}\Big[1\,-\,{g_2\over 2b-1}\,{1\over x_*^2}\,+\,\cdots\Big]\,\,.
\eeq
Then, one can check that:
\beq
E_{q\bar q}\approx\kappa\,r_q\,x_*^{{1\over b}}\Bigg[-{\sqrt{2\pi}\over 
\big[\Gamma\big({1\over 4}\big)\big]^2}+
{1\over x_*^2}\Bigg({\sqrt{2\pi}\over \big[\Gamma\big({1\over 4}\big)\big]^2}\,
{h_2\over 4}+{2b+1\over 4\sqrt{\pi}(2b-1)}
\Big(g_2+(1-2b)\,{h_2\over 4}\Big)\,G(b)\Bigg)\Bigg]\,\,.
\eeq
Let us write $E_{q\bar q}$ as a function of the $q\bar q$ separation $d$. For this purpose we have to eliminate $x_*$ in favor of $d$. By inverting (\ref{d-x_*}), we get:
\bear
&&\kappa\,r_q\,x_*^{{1\over b}}\approx {2\sqrt{2}\pi^{{3\over 2}}\over 
 \big[\Gamma\big({1\over 4}\big)\big]^2}\,\,{L_0^2\over d}\,+\,\rc\rc
 &&\qquad\qquad
 +\,
 r_q\,{\sqrt{\pi}\over 2}\,
 \Bigg[{\kappa\,\big[\Gamma\big({1\over 4}\big)\big]^2
 \over 2\sqrt{2}\,\pi^{{3\over 2}}}\Bigg]^{2b}\,
 \Bigg({\sqrt{2}\,\pi\over \big[\Gamma\big({1\over 4}\big)\big]^2}\,
h_2+\Big(g_2+(1-2b)\,{h_2\over 4}\Big)\,G(b)\Bigg)
\Bigg({r_q \,d\over L_0^2}\Bigg)^{2b-1}.
\qquad\qquad
 \eear
Using this result, we get the following dependence of $E_{q\bar q}$ with the distance $d$
\beq
{E_{q\bar q}\over r_q}\approx-
 {4\pi^{2}\over 
 \big[\Gamma\big({1\over 4}\big)\big]^4}\,\,{L_0^2\over r_q\,d}\,+\,
 {\sqrt{2}\,\pi\over  \big[\Gamma\big({1\over 4}\big)\big]^2}\,
  \Bigg[{\kappa\,\big[\Gamma\big({1\over 4}\big)\big]^2
 \over 2\sqrt{2}\,\pi^{{3\over 2}}}\Bigg]^{2b}\,
 \Big({g_2\over 2b-1}-{h_2\over 4}\Big)\,G(b)
\,\Bigg({r_q\, d\over L_0^2}\Bigg)^{2b-1},
\label{potl_lead_and_sublead}
 \eeq
where we have represented this relation in terms of the rescaled quantities 
$E_{q\bar q}/ r_q$ and $r_q \,d/ L_0^2$. Notice that the leading term (the first term on the right-hand side of (\ref{potl_lead_and_sublead})) is given by the potential of the massless flavored background, as expected. 

\subsection{IR potential}
\label{qq_potential_IR}
We now estimate the $q\bar q$ potential for large separations. In this case we will content ourselves to compute the leading order contribution. For large $q\bar q$ separations the hanging fundamental string penetrates deeply in the geometry, which is equivalent to saying that $x_*$ is small.  We will follow an approach similar to the one in Sections \ref{IR_limit_disk_entanglement}  and \ref{Strip_entanglement_S_IR} and we will split the $[x_*,\infty]$ interval of the integrals (\ref{d_x-integral}) and (\ref{E_x-integral}) as  $[x_*,\infty]=[x_*,x_a]\cup [x_a,\infty]$ with $x_a<1$ being  small. We will assume that $x_a$ is small enough so that the background functions are well approximated by (\ref{functions_deep_IR}) in the interval $[x_*,x_a]$ . Then, we can estimate the integral (\ref{d_x-integral}) for $d$ as:
\beq
d\,\approx\,{2\,L_{ABJM}^2\,x_*^2\over r_q}\,\,
\int_{x_*}^{x_a}\,{dx\over x^2\,\sqrt{x^4-x_*^4}}\,+\, {2\over \sqrt{h_*}}\,
\int_{x_a}^{\infty}\,{e^g\,h\over x}\,dx\,\,,
\label{d_deepIR}
\eeq
where $L_{ABJM}$ is the unflavored $AdS$ radius (\ref{ABJM-AdSradius}) and we have used the fact that for $x_*\ll 1$ we have that $h_*\gg h(x)$  when $x\in  [x_a,\infty]$ and, therefore, in this interval we can neglect  $h(x)$ in the square root of the denominator of the integrand in (\ref{d_x-integral}). The first integral in (\ref{d_deepIR}) can be done analytically, yielding the result:
\beq
d\,\approx\,{2\,L_{ABJM}^2\over x_*\, r_q}\,\,\Bigg[
{\sqrt{2}\,\pi^{{3\over 2}}\over  \big[\Gamma\big({1\over 4}\big)\big]^2}\,-\,
{1\over 3}\,\Big({x_*\over x_a}\Big)^3\,
{}_2F_1\Big({1\over 2}, {3\over 4}; {7\over 4}; \Big({x_*\over x_a}\Big)^4\Big)\Bigg]\,+\,
{2\,x_*^2\, r_q^2\over L_{ABJM}^2}\,\int_{x_a}^{\infty}\,{e^g\,h\over x}\,dx\,\,.
\label{d_deepIR_explicit}
\eeq
For small $x_*$, at leading order, we get from (\ref{d_deepIR_explicit}):
\beq
d\,\approx\,{L_{ABJM}^2\over x_*\, r_q}\,\,
{2\sqrt{2}\,\pi^{{3\over 2}}\over  \big[\Gamma\big({1\over 4}\big)\big]^2}\,\,,
\label{d_deepIR_approx}
\eeq
which confirms that $r_q\,d$ is large when $x_*$ is small. We can make similar approximations in the integral (\ref{E_x-integral}) with the result:
\beq
E_{q\bar q}\approx {r_q\over \pi}\,\Bigg[\,
\int_{x_*}^{x_a}\,
\Big[\,{x^2\over \sqrt{x^4-x_*^4}}\,-\,1\Big]dx\,-\,x_*\,\Bigg]\,\,.
\eeq
Performing explicitly the integral, we get:
\beq
E_{q\bar q}\approx - {x_*\,r_q\over \pi}\,\Bigg[\,
{\sqrt{2}\,\pi^{{3\over 2}}\over  \big[\Gamma\big({1\over 4}\big)\big]^2}\,+\,
{x_a\over x_*}\,-\,{x_a\over x_*}\,
{}_2F_1\Big(-{1\over 4}, {1\over 2}; {3\over 4}; \Big({x_*\over x_a}\Big)^4\Big)\,\Bigg]\,\,,
\eeq
which, at leading order, becomes:
\beq
E_{q\bar q} \approx-
{\sqrt{2\pi}\over  \big[\Gamma\big({1\over 4}\big)\big]^2}\,\,
x_*\,r_q\,\,.
\eeq
Eliminating $x_*$ by using (\ref{d_deepIR_approx}) we arrive at the estimate (\ref{leading_IR_pot}).

\vskip 1cm
\renewcommand{\theequation}{\rm{E}.\arabic{equation}}
\setcounter{equation}{0}
\medskip

\section{Asymptotics of the two-point functions}
\label{asymp_two-point_functions}
In this appendix we study the renormalized geodesic distance, and the corresponding two-point function of bulk operators with large conformal dimensions, in the UV limit of small separation $l$ and in the IR regime in which $r_q\,l\to \infty$. In the former case, our study will serve to fix the normalization constant ${\cal C}$ of (\ref{renormalized_geodesic}), as well as the analytic form of the correlator near the UV fixed point. This is the case we analyze first in the next subsection.

\subsection{UV behavior}
\label{UV_two-point_appendix}
We will now obtain the form of the correlator in the UV limit in which the turning point $x_*$ is large and the geodesic does not penetrate much into the bulk of the  geometry. In order to study this limiting case, it is convenient to perform a change of variables in the integral (\ref{l_general}) and write it as:
\beq
l\,=\,2\,x_{*}\,\int_{1}^{\infty}\,dz
{\sqrt{G(x_{*}\, z)}\over
\sqrt{e^{{1\over 2}\,(\phi_*-\phi(x_{*}\, z))}\,\,\big({h_*\over h(x_{*}\, z)}\big)^{1\over 2}\,-\,1}
}\,\,.
\label{l_xstar_z}
\eeq
Similarly, the renormalized geodesic length can be represented as:
\beq
{\cal L}_r\,=\,2\,x_{*}\,\int_{1}^{z_{max}}\,dz
{e^{-{\phi(x_{*}\, z)\over 4}}\,\big[h(x_{*}\, z)\big]^{-{1\over 4}}\,\,\sqrt{G(x_{*}\, z)}
\over
\sqrt{1-e^{{1\over 2}\,(\phi(x_{*}\, z))-\phi_*)}\,\,\big({h(x_{*}\, z)\over h_*}\big)^{1\over 2}}
}\,-\,{2\,L_0\,e^{-{\phi_0\over 4}}\over b}\,\log ({\cal C}\,x_{*}\, z_{max})\,\,,
\label{Lr_xstar_z}
\eeq
where $z_{max}=x_{max}/x_*$. When $x_*$ is large the argument of the functions in the integrals (\ref{l_xstar_z}) and (\ref{Lr_xstar_z}) is large and one can use the UV asymptotic expressions  (\ref{UVexpansion_h_phi}) and (\ref{H_G_UV}). Therefore, we can approximate $l$ as:
\beq
l\approx 2 \sqrt{G_{\infty}}\,x_{*}^{-{1\over b}}\,
\int_{1}^{\infty}\,{dz\over z^{1+{1\over b}}\,
\sqrt{z^{{2\over b}}-1}}\,\,.
\label{l_UV_integral}
\eeq
The integral  in (\ref{l_UV_integral})  just gives $b$ and, thus, we have:
\beq
l\approx 2\, b \,\sqrt{G_{\infty}}\,x_{*}^{-{1\over b}}\,=\,{2\over \kappa}\,{L_0^2\over r_q\,x_*^{{1\over b}}}\,\,.
\eeq
It follows that, when $x_*$ is large $r_q\,l$ is small. Thus the UV limit we are studying corresponds to $r_q\,l\to 0$. Similarly, the UV limit of ${\cal L}_r$ is:
\beq
{\cal L}_r\,\approx\,{2\,L_0\,e^{-{\phi_0\over 4}}\over b}\,
\Bigg[\int_{1}^{\infty}\,{dz\over z}\,\Big[{z^{{1\over b}}\over \sqrt{z^{{2\over b}}-1}}\,-\,1
\Big]\,-\,\log({\cal C}\,x_{*})\Bigg]\,\,.
\eeq
The integral in this last equation is:
\beq
\int_{1}^{\infty}\,{dz\over z}\,\Big[{z^{{1\over b}}\over \sqrt{z^{{2\over b}}-1}}\,-\,1
\Big]\,=\,b\,\log 2\,\,.
\eeq
On the other hand, the UV conformal dimension $\Delta_{UV}$  of a bulk field of mass $m$ has been written  in (\ref{DeltaUV}). Taking these facts into account, we can write:
\beq
e^{-m\,{\cal L}_r}\,\approx {\Big( b\,\sqrt{G_{\infty}}\,\,{\cal C}^{{1\over b}}\,
\Big)^{2\Delta_{UV}}\over l^{2\Delta_{UV}}}\,\,.
\label{emLr_UV}
\eeq
If the operator ${\cal O}$  is canonically normalized in the short-distance $r_q\,l\to 0$ limit, the coefficient in the numerator of (\ref{emLr_UV}) should be chosen to be one as in (\ref{UV_two-point}). Therefore, it follows that the constant ${\cal C}$ is fixed to
\beq
{\cal C}^{{1\over b}}={\sqrt{\lambda}\over r_q\, b\,\sqrt{G_{\infty}}}\,=\,{\kappa\sqrt{\lambda}\over L_0^2}\,\,.
\label{calC_value}
\eeq

Let us now evaluate the first correction of the two-point correlator around the UV fixed point. Let us expand the function $G(x)$ for large $x$ as in (\ref{H_G_rho_UVexpansion}) and (\ref{deltaHG_UV}). Then, for large $x_*$ we get:
\beq
\sqrt{G(x_{*}\, z)}\approx \sqrt{G_{\infty}}\,\,x_*^{-1-{1\over b}}\,
z^{-1-{1\over b}}\,\Big[1\,+\,{G_2\over 2}\,x_*^{-2}\,z^{-2}\Big]\,\,,
\eeq
where $G_2=h_2+2g_2$.  Similarly:
\beq
\sqrt{e^{{1\over 2}\,(\phi_*-\phi(x_{*}\, z))}\,\,\Big({h_*\over h(x_{*}\, z)}\Big)^{1\over 2}\,-\,1}\approx
\sqrt{z^{{2\over b}}-1}\,\Big[1+{\phi_2+h_2\over 4 x_*^2}\,\,{
z^{{2\over b}-2}\,(z^2-1)\over z^{{2\over b}}-1}\,\Big]\,\,.
\eeq
It follows that the separation $l$ of the two points of the correlator can be written as:
\beq
l\,\approx\,2\,b\,\sqrt{G_{\infty}}\,x_*^{-{1\over b}}
\Big[1\,+\,{1\over 2x_*^2}\,\big(G_2\,J_1\,-\,{\phi_2+h_2\over 2}\,J_2\big)\Big]\,\,,
\label{l_xstar}
\eeq
where $J_1$ and $J_2$ are the following integrals:
\bear
&&J_1\,\equiv\,{1\over b}\,\int_{1}^{\infty}\,{dz\over z^{3+{1\over b}}\,\sqrt{z^{{2\over b}}-1}}\,=\,
{\sqrt{\pi}\over 2}\,{\Gamma\big(1+b\big)
\over 
\Gamma\Big({3\over 2}+b\Big)}
\,\,,\rc\rc
&&J_2\,\equiv\,{1\over b}\,\int_{1}^{\infty}\,{z^2-1\over z^{3-{1\over b}}\,
\big(z^{{2\over b}}-1\big)^{{3\over 2}}}\,dz\,=\,
\sqrt{\pi}\,{\Gamma\big(1+b\big)\over 
\Gamma\Big({1\over 2}+b\Big)}\,-\,1\,\,.
\label{J1_J2}
\eear
Let us invert the relation (\ref{l_xstar}) at first order and write $x_*$ as a function of $l$. We get:
\beq
x_*\,\approx\, \Big[{2\sqrt{G_{\infty}}\,b\over l}\Big]^b\,\Big[1+c_{x_*}\,l^{2b}\Big]\,\,,
\label{xstar_l_fo}
\eeq
where $c_{x_*}$ is a constant given by:
\beq
c_{x_*}\,=\,{b\over \big(2\sqrt{G_{\infty}}\,b\big)^{2b}}\,\,
\Bigg[\Big({g_2\over 2b+1}\,-\,{2b-1\over 2b+1}\,{h_2\over 4}\,-\,{\phi_2\over 4}\Big)\,
\,{\sqrt{\pi}\,\Gamma\big(1+b\big)\over 
\Gamma\Big({1\over 2}+b\Big)}\,+\,{\phi_2+h_2\over 4}\,\Bigg]\,\,.
\eeq
Moreover,  the  renormalized geodesic length at first order takes the form:
\beq
{\cal L}_r\,\approx\,2\,L_0\,e^{-{\phi_0\over 4}}\,\Bigg[\,
\log\Big({2b \,r_q\,\,\sqrt{G_{\infty}}\over x_*^{{1\over b}}\sqrt{\lambda}}\Big)\,+\,
{1\over x_*^2}\,\Big[\,\Big(g_2\,-\,{\phi_2\over 4}\,+\,{h_2\over 4}\Big)\,J_3\,-\,
{\phi_2+h_2\over 4}\,J_2\,\Big]\Bigg]\,\,,
\eeq
where $J_2$ is the integral defined in (\ref{J1_J2}) and $J_3$ is:
\beq
J_3\,\equiv {1\over b}\,
\int_{1}^{\infty}\,{dz\over z^{3-{1\over b}}\,\sqrt{z^{{2\over b}}-1}}\,=\,
{\sqrt{\pi}\over 2}\,{\Gamma\big(b\big)
\over 
\Gamma\Big({1\over 2}+b\Big)}\,\,.
\eeq
Let us now write ${\cal L}_r$ in terms of $l$. By using (\ref{xstar_l_fo}) to eliminate $x_*$ in terms of $l$ we arrive at:
\beq
{\cal L}_r\,\approx\,2\,L_0\,e^{-{\phi_0\over 4}}\,\Big[\,\log \Big({r_q\, l\over \sqrt{\lambda}}\Big)\,+\,c_{{\cal L}}\,\Big({r_q\, l\over \sqrt{\lambda}}\Big)^{2b}\,\,\Big]\,\,,
\eeq
where the coefficient $c_{{\cal L}}$ is given by:
\beq
c_{{\cal L}}\,=\,\Bigg({\sqrt{\lambda}\over 2b\,r_q\,\sqrt{G_{\infty}}}\Bigg)^{2b}\,
\,{\sqrt{\pi}\,\Gamma\big(b\big)
\over 
\Gamma\Big({1\over 2}+b\Big)}\,\,
\Bigg[\,{g_2\over 2(2b+1)}\,-\,{\phi_2\over 8}\,-\,{1\over 8}\,{2b-1\over 2b+1}\,h_2\,\Bigg]\,\,.
\eeq
Therefore, the two-point correlator near the UV fixed point can be written as:
\beq
\Big\langle {\cal O}(t, l)\,{\cal O}(t, 0)\Big\rangle\,=\,
{f_{\Delta}(r_q\,l/\sqrt{\lambda})\over (r_q\, l/\sqrt{\lambda})^{2\Delta_{UV}}}\,\,,
\label{nearUV_corr}
\eeq
where the function $f_{\Delta}$ parameterizes the deviation from the conformal behavior near the UV fixed point and is given, at first order, by
\beq
f_{\Delta}(r_q\,l/\sqrt{\lambda})\,\approx\,1\,+\,c_{\Delta}\,\Big({r_q\,l\over \sqrt{\lambda}}\Big)^{2b}\,\,,
\label{fDelta_rewrite}
\eeq
where the new constant $c_{\Delta}$ is just:
\beq
c_{\Delta}\,=\,-{\Delta_{UV}\over 4}\,
{\sqrt{\pi}\,\Gamma\big(b\big)
\over \Gamma\Big({1\over 2}+b\Big)}\,
\Big({\kappa\over 2 \sqrt{2}\,\pi \,\sigma}\Big)^{2b}\,
\Big[\,{4\over 2b+1}\,g_2\,-\,{2b-1\over 2b+1}\,h_2\,-\,\phi_2\,\Big]\,\,.
\label{cDelta}
\eeq

\subsection{IR behavior}
\label{IR_two-point_appendix}

Let us now analyze the two-point functions in the IR limit, which corresponds to taking $x_*\to 0$. In this case the geodesic penetrates deeply in the bulk. 
We will proceed similarly to what we did in Section \ref{qq_potential_IR} for the $q\bar q$ potential in the IR. Accordingly, we split the interval $[x_*, x_{max}]$ as
$[x_*, x_{max}]=[x_*, x_{a}]\cup [x_{a}, x_{max}]$ for some $x_a<1$ small. 
Near the turning point, \ie, when $x_*\le x\le x_a$, we can approximate the functions $h(x)$ and $g(x)$ as:
\beq
h(x)\approx  \Big({c\,L_{IR}\over \gamma}\Big)^4\,{1\over x^4}\,\,,
\qquad\qquad
e^{g}\approx {\gamma\over c}\,\,x\,\,,
\label{h_g_deepIR}
\eeq
where $L_{IR}=L_{ABJM}$ is the AdS radius of the unflavored ABJM solution and $\gamma$ and $c$ are the parameters of the unflavored running solution.  It follows that, in this IR region,  the function $G(x)$ behaves as:
\beq
G(x)\approx {c^2\over \gamma^2}\,{L_{IR}^4\over x^4}\,\,.
\eeq
Moreover, the dilaton $\phi_{IR}$ is constant and given by the ABJM value $\phi_{ABJM}$ written in (\ref{ABJMdilaton}). 

Suppose that we have chosen some $x_a\gg x_*$.  Let us then split the integral (\ref{l_general}) for $l$ as:
\beq
l\,=\,2\,\Big[\int_{x_*}^{x_a}dx+\int_{x_a}^{x_{max}}dx\Big]\,\,
{\sqrt{G(x)}\over
\sqrt{e^{{1\over 2}\,(\phi_*-\phi(x))}\,\,\big({h_*\over h(x)}\big)^{1\over 2}\,-\,1}
}\,\equiv\,l_{IR}+l_{UV}\,\,,
\label{l_IRplusUV}
\eeq
where $x_{max}$ should be sent to $+\infty$ at the end of the calculation. Let us approximate the first integral in (\ref{l_IRplusUV}) by taking the functions in the deep IR, as in (\ref{h_g_deepIR}). We obtain:
\beq
l_{IR}\,\approx\,{2c\over \gamma}\,L_{IR}^2\,\int_{x_*}^{x_a}\,{dx\over x^2\,
\sqrt{{x^2\over x_*^2}-1}}\,=\,
{2c\over \gamma}\,{L_{IR}^2\over x_*}\,\sqrt{1-{x_*^2\over x_a^2}}\,\,.
\eeq
For small $x_*/x_a$ we can approximate this integral as:
\beq
l_{IR}\,\approx\,{2c\over \gamma}\,{L_{IR}^2\over x_*}\,\,.
\eeq
To evaluate $l_{UV}$ approximately we notice that $h_{*}\propto x_*^{-4}$ and, therefore, it is large for $x_*\to 0$. Then, we can neglect  the one inside the square root and approximate  $l_{UV}$ as:
\beq
l_{UV}\approx x_*\,{\gamma\,e^{-{\phi_{IR}\over 4}}\over c\,L_{IR}}\,\,
\int_{x_a}^{\infty} dx\,e^{{\phi(x)\over 4}}\,\big[h(x)\big]^{{1\over 4}}\,\sqrt{G(x)}\,\,.
\label{l_UV_estimate}
\eeq
Since the integral in (\ref{l_UV_estimate})  converges and is independent of $x_*$, it follows that $l_{UV}\sim x_*$ and, therefore, it can be neglected with respect to the large value of $l_{IR}\sim 1/x_*$.  Thus, we take
\beq
l\approx {\hat\gamma-1\over \gamma}\,{L_{IR}^2\over r_q\,x_*}\,\,.
\label{l_estimate_smallxs}
\eeq
Notice that it follows from this equation that $r_q\,l$ is large if $x_*$ is small, as it should be in the IR regime.  Let us next perform a similar analysis for the renormalized geodesic length ${\cal L}_r$. First, we consider the IR integral:
\bear
&&\int_{x_*}^{x_{a}}\,dx
{e^{-{\phi(x)\over 4}}\,\big[h(x)\big]^{-{1\over 4}}\,\,\sqrt{G(x)}
\over
\sqrt{1-e^{{1\over 2}\,(\phi(x))-\phi_*)}\,\,\big({h(x)\over h_*}\big)^{1\over 2}}}\,\approx\,
e^{-{\phi_{IR}\over 4}}\,L_{IR}\,\int_{x_*}^{x_{a}}\,
{dx\over x\sqrt{1-{x_*^2\over x^2}}}\,=\,\rc\rc
&&\qquad\qquad
=\,e^{-{\phi_{IR}\over 4}}\,L_{IR}\,\log\Big({x_a+\sqrt{x_a^2-x_*^2}\over x_*}\Big)
\,\approx\,e^{-{\phi_{IR}\over 4}}\,L_{IR}\,\log\big({2x_a\over x_*}\big)\,\,.
\eear
To evaluate the UV integral we proceed similarly to what we did for the integral for $l$ and, in this case,  we only keep the one inside the square root. Then:
\beq
\int_{x_{a}}^{x_{max}}\,dx
{e^{-{\phi(x)\over 4}}\,\big[h(x)\big]^{-{1\over 4}}\,\,\sqrt{G(x)}
\over
\sqrt{1-e^{{1\over 2}\,(\phi(x))-\phi_*)}\,\,\big({h(x)\over h_*}\big)^{1\over 2}}}\,\approx\,
\int_{x_{a}}^{x_{max}}\,dx\,
e^{-{\phi(x)\over 4}}\,\big[h(x)\big]^{-{1\over 4}}\,\,\sqrt{G(x)}\,\,.
\eeq
Therefore:
\beq
m\,{\cal L}_r\,\approx\,2\,\Delta_{IR}\log\big({2x_a\over x_*}\big)\,+\,2m\,
\int_{x_a}^{x_{max}}\,dx\,\,e^{-{\phi(x)\over 4}}\,\big[h(x)\big]^{-{1\over 4}}\,\,\sqrt{G(x)}\,
-\,2\,\Delta_{UV}\log\big({\kappa\,\sqrt{\lambda}\,x_{max}^{{1\over b}}\over L_0^2}\big)\,\,,\qquad
\eeq
where $\Delta_{IR}$ is the conformal dimension of the operator ${\cal O}$ in the IR conformal point, given by:
\beq
\Delta_{IR}\,=\,m\,L_{IR}\,e^{-{\phi_{IR}\over 4}}\,\,,
\eeq
which is just (\ref{Delta_IR}). 
Let us next define the following quantities:
\bear
&& {\cal I}_{IR}\,\equiv\,-2m\,\int_{x_a}^{1}\,dx\, 
e^{-{\phi(x)\over 4}}\,\big[h(x)\big]^{-{1\over 4}}\,\,\sqrt{G(x)}\,-\,2\, \Delta_{IR}\,\log x_a\,\,,\rc\rc
&& {\cal I}_{UV}\,\equiv\,-2m\,\int_{1}^{x_{max}}\,dx\, 
e^{-{\phi(x)\over 4}}\,\big[h(x)\big]^{-{1\over 4}}\,\,\sqrt{G(x)}\,+\,
{2\, \Delta_{UV}\over b}\,\log x_{max}\,\,.
\label{I_IR_UV_def}
\eear
Then, after using the relation (\ref{l_estimate_smallxs}) to eliminate $x_*$ in favor of $l$, we get:
\beq
m\,{\cal L}_r\,\approx\,\log (r_q\,l/\sqrt{\lambda})^{2\Delta_{IR}}\,+\,
\log\Big[\,\Big({2\gamma\over \hat\gamma-1}\,{\sqrt{\lambda}\over L^2_{IR}}\Big)^{2\Delta_{IR}}
\Big({\kappa\over L_0^2}\Big)^{-2\Delta_{UV}}\Big]\,-\, {\cal I}_{IR}\,-\, {\cal I}_{UV}\,\,.
\eeq
Then, it follows that the IR limit $r_q\, l\to \infty$ of the two-point correlator is as in (\ref{VEV_IR}),  where ${\cal N}$ is the normalization constant given by:
\beq
{\cal N}\,=\,{\Big({\kappa\,\sqrt{\lambda}\over L_0^2}\Big)^{2\Delta_{UV}}\over
\Big({2\gamma\over \hat\gamma-1}\,{\sqrt{\lambda}\over L^2_{IR}}\Big)^{2\Delta_{IR}}}\,\,
\exp\big[\, {\cal I}_{IR}\,+\,{\cal I}_{UV}\big]\,\,.
\label{Cal_N}
\eeq
It turns out that, in the expression of the integrals in (\ref{I_IR_UV_def}) we can take the limits $x_a\to 0$  (in ${\cal I}_{IR}$) and  $x_{max}\to\infty$ (in ${\cal I}_{UV}$). Actually, it can be easily proved that, after taking these limits,  ${\cal I}_{IR}$ and ${\cal I}_{UV}$ can be recast as:
\bear
&& {\cal I}_{IR}\,=\,2\Delta_{IR}\,\int_{0}^{1}\,
{dx\over x}\,\Bigg[\,1-\,e^{{\phi_{IR}-\phi(x)\over 4}}\,
\Big[ {h_{IR}(x)\over h(x)}\Big]^{{1\over 4}}\,
\sqrt{{G(x)\over G_{IR}(x)}}\,\Bigg]\,\,,\rc\rc
&& {\cal I}_{UV}\,=\,{2\Delta_{UV}\over b}\,\int_{1}^{\infty}\,
{dx\over x}\,\Bigg[\,1\,-\,e^{{\phi_{0}-\phi(x)\over 4}}\,
\Big[ {h_{UV}(x)\over h(x)}\Big]^{{1\over 4}}\,
\sqrt{{G(x)\over G_{UV}(x)}}\,\Bigg]\,\,.
\eear
Notice that the form of the correlator is consistent with the fact that the conformal symmetry is recovered in the IR limit $r_q\,l\to\infty$. The non-canonical normalization factor ${\cal N}$ is due to the fact that we chose to renormalize ${\cal L}$ in such a way that the correlator is canonically normalized in the opposite UV limit $r_q\,l\to 0$.

\vskip 1cm
\renewcommand{\theequation}{\rm{F}.\arabic{equation}}
\setcounter{equation}{0}
\medskip

\section{WKB mass levels}
\label{appendix_WKB_masses}

Consider the following differential equation for the function $R(x)$:
\beq
\partial_{x}\,\Big[\,P(x)\,\partial_{x} R\,\Big]\,+\,
m^2\,Q(x)\,R\,=\,0\,\,,
\label{second_order_ODE}
\eeq
where $x$ takes values in the range  $x_*\le x\le \infty$, 
$ m$ is the mass parameter and $P(x)$ and $Q(x)$  are two
arbitrary functions that are independent of $ m$. We will assume that
near $x\approx x_*,\infty$ these functions behave as:
\bear
&&P\approx P_1(x-x_*)^{s_1}\,\,,
\,\,\,\,\,\,\,\,\,\,\,\,\,\,
Q\approx Q_1(x-x_*)^{s_2}\,\,,
\,\,\,\,\,\,\,\,\,\,\,\,\,\,{\rm as}\,\,x\to x_*\,\,,\rc\rc
&&P\approx P_2\,x^{r_1}\,\,,
\,\,\,\,\,\,\,\,\,\,\,\,\,\,
\qquad\,\,\,\,\,
Q\approx Q_2\, x^{r_2}\,\,,
\,\,\,\,\,\,\,\,\,\,\,\,\,\,{\rm as}\,\,x\to \infty\,\,,
\eear
where $P_i$, $Q_i$,  $s_i$, and $r_i$ are constants. By a of suitable change of variables, the differential equation (\ref{second_order_ODE}) can be converted into a Schr\" odinger equation, which only admits a discrete set of normalizable solutions for a set of values of $m$ parameterized by a quantum number $n\ge 0$. The mass levels for large values of  $n$ can be evaluated in the WKB approximation \cite{RS}, with the result:
\beq
m_{WKB}\,=\,{\pi\over \xi}\,
\sqrt{(n+1)\,\Big(\,n+{|s_1-1|\over s_2-s_1+2}+{|r_1-1|\over r_1-r_2-2}\,\Big)}\,\,,
\eeq
where $\xi=\xi(x_*)$ is the integral:
\beq
\xi\,=\,\int_{x_*}^{\infty}\,dx\,\sqrt{Q(x)\over P(x)}\,\,.
\eeq
In our case, the functions $P(x)$ and $Q(x)$ which correspond to the fluctuation equation (\ref{meson_fluct_eq}) are:
\beq
P(x)\,=\,{h^{{1\over 4}}\,e^{2f-\phi}\over x}\,(x^2-x_*^2)\,\,,
\qquad\qquad
Q(x)\,=\,{h^{{5\over 4}}\,e^{2f+2g-\phi}\over x}\,\,,
\eeq
and it is  immediate to extract the exponents $s_1$, $s_2$, $r_1$, and $r_2$:
\beq
s_1\,=\,1\,\,,\qquad\qquad
s_2\,=\,0\,\,,\qquad\qquad
r_1\,=\,1+{1\over b}\,\,,\qquad\qquad
r_2\,=-1-{1\over b}\,\,.
\eeq
Using these values one immediately finds that the WKB mass levels are given by (\ref{WKB_masses}).

\subsection{Asymptotic spectra}

Let us now evaluate analytically the meson spectrum in the two limiting cases in which the sea quark mass $m_q=r_q/2\pi$ is small or large compared to the valence quark mass $\mu_q$. Notice that $e^{g}\propto m_q$ (see eqs. (\ref{background_functions_xle1}) and (\ref{background_functions_xge1})). Therefore, (\ref{muq}) can be regarded as giving the relation $\mu_q/m_q$ as a function of $x_*$. 

Let us consider first the case in which $m_q$ is small. Clearly, when $m_q\to 0$ for fixed $\mu_q$ one necessarily   must have $x_*$  large. Actually, integrating (\ref{muq}) by using the asymptotic expression of $g$ written in (\ref{g_f_UVexpansions_x}) at leading order, we get:
\beq
\mu_q\,\approx\,\kappa\,\,
m_q\,\,x_*^{{1\over b}}\,\,,
\eeq
where the constant  $\kappa$ has been defined in (\ref{kappa_def}). 
Thus, when $m_q\to 0$ for fixed $\mu_q$ the $x$ coordinate of the tip of the flavor brane increases as:
\beq
x_*\,\sim\,m_q^{-b}\,\,\,,
\qquad\qquad
(m_q\to 0)\,\,.
\eeq
Let us now evaluate $\xi(x_*)$  when $x_*$ is large. First of all we perform a change of variables in the integral (\ref{WKB_xi}) and rewrite $\xi(x_*)$ as:
\beq
\xi(x_*)\,=\,\int_{1}^{\infty}\,dz\,{e^{g(x_*\,z)}\sqrt{h(x_*\,z)}\over \sqrt{z^2-1}}\,\,.
\label{WKB_xi_asymp}
\eeq
If $x_*$ is large,  the argument of the functions $g$ and $h$ in (\ref{WKB_xi_asymp}) is large and we can use their UV asymptotic expressions (\ref{g_f_UVexpansions_x}) and (\ref{UVexpansion_h_phi}) to evaluate them. At leading order, we get:
\beq
e^{g(x_*\,z)}\approx {r_*\over b}\,z^{{1\over b}}\,\,,
\qquad\qquad
h(x_*\,z)\approx {L_0^4\over r_*^4}\, z^{-{4\over b}}\,\,,
\eeq
where $r_*$ is the value of the $r$ coordinate corresponding to $x=x_*$. Therefore, for $m_q\to 0$ for fixed $\mu_q$ the $\xi$ is approximately
\beq
\xi\,\approx\,{L_0^2\over b\,r_*}\,\int_{1}^{\infty}\,{dz\over z^{{1\over b}}\,
\sqrt{z^2-1}}\,=\,{L_0^2\over \,r_*}\,\sqrt{\pi}\,\,
{\Gamma\Big({2b+1\over 2b}\Big)\over 
\Gamma\Big({b+1\over 2b}\Big)}\,\,.
\eeq
Using this result in (\ref{WKB_masses}) we get the following mass spectrum in the UV limit:
\beq
m_{_{WKB}}^{^{(UV)}}\,=\,{\sqrt{\pi}\over \sqrt{2}}\,{r_*\over L_0^2}\,
{\Gamma\Big({b+1\over 2b}\Big)\over 
\Gamma\Big({2b+1\over 2b}\Big)}\,\,
\sqrt{(n+1)(2n+1)}\,\,,
\qquad\qquad (\mu_q/m_q\,\to \infty)\,\,.
\eeq
Let  us write this result in terms of physical quantities. Recall that $L_0$ and $r_*$ are related to the 't Hooft coupling $\lambda$ and to the valence quark mass $\mu_q$ as:
\beq
L_0^2\,=\,\pi\sqrt{2\lambda}\,\sigma\,\,,
\qquad\qquad
r_*\,=\,2\pi\alpha'\,\mu_q\,\,,
\eeq
where $\sigma$ is the screening function written in (\ref{screening-sigma}). 
Then, for  large  $\mu_q/m_q$, we have:
\beq
m_{_{WKB}}^{^{(UV)}}\,=\,
{\sqrt{\pi}\,\mu_q\over \sigma\,\sqrt{\lambda}}\,\,
{\Gamma\Big({b+1\over 2b}\Big)
\over 
\Gamma\Big({2b+1\over 2b}\Big)}\,\,
\sqrt{(n+1)(2n+1)}\,\,,
\qquad\qquad (\mu_q/m_q\,\to \infty)\,\,,
\label{m_WKB_UV}
\eeq
where we have taken $\alpha'=1$. Eq. (\ref{m_WKB_UV}) is exactly the WKB mass spectrum one gets for vector mesons in the massless  flavored background of ref. 
\cite{Conde:2011sw}.

Let us next consider the limit in which $\mu_q/m_q$ is small.  Since $e^g\sim m_q$, it follows from (\ref{muq}) that $x_*$ must be small. Actually, we can estimate the relation  between $\mu_q/m_q$ and $x_*$ by extracting from (\ref{functions_running_x}) and (\ref{background_functions_xle1}) the approximate expression of $e^{g}$ near $x=0$. We get:
\beq
e^{g}\approx {\gamma\over c}\,x\,\approx\,\pi\, (\hat\gamma+1)\,m_q\,x\,\,.
\eeq
Then, we have:
\beq
\mu_q\approx {\hat\gamma+1\over 2}\,m_q\,x_*\,\,,
\eeq
and it follows that, for fixed $\mu_q$ and large $m_q$, the coordinate of the tip of the flavor brane behaves as:
\beq
x_*\,\sim\,{1\over m_q}\,\,,
\qquad\qquad
(m_q\to\infty)\,\,.
\eeq
Thus, for large sea quark mass $x_*\to 0$ and thus 
the dynamics of the fluctuating flavor brane is dominated by the IR, where the solution corresponds to the running solution of the unflavored system. In this case, we have at leading order near $x\sim 0$:
\beq
e^{g(x)}\approx {\gamma\over c}\,x\,\approx\,r_*\,{x\over x_*}\,\,,
\qquad\qquad
h(x)\,\approx\,{2\pi^2 \,N\over k}\,{c^4\over \gamma^4}\,{1\over x^4}\,\approx\,
{2\pi^2 \,N\over k}\,{1\over r_*^4}\,\,\Big({x_*\over x}\Big)^4\,\,,
\eeq
where, in the last step, we used that $c/\gamma\approx x_{*}/r_*$. Thus,
\beq
e^{g(x_*\,z)}\sqrt{h(x_*\,z)}\approx {\sqrt{2\pi^2\lambda}\over r_*}\,{1\over z}
\,\,,
\eeq
and, after evaluating the integral (\ref{WKB_xi_asymp}) and writing the result in terms of the 't Hooft coupling $\lambda$  and the valence quark mass $\mu_q$ ,  we have:
\beq
m_{_{WKB}}^{^{(IR)}}\,=\,
{2\mu_q\over \sqrt{\lambda}}\,
\sqrt{(n+1)(2n+1)}\,\,,
\qquad\qquad (\mu_q/m_q\,\to 0)\,\,,
\label{m_WKB_IR}
\eeq
which is exactly the mass spectrum of vector mesons in the unflavored ABJM model \cite{Jensen:2010vx}. By combining (\ref{m_WKB_UV}) and (\ref{m_WKB_IR}) we obtain the UV/IR mass relation written in (\ref{UV-IR-mass_ratio}). It is now straightforward to obtain  this ratio  for small and large values of the deformation parameter. Indeed, for small  $\hat\epsilon$ one can expand $m_{_{WKB}}^{^{(UV)}}/ m_{_{WKB}}^{^{(IR)}}$ as:
\beq
{m_{_{WKB}}^{^{(UV)}}\over m_{_{WKB}}^{^{(IR)}}}\,=\,1\,+\,
{3-\log 2\over 4}\,\,\hat\epsilon\,+\,{1\over 384}\,
\Big[12\Big(\log 2\big(\log 2+3\big)\,-\,3\Big)-\pi^2\Big]\,\hat\epsilon^2\,+\,\cdots\,\,.
\label{UV/IR_mass_ratio_lowNf}
\eeq
Moreover, in the opposite    limit in which $\hat\epsilon$ is very large, we have:
\beq
{m_{_{WKB}}^{^{(UV)}}\over m_{_{WKB}}^{^{(IR)}}}\,\approx\, 
{8\sqrt{2\pi}\over 5\sqrt{15}}\,
{\Gamma\Big({9\over 10}\Big)
\over 
\Gamma\Big({7\over 5}\Big)}\,\,\sqrt{\hat \epsilon}\,\,.
\label{UV/IR_mass_ratio_largeNf}
\eeq


\end{document}